\documentclass[10pt]{article}

\def \ri{\right}
\def\ZC{{\cal Z}_C}

\def\slint{\setminus \hspace{-9.5pt} \int }

\def\ds{\displaystyle}

\def\F{{\cal F}}
\def\tF{{\tilde{\cal F}}}

\def \be{\begin{equation}}
\def \ee{\end{equation}}
\def \bea{\begin{eqnarray}}
\def \eea{\end{eqnarray}}
\def\A{{\cal A}}
\def\B{{\cal B}}
\def\x{{x}}
\def\X{{\mathcal{X}}}
\def\pa{\partial}
\def\calC{{\mathcal{C}}}
\def\calM{{\mathcal{M}}}

\def\th{{\theta}}
\newcommand{\Remark}[1]{{\noindent {\footnotesize {\bf Remark:} #1}}}

\usepackage{wrapfig,cancel}
\usepackage{amsmath}%
\usepackage{amsfonts}%
\usepackage{amssymb}%
\usepackage[dvips]{graphicx}
\usepackage[usenames]{color}
\numberwithin{equation}{section}
\newcommand \mathbox[1]{
	\vbox{\hrule
		\hbox{\vrule\kern8pt
			\vbox{\kern8pt 
				\hbox{
					$\displaystyle #1$
				}\kern8pt
			}\kern8pt\vrule
		}\hrule
	}
}
\def\omit#1{{}}
\newcommand{\tops}[2]{{#1}}
\definecolor{darkblue}{rgb}{0,0,.8}
	
\definecolor{red}{rgb}{1,0,0}
	
\def\BK#1{ \left\langle \rule{0pt}{10pt} #1 \right\rangle}

\newcommand{\Res}{\operatorname{Res}\displaylimits}

\def\e#1{\text{e}^{#1}}

\def\pd{\partial}
\def\tr#1{\text{tr}\left(#1\right)}
\def\Tr#1#2{{\text{tr}}_{{#1}}\left(#2\right)}
\def\det#1{\text{det}\left(#1\right)}
\def\d#1{\mathrm{d}{#1}}

\def\OO#1{{\cal{O}}\left(#1\right)}

\def \id {\mathbb{I}}

\def \A {{\cal{A}}}
\def \C {\mathbb{C}}

\def \R {\mathbb{R}}

\def \W {{\cal W}}
\def \w {{\omega}}
\def \X {{\cal{X}}}
\def \Z {{\cal Z}}

\def \ep {\epsilon}
\def \and {\text{and }}

\begin{document}
\baselineskip 15pt plus 1pt minus 1pt

\vspace{0.2cm}
\begin{center}
\begin{Large}
\fontfamily{cmss}
\fontsize{17pt}{27pt}
\selectfont
\textbf{New recursive residue formulas for the\\ \vspace*{3pt}
 topological expansion of the Cauchy Matrix Model}
\end{Large}\\
\bigskip
\begin{large} {
A. Prats Ferrer$^{\sharp}$\footnote{pratsferrer@crm.umontreal.ca}}
\end{large}
\\
\bigskip
\begin{small}
$^{\ddagger}$ {\em Department of Mathematics and
Statistics, Concordia University\\ 1455 de Maisonneuve W., Montr\'eal, Qu\'ebec,
Canada H3G 1M8} \\ $^{\sharp}$ {\em Centre de recherches math\'ematiques, Universit\'e\ de Montr\'eal\\ 2920 Chemin de la tour, Montr\'eal, Qu\'ebec, Canada H3T 1J4.}
\end{small}

\bigskip
{\bf Abstract}
\end{center}
In a recent work \cite{BePr-09.1} we consider the topological expansion for the \emph{non-mixed} observables (including the free energy) for the formal Cauchy matrix model. The only restriction in \cite{BePr-09.1} was the fact that all the branch points have to be simple. This excludes a very interesting situation not encountered in the literature before, namely the case in which two branch points merge in such a way that no cycle is pinched.
In this work we focus on this situation and derive new formulas that apply to non-simple and non-singular branch-points.

\vspace{0.7cm}
\tableofcontents

\section{Introduction}

The fat-graph interpretation of matrix models has attracted a lot of attention since it was first noticed (see \cite{diFGiZiJ-95.1,BrItPaZu-78.1,Da-85.1,KaKoMi-85.1} for a short selection of early important works). The Feynman graph expansion of matrix models produces graphs with certain characteristics depending on the model (like orientability or colorability of the graph), and they are weighted in such a way that one can organize the expansion in terms of many of its components and properties. The graphs are naturaly organized by the Euler characteristics of the Riemann surface on which it can be embedded. 
The free energy (the logarithm of the partition function) then is a generating function for the counting numbers of connected graphs that can be embedded in a compact surface with a given genus.
In fact any observable of a matrix model has an interpretation in terms of combinatorics of graphs, for example the one-point resolvent generates counting numbers of connected graphs that can be embedded in a compact surface of a given genus with one border to which a given number of edges of the graph are connected.

The topological expansion is an immediate consequence of the Feynman graph interpretation of the matrix integral, and so it is inevitably connected with the counting of graphs. That is one of the reasons why the topological expansion of matrix models has been so much investigated.

In the past years a lot of progress has been achieved (see \cite{Ey-04.1,ChEyOr-06.1,EyPr-08.1,ChEy-06.2,BePr-09.1} and references therein), and the full topological expansion for a considerably large amount of models has been found. The whole scheme has been translated to a framework without matrix models as well \cite{EyOr-07.1} giving it an identity of its own and an interest beyond matrix models. This evolution has produced noticeable results in fields like String Theory \cite{eynard-2008,EyMaOr-07.1} and algebraic geometry \cite{EyOr-07.2,Ey-07.1,eynard-2007,EyOr-08.2}. 

The whole construction is based on an algebraic curve (called spectral curve whenever a matrix model is related to it) under the generic assumption that all branch-points are simple. This constraint is sufficient to ensure that the algebraic curve is not singular, a case where the whole topological expansion stops making sense. However, it is not a necessary constraint: non-simple branch-points can appear without making the algebraic curve singular.

In this work we find the first example of such a situation in the case of the Cauchy matrix model introduced in \cite{BeGeSz-08.1} and explored further in \cite{BeGeSz-08.2}. In \cite{BePr-09.1} it was studied as a formal model.
The Cauchy matrix model with partition function 
\begin{equation}
	\ZC=\int_{M_i=M_i^\dag>0}\hspace{-15pt}\d{M_1} \d{M_2} \frac{\e{-\frac{N}{T}
	\tr{V_1(M_1)+V_2(M_2)}}}{\det{M_1+M_2}^N}=\e{-\frac{N^2}{T^2} \F}
\end{equation}
consists of two positive definite hermitian matrices with external potentials $V_1$ and $V_2$ and an attractive interaction with each other determined by the determinant in the denominator. The eigenvalue density distributions in encoded in the cuts of a three sheeted covering of $\C\cup\{\infty\}$ (see figure \ref{fig:3sheet}). In particular, the eigenvalue density distribution for the first matrix $\rho_1(x)$ is encoded in the branch cuts joining the sheets $\X_1$ and $\X_0$ along pieces of $\R_+$ while the eigenvalue distribution for the second matrix $\rho_2(x)$ is encoded in the branch cuts joining the sheets $\X_2$ and $\X_0$ along pieces of $\R_-$.
\begin{figure}[ht!]%
\centering
\resizebox{0.6\textwidth}{!}{\input{SheetedCover-1.pstex_t}}
\caption{Three sheeted covering of $\C\cup\{\infty\}$ related to the Cauchy matrix model. In blue the cuts corresponding to the first matrix eigenvalue density distribution, in red the cuts corresponding to the one for the second matrix.}%
\label{fig:3sheet}%
\end{figure}

In \cite{BePr-09.1} the authors developed the topological expansion for the Cauchy matrix model with the condition that neither of the eigenvalue density distributions met the hard edge $x=0$. This constraint allowed the authors to remain within the framework of topological expansion that has become standard until now. In fact, the only situation in which this standard framework does not work for the Cauchy Matrix Model is when the density distributions of both matrices meet the hard edge at the origin in which case the branch point $x=0$ is no longer simple. The resulting algebraic curve in not singular due to the fact that the merging branch points connect different pairs of sheets, in other words, no cycle is pinched due to this branch point coalescence.

The main result of this work is the extension of the Eynard-Orantin residue formulas for the topological expansion to the situation we have described where the spectral curve contains one non-singular branch-point with branching number $n_b=2$, i.e. the simpler non-simple branch-point. 
The non-singularity of the algebraic curve is consequence of the fact that no cycle degenerates when the simple branch-points merge at the hard edge to form the multiple branch-point, in other words the algebraic curve does not experience any critical topological change in the formation of the multiple branch-point.
In terms of the recursive residue formulas for the topological expansion, the addition of a new four-legged vertex together with the standard three-legged one is the most remarkable result. Graphically, the recursion takes the form
\begin{equation}
	\begin{split}
		\begin{array}{l}
			\resizebox{0.2\textwidth}{!}{\input{wk+1h.pstex_t}}
		\end{array}&=
		\begin{array}{l}
			\resizebox{0.2\textwidth}{!}{\input{ww2hk.pstex_t}}
		\end{array}+
		\sum_{{\bf J}\subset {\bf K}}\sum_{m=0}^{h-1}
		\begin{array}{l}
			\resizebox{0.2\textwidth}{!}{\input{w1hw1hk.pstex_t}}
		\end{array}\\
		&+
		\begin{array}{l}
			\resizebox{0.2\textwidth}{!}{\input{ww3hk.pstex_t}}
		\end{array}
		+\sum_{{\bf J_1},{\bf J_2}\subset {\bf K}}\sum_{m=0}^{h-2}\sum_{n=0}^{h-m-1}
		\begin{array}{l}
			\resizebox{0.3\textwidth}{!}{\input{w1hw1hw1hk.pstex_t}}
		\end{array}
		\\+\sum_{{\bf J}\subset {\bf K}}\sum_{m=0}^{h-1}&\left(
		\begin{array}{l}
			\resizebox{0.2\textwidth}{!}{\input{w1hw2hpk.pstex_t}}
		\end{array}+
		\begin{array}{l}
			\resizebox{0.2\textwidth}{!}{\input{w1hw2hqk.pstex_t}}
		\end{array}+
		\begin{array}{l}
			\resizebox{0.2\textwidth}{!}{\input{w1hw2hrk.pstex_t}}
		\end{array}\right),
	\end{split}
	\nonumber
\end{equation}
that we prove in section \ref{Sec:ALOQ} and express graphicaly in \ref{Sec:DRwkh}. This recursion equation gives us a glimpse of a larger structure for higher order non-singular multiple branch-points that we leave for a future work.

The paper will be organized as follows:
\begin{itemize}
	\item {In section \ref{Sec:M} we introduce the model and explain very basic facts that are already known in 
	the literature.}
	\item{In section \ref{Sec:LESC} we introduce the two master loop equations for this model. One of them contains all the information about the algebraic curve of the model. We explain the structure of the spectral curve and introduce the necessary definitions. We also study how the curve is modified when we move in the moduli space. Finally we extract from the loop equations the recursion equations that is the starting point of our solution.}
	\item{In section \ref{Sec:SolRR} we solve the recursion relations found in the previous section by using residue formulas and the Abel differentials on the spectral curve.  Then we find how the variations on the curve affect the observables of the model.}
	\item{In section \ref{TEFE} the topological expansion of the free energy of the model we found by inverting the loop insertion operator.}
	\item{Appendix \ref{App:LE&D} gives some extra information about the meaning and how to derive the loop equations found earlier in the paper.}
	\item{Appendix \ref{App:AC} summarizes the definitions and properties of all the algebro-geometric objects we need in the topological expansion construction.}
	\item{Appendix \ref{app:Hrules} contains the detailed account of some of the more tedious calculations.}
\end{itemize}

\section{The Cauchy Matrix Model}\label{Sec:M}

The model of interest here is the Cauchy Matrix Model introduced in \cite{BeGeSz-08.1,BeGeSz-08.2} and whose topological expansion was found in \cite{BePr-09.1} for all situations except a small subset.
The partition function for this model is
\begin{equation}
	\ZC=\int_{M_i=M_i^\dag>0}\hspace{-15pt}\d{M_1} \d{M_2} \frac{\e{-\frac{N}{T}
	\tr{V_1(M_1)+V_2(M_2)}}}{\det{M_1+M_2}^N}=\e{-\frac{N^2}{T^2} \F},
	\label{eq:CMM}
\end{equation}
where the domain of integration (as indicated) is the ensemble of $N\times N$ Hermitian {\em positive definite} matrices. 
Consider the matrices $M_j$ as random variables with the probability measure 
\begin{equation}
{\rm d}P(M_1,M_2)=\frac 1{\ZC }\d{M_1} \d{M_2} \frac{\e{-\frac{N}{T}
	\tr{V_1(M_1)+V_2(M_2)}}}{\det{M_1+M_2}^N}.
\end{equation}
The {\it potentials} $V_i$ parametrize the measure associated to each matrix, and must satisfy a suitable growth condition at infinity  for the integral to be convergent; this condition is usually expressed as $\displaystyle V_i(x)/\ln (x) \mathop{\longrightarrow}_{x\to+\infty}+\infty$. However, we are interested in the formal expansion of the integral (the topological expansion) so this condition can be disregarded completely.

We restrict this paper to the class of polynomial potentials of degree $d_i$
\begin{equation}
	V_i(x)=\sum_{k=1}^{d_i} t_k^{(i)} x^k\ ,\ \ t_{d_i}^{(i)} >0,\ \ i=1,2,
\end{equation} 
although later we allow {\it formal} variations with respect to the infinite number of parameters $t_k^{(i)}$.
The logarithm of the partition function $\Z$ is called the free energy $\F$ of the matrix model and the parameter $T$ is called the total charge of the model. 

With the Harnad-Orlov formula \cite{HaOr-06.1} and following \cite{BeGeSz-08.2} we can write \eqref{eq:CMM} in eigenvalue representation 
\begin{equation}
	\ZC=\int_{\R_+^N} \left(\prod_{i=1}^N\d{x_i} \d{y_i}\right)\Delta^2(x_i)\Delta^2(y_i) 
	\frac{\e{-\frac{N}{T}\sum_{i=1}^N \left(V_1(x_i)+V_2(y_i)\right)}}
	{\prod_{i,j=1}^N(x_i+y_j)},
	\label{eq:evCMM}
\end{equation}
where $x_i$ (resp. $y_i$) are the (positive) eigenvalues of $M_1$ (resp. $M_2$), and $\Delta(a_i)=\det{a_i^{j-1}}_{1\le i,j\le N}=\prod_{i<j}^N(a_i-a_j)$ is the Vandermonde determinants. 

In the large $N$ limit the eigenvalues distribute along segments of the real positive line with eigenvalue distributions
\begin{equation}
	\rho_1(x)=\lim_{N\to\infty}\frac{1}{N}\sum_{i=1}^N \delta(x-x_i) \quad\,,\quad \qquad \rho_2(y)=\lim_{N\to\infty}\frac{1}{N}\sum_{i=1}^N \delta(y-y_i).
\end{equation}
In the generic case the eigenvalue distributions may reach the hard edge at $x=0$. This implies a generic behavior $\rho_k(x)\sim x^{-\frac{1}{2}}$. When both eigenvalue distributions reach the hard edge, the attraction to the eigenvalues of the other matrix modifies this law to $\rho_k(x)\sim x^{-\frac{2}{3}}$.

Note that we can easily rewrite equation \eqref{eq:evCMM} back into matrix representation with a different (but equivalent as long as we look at non-mixed correlators) interaction, namely
\begin{equation}
	\ZC=\int_{M_i>0}\hspace{-15pt}\d{M_1} \d{M_2} 
	\frac{\e{-\frac{N}{T}\tr{V_1(M_1)+V_2(M_2)}}}
	{\det{M_1\otimes\id+\id\otimes M_2}},
	\label{eq:tpCMM}
\end{equation}
where now the determinant is acting on a $N^2\times N^2$ size matrix, and $\id$ represents the $N\times N$ identity matrix.
Remark that the exponent $N$ in the determinant in \eqref{eq:CMM} is not present in \eqref{eq:tpCMM}.

We will consider here the formal perturbative expansion of the matrix integral around one local minimum generically constituted of a finite number of disconnected supports for the eigenvalue distributions of both $M_1$ and $M_2$\footnote{Note that we don't ask it to be the absolute minimum}. We don't extend here defining precisely the formal perturbative expansion of matrix models since it has already been defined in great generality elsewhere --see \cite{Ey-06.1,Or-07.1} for more details--. For the sake of completeness we reproduce the main idea of the formal matrix model definitions in appendix \ref{App:FMM}.

The goal of the this paper is to extend the results of \cite{BePr-09.1} to the case where the supports of the eigenvalue densities for the two matrices meet the hard edge at the origin.

\section{Loop Equations, spectral curve and recurrence equations}\label{Sec:LESC}

The two loop equations essential for the Cauchy matrix model have been derived in \cite{BePr-09.1}. We have reproduced the main steps in appendix \ref{App:LE&D}. Following the indications in the appendix and in \cite{BePr-09.1} the reader can reproduce these loop equations.

\subsection{Definitions}

In order to write the loop equations in a manageable form we need some definitions. Let us define the matrix average notation  
\begin{equation}
	\BK{(\cdots)}=\frac{1}{\ZC}\int_{M_i>0}\hspace{-15pt}\d{M_1} \d{M_2} \,(\cdots)\, 
	\frac{\e{-\frac{N}{T}\tr{V_1(M_1)+V_2(M_2)}}}{\det{M_1\otimes\id+\id\otimes M_2}}.
\end{equation}
We also define the one point resolvent functions, the $U$-potentials and the $Y$-functions, the main objects of the next section
\begin{equation}
\begin{split}
	W_1(x)&=\frac{T}{N}\tr{\frac{1}{x-M_1}},\\
	Y_1(\x)&=U^\prime_1(\x)-W_1(\x)\,,\quad U^\prime_1(\x)=\frac{2V_1^\prime(\x)-V_2^\prime(-\x)}{3},\\
	{W}_2(x)&=-\frac{T}{N}\tr{\frac{1}{x+M_2}},\\
	{Y}_2(\x)&=U^\prime_2(\x)-{W}_2(\x)\,,\quad U^\prime_2(\x)=\frac{-V_1^\prime(\x)+2V_2^\prime(-\x)}{3},\\
	W_0(\x)&=-W_1(\x)-{W}_2(\x),\\
	Y_0(\x)&=-Y_1(\x)-{Y}_2(\x)=U^\prime_0(\x)-W_0(\x),
\end{split}	
\label{eq:defW}
\end{equation}
where $U^\prime_0$ is defined implicitely through the definitions of $Y_1$, $Y_2$, $U^\prime_1$ and $U^\prime_2$.
The one point resolvent functions are related to the eigenvalue density distribution by\footnote{Note that for $W_2(x)$ is defined in such a way that the resolvent function's poles are all on the negative axis and their residue is negative. This explains the signs in the definitions of $\rho_2$.}
\begin{equation}
\begin{split}
 2 \pi i \rho_1(x) &= \lim_{\epsilon\to 0}\lim_{N\to\infty} \BK{W_1(x-i \epsilon)}-\BK{W_1(x+i \epsilon)},\\
 2 \pi i \rho_2(-x) &= -\lim_{\epsilon\to 0}\lim_{N\to\infty} \BK{W_2(x-i \epsilon)}-\BK{W_2(x+i \epsilon)}.
\end{split}
\end{equation}

\subsection{The loop equations}

The first of the two master loop equations (whose derivation is more explicit in appendix \ref{App:LE&D}) is quadratic in both $Y_1$ and $Y_2$ and has the form
\begin{equation}
	\BK{(Y_1(x))^2}+\BK{(Y_2(x))^2}+\BK{Y_1(x)Y_2(x)} = \BK{R(x)},
\label{eq:QLE.1}
\end{equation}
where $\BK{R(x)}$ is an unknown meromorphic function of $x$ with a pole of order $1$ at most at $x=0$ and a polynomial part, as can be seen from the explicit expression (see appendix \ref{App:LE&D})
\begin{equation}
\begin{split}
	R(\x)&=\frac{1}{3}(V_1^\prime(\x)^2+V_2^\prime(-\x)^2-V_1^\prime(\x)V_2^\prime(-\x))
	-P_1(\x)-{P}_2(\x)-\frac{1}{x}\BK{I_1+I_2}\\
	&=(U^\prime_1(\x))^2+(U^\prime_2(\x))^2+U^\prime_1(\x)U^\prime_2(\x)-P_1(\x)-{P}_2(\x) 	
	-\frac{1}{x}\BK{I_1+I_2},
\end{split}
	\label{eq:RDef}
\end{equation}
where the polynomials $\BK{P_1(x)}$ and $\BK{P_2(x)}$ as well as the constants $\BK{I_1}$ and $\BK{I_2}$ are defined in terms of matrix averages in appendix \ref{App:LE&D}.
For future purposes we rewrite equation \eqref{eq:QLE.1} using the definition of $Y_0(x)$
\begin{equation}
	\sum_{k=0}^2\BK{Y_k(x)^2}=2\BK{R(x)}.
	\label{eq:QLE}
\end{equation}
The second of the two master loop equations is a cubic equation that both $Y_1$, $Y_2$ and $Y_0$ satisfy identically  
\begin{equation}
\begin{split}
	\BK{Y_k(\x)^3}-\BK{R(\x)Y_k(\x)}
	-{\frac{T^2}{N^2}}\left(\frac{1}{2}\frac{\d{}^2}{\d{\x}^2}+\frac{1}{\x}\frac{\d{}}{\d{\x}}\right)
	\BK{W_k(\x)}&=\BK{D(\x)}\,,\quad \text{for $k=0,1,2$},
\end{split}
	\label{eq:CLE.a}
\end{equation}
where $\BK{D(x)}$ is a meromorphic function of $x$ with a pole of order at most $2$ at $x=0$ and a polynomial part and can be expressed as
\begin{equation}
\begin{split}
	\BK{D(\x)}& 
	=-U^\prime_0(\x)U^\prime_1(\x)U^\prime_2(\x)-U^\prime_1(\x)\BK{{P}_2(\x)}-U^\prime_2(\x)\BK{P_1(\x)}
	+\BK{Q_1(\x)+{Q}_2(\x)}\\
	&\qquad +\frac{1}{\x}\left(\BK{K_1}-U^\prime_1(\x)\BK{{I}_2}+U^\prime_2(x)\BK{I_1}\right)+\frac{1}{x^2}\BK{K_2},
\end{split}
	\label{eq:DDef}
\end{equation}
where the polynomial $\BK{Q_1(x)}$ and $\BK{Q_2(x)}$ and the constants $\BK{K_1}$ and $\BK{K_2}$ are defined in terms of matrix averages in appendix \ref{App:LE&D}.
Equation \eqref{eq:CLE.a} is essential to obtain the spectral curve that determines all the properties of the model.
If we combine the cubic equation for all three $Y_k$ we obtain
\begin{equation}
	\sum_{k=0}^2\BK{Y_k(x)^3}=3\BK{D(x)}.
	\label{eq:CLE.b}
\end{equation}
This equation together with equation \eqref{eq:QLE} is all we need to produce the recursion equations that allow us to compute the topological expansion.

\subsection{Spectral curve}

From equation \eqref{eq:CLE.a} we can extract the spectral curve that defines the leading order of $\BK{Y_k(x)}$.
Consider the leading terms in the topological expansion of the objects in \eqref{eq:CLE.a}
\begin{equation}
 \begin{split}
  \BK{Y_k(x)}&=y_k(x)+{\cal O}\left(\frac{T^2}{N^2}\right),\\
  \BK{{R}(x)}&=R^{(0)}(x)+{\cal O}\left(\frac{T^2}{N^2}\right),\\
  \BK{D(x)}&=D^{(0)}(x)+{\cal O}\left(\frac{T^2}{N^2}\right),
 \end{split}
\end{equation}
and extract the leading part of the equation \eqref{eq:CLE.a}
\begin{equation}
 \begin{split}
  y_k(x)^3-R^{(0)}(x) y_k(x) = D^{(0)}(x) \,\ \ \text{for $i=0,1,2$}.
 \end{split}
\end{equation}
The three functions $y_k(x)$ are the three solutions to the algebraic curve $E(x,y)=y^3-R^{(0)}(x) y -D^{(0)} =0$. This algebraic curve is called the spectral curve of the matrix model.
The algebraic curve can be parametrized by a Riemann Surface $\Sigma$. For every solution of the equation $E(x,y)=0$ there is a point $p\in\Sigma$, thus there is a parameterization $x(p)$ and $y(p)$ for the ensemble of solutions. Since there are three possible solutions of the algebraic curve for a given $x$, the curve can be represented as a three sheeted covering of $\C\cup\{\infty\}$. In other words, for a generic $x$, there are three and only three different points $p,q,r\in\Sigma$ with the same $x$-projection $x(p)=x(q)=x(r)=x$, and whose $y$-projections are $y_k(x),\,k=0,1,2$. Reversing this argument we can construct the meromorphic function on the algebraic curve
\begin{equation}
 \begin{split}
  y(p)=y_k(x) \,,\quad \text{whenever $p\in\X_k$},
 \end{split}
\end{equation}
where $\X_k$ is the $x$-based sheet of the algebraic curve defined by $y_k(x)$. 
From the definitions \eqref{eq:defW} we can see that $y(p)$ have cuts joining $\X_1$ and $\X_0$ along the positive real line $\R_+$ (those coming from the eigenvalue distribution of $M_1$), and cuts joining $\X_2$ and $\X_0$ along the negative real line $\R_-$ like represented in figure \ref{fig:3sheet}. We call\footnote{Note that though the maximum genus is reached with $d_1$ cuts $A_i^{(1)}$ and $d_2$ cuts $A_i^{(2)}$, we allow a smaller number of cuts.} $A_i^{(1)},\,i=0,\dots,\bar{d}_1-1$ the cuts corresponding to the eigenvalue distributions of $M_1$, and we call $A_i^{(2)},\,i=0,\dots,\bar{d}_2-1$, those corresponding to the eigenvalue distributions of $M_2$ with $\bar{d}_k\leq d_k$. We call $\alpha_i^{(k)}$ the branch-points and $a_i^{(k)}=x(\alpha_i^{(k)})$ its $x$-projection, so that $A_i^{(k)}=[a_{2i}^{(k)},a_{2i+1}^{(k)}]$ is the support for the cuts.
\begin{figure}[ht!]%
\centering
\resizebox{0.6\textwidth}{!}{\input{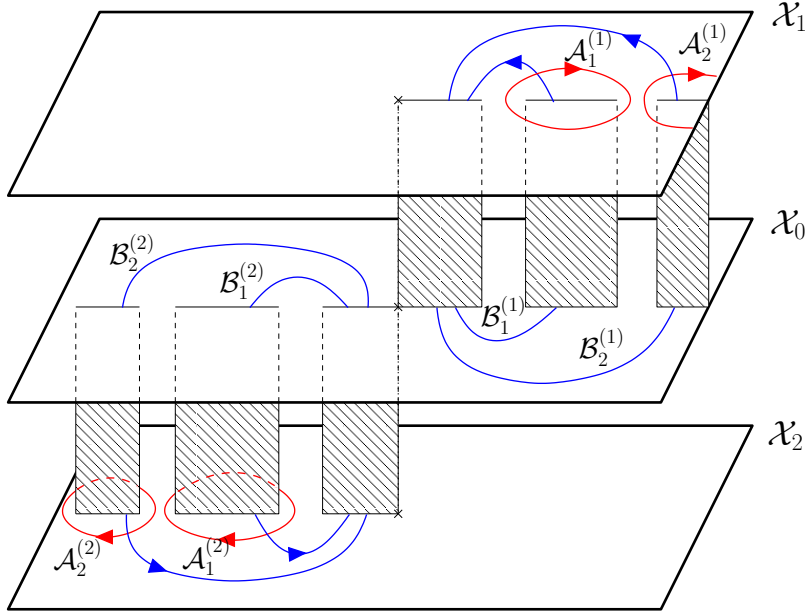}}
\caption{Three sheeted covering of $\C\cup\{\infty\}$ with a branch point at $x=0$ with branching number $2$. Red cycles are the $\A$ cycles, blue cycles are the $\B$ cycles.}%
\label{fig:3sheet.a}%
\end{figure}

Denote by $\A_i^{(k)},\,i=0,\dots,\bar{d}_k-1,\,k=1,2$ small cycles on $\X_k$ containing $A_i^{(k)}$, and $\B_i^{(k)}$ cycles that go through $A_0^{(k)}$ and $A_i^{(k)}$ and whose $x$-projection encircles the branch-points $a_1^{(k)},\dots,a_{2i}^{(k)}$. The cycles $\A_i^{(k)},\B_i^{(k)}$ --for $i=1,\dots,\bar{d}_k-1$, and $k=1,2$-- form a canonical homology basis on the algebraic curve with the intersection property $\A_i^{(k)}\cap\B_j^{(l)}=\delta_{i,j}\delta_{k,l}$ (see fig.~\ref{fig:3sheet.a}).
We are interested in a situation where cuts of both species reach the hard edge at $x=0$ like in fig.~\ref{fig:3sheet.a}. It is clear from our definitions that the convergence of two simple branch-points and the corresponding formation of a branch-point with branching number $2$ (connecting all three sheets) do not imply the degeneration of a cycle or a change in the genus of the Riemann surface $g=\bar{d}_1+\bar{d}_2-2$.

To specify one minimum for the topological expansion it is sufficient to fix the number of eigenvalues that belong to each disconnected component of the eigenvalue support. This is called the fixed filling fraction condition. 
The expansion itself cannot modify the minimum originally chosen, or equivalently the filling fractions do not receive corrections in $N$.
The leading part of the fixed filling fraction condition can be written as
\begin{equation}
	\begin{split}
		\oint_{\A_i^{(k)}}y(p)\d x(p) &= 2 \pi i \epsilon_i^{(k)}\quad,\quad\qquad 
		k=1,2\,,\,i=1,\dots,\bar{d}_k-1,\\
	\end{split}
	\label{eq:LFFC}
\end{equation}
and the rest of the expansion of the fixed filling fraction condition is
\begin{equation}
	\begin{split}
		\oint_{\A_i^{(k)}}w_1^{(h)}(p) &=0\quad,\quad\qquad k=1,2\,,\,i=1,\dots,\bar{d}_k-1,\\
	\end{split}
\end{equation}
where $w_1^{(h)}(p)$ --that we define in more detail later-- is related to the $h$ order term in the topological expansion of $\BK{W_k(x)}$.
Although $R^{(0)}(x)$ and $D^{(0)}(x)$ are unknown a priori, choosing values for the filling fractions should amount to determine those two functions\footnote{We live it as an unproven statement here that the amount of unknown independent parameters in $R^{(0)}$ and $D^{(0)}$ are exactly equal to $d_1+d_2-2$ and that the conditions \eqref{eq:LFFC} plus the fact that the genus is $g=\bar{d}_1+\bar{d}_2-2$ are enough information to determine all these unknowns.}.

For convenience we denote by $p^{(i)}$ ($i=0,1,2$) the three different points of the curve with the same $x$-projection as $p$, i.e. $x(p^{(i)})=x(p)$ (obviously $p=p^{(i)}$ for some value of $i$.
We also denote the three points that project to $x=\infty$ by $\infty_i\in\X_i$ for $i=0,1,2$.\newline
Consider also automorphisms defined locally around a branch-point $\alpha$ by 
\begin{equation}
	\begin{split}
		\vartheta_\alpha^{i}:\quad \Sigma \quad &\longrightarrow \quad \Sigma\\
		p \quad &\longrightarrow \quad \vartheta_\alpha^{i}(p)
	\end{split}
\end{equation}
that results from moving the point $p$ around the branch point $\alpha$ in such a way that its $x$-projection turns counterclockwise exactly $i$ times around $a=x(\alpha)$ and such that $x(\vartheta_\alpha^i(p))=x(p)$. It is easy to see that $\vartheta_\alpha^2(p)=p$ if the branching number is $1$, $\vartheta_\alpha^3(p)=p$ if the branching number is $2$, and so on. We will drop the subindex $\alpha$ from now on unless the expression becomes ambiguous.

\subsubsection{Moduli of the model.}

The algebraic curve is parametrized by the data from the model
\begin{itemize}
\item{The potential parameters $t_j^{(k)}$;}
\item{The total charge $T$;}
\item{The filling fractions $\epsilon_i^{(k)}$ defined below in eq. (\ref{eq:FFFC}).}
\end{itemize}
We denote a generic setting of these parameters as $\calM=\{t_j^{(k)},T,\epsilon_i^{(k)}\}$.
We will be interested in computing variations (derivatives) with respect to these parameters. For the potential parameters, we introduce a {\it generating operator} of such variations called the loop insertion operator in the matrix model literature. These operators are very important as they define the one point resolvent when applied to the free energy, as well as the connected $n+1$-point resolvent when applied to the connected $n$-point resolvent. In our model we have two of them corresponding to the two sets of times $t_k^{(i)}$, $i=1,2$
\begin{equation}
	\begin{split}
		\frac{\pd{}}{\pd{}V_1(x)}&=-\sum_k \frac{1}{x^{1+k}}\frac{\pd{}}{\pd{}t_k^{(1)}},\\
		\frac{\pd{}}{\pd{}V_2(x)}&=-\sum_k \frac{1}{(-x)^{1+k}}\frac{\pd{}}{\pd{}t_k^{(2)}}.
	\end{split}
	\label{eq:LIO}
\end{equation}
With these two operators we will define the global loop insertion operator acting on functions on the algebraic curve
\begin{equation}
	\frac{\pd{}}{\pd{}V(p)}=\left\{
	\begin{array}{ll}
		\frac{\pd{}}{\pd{}V_1(x)} & \text{whenever } p\in\X_1\\
		-\frac{\pd{}}{\pd{}V_1(x)}-\frac{\pd{}}{\pd{}V_2(x)} & \text{whenever } p\in\X_0\\
		\frac{\pd{}}{\pd{}V_2(x)} & \text{whenever } p\in\X_2
	\end{array}  
	\right.
	\label{eq:LIOP}
\end{equation}
We define as well the function\footnote{Remark that from the above definitions it follows by a direct computation that 
\begin{equation}
	 -\frac{\pd}{\pd V(p)}U^\prime(p^\prime) = \frac{2}{3}\frac{1}{(x(p)-x(p^\prime))^2}.
\end{equation}
}
 $U^\prime(p)$ on $\Sigma$
\begin{equation}
	U^\prime(p)=U^\prime_k(x(p)) \qquad \text{whenever } p\in\X_k.
	\label{eq:defWU}
\end{equation}
The loop insertion operators \eqref{eq:LIO} can be used to generate all the connected resolvent observables from the free energy $\F$
\begin{equation}
		\BK{\prod_{k=1}^n W_{i_k}(x_k)}_c = -\left(\prod_{k=1}^n \frac{\partial}{\partial{V_{i_k}(x_k)}}\right)\F.
\end{equation}

\subsubsection{Fundamental differential of the algebraic curve}

The algebraic curve can be parametrized equally well by $\calM$ or by the meromorphic functions $y(p)$ and $x(p)$. In fact we will see that the differential $y(p)\d x(p)$ on $\Sigma$ contains the information in $\calM$.
Indeed,we can extract $t_k^{(j)}$ for $k>0$ from the behavior of $y(p)\d x(p)$ around\footnote{The behavior around the third infinity point $p=\infty_0$ is not independent due to the relation $y_1(x)+y_2(x)+y_3(x)=0$.} $\infty_{1,2}$
\begin{equation}
	\begin{split}
		\Res_{p\to\infty_1} \frac{y(p)\d x(p)}{(x(p))^{k}} &= \frac{2t_k^{(1)}-(-1)^k t_k^{(2)}}{3},\\
		\Res_{p\to\infty_2} \frac{y(p)\d x(p)}{(x(p))^{k}} &= \frac{-t_k^{(1)}+(-1)^k 2t_k^{(2)}}{3}.
	\end{split}
	\label{eq:PotPar}
\end{equation}
Also the filling fractions can be explicitly written as the $\A$ cycles of the fundamental differential $y(p)\d x(p)$ as we have already seen above
\begin{equation}
	\frac{1}{2\pi i}\oint_{\A_j^{(k)}} y(p) \d x(p) = 
	\epsilon_j^{(k)}=(-1)^{k}\frac{N_j^{(k)}T}{N},
	\label{eq:FFFC}
\end{equation}
where $N_j^{(k)}$ is the number of eigenvalues belonging to the cut $A_j^{(k)}$.
The sign in \eqref{eq:FFFC} appears due to the definition of $W_2(x)$.
The total charge $T$ can be represented in two different ways
\begin{equation}
	\begin{split}
		\Res_{p\to\infty_{1}} y(p)\d x(p)=&T,\\
		\Res_{p\to\infty_{2}} y(p)\d x(p)=&-T.
	\end{split}
\end{equation}
The same expression evaluated at $\infty_0$ reflects the condition $\sum_{i=0}^2 y(p^{(i)})=0$
\begin{equation}
	\begin{split}
		\Res_{p\to\infty_{0}} y(p)\d x(p)=&0.
	\end{split}
\end{equation}

\subsection{Variations of the spectral curve with respect to the moduli}
Consider the formulas from the previous section, it is clear that when we modify the moduli of the spectral curve, these changes must be reflected on the fundamental differential $y(p)\d x(p)$.
We are going to study how $y(p)\d x(p)$ changes under the variations of the moduli. 
Since we have the constraint $E(x,y)=0$, $x(p), y(p)$ and $p$ are not independent, we have to keep fixed one of the three while computing the variations. For convenience all variations are taken at $x(p)$ fixed. We will generically call this variations $\d\Omega$.
The variations will be written in terms of standard objects in algebraic geometry that we define in appendix \ref{App:AC}.

\subsubsection{Variation with respect to the filling fractions}
Not all filling fractions are independent due to the restrictions pointed out in the previous section.
As with $\A_j^{(k)}$ we consider as independent the $\epsilon_j^{(k)}$ for $j=1,\dotsc,d_k-1$. 
Since these are independent we have $\frac{\pd \epsilon_i^{(k)}}{\pd \epsilon_j^{(k^\prime)}}=\delta_{i,j}\delta_{k,k^\prime}$ for $i,j\not=0$.

Let us compute the variation of $y(p)\d x(p)$. We have that
\begin{equation}
	\frac{\pd}{\pd \epsilon_j^{(k)}} y(p)\d x(p) =\frac{\pd}{\pd \epsilon_j^{(k)}} (U^\prime(p)-W(p))\d x(p)= \OO{x(p)^{-2}}\d x(p)
	\label{eq:Ve1}
\end{equation}
when $x(p)\to\infty$. Indeed the potential $U^\prime(p)$ does not depend on the $\epsilon$'s by hypothesis, and $W(p)$ gives us the behavior at $\infty$.
On the other side we can compute the $\A$ cycles of the above differential
\begin{equation}
	\begin{split}
		\frac{1}{2\pi i}\oint_{\A_j^{(k)}} \frac{\pd}{\pd \epsilon_{j^\prime}^{(k^\prime)}} y(p)\d x(p)&=
		\frac{\pd}{\pd \epsilon_{j^\prime}^{(k^\prime)}} \frac{1}{2\pi i}\oint_{\A_j^{(k)}} y(p)\d x(p)\\
		&=\frac{\pd \epsilon_j^{(k)}}{\pd \epsilon_{j^\prime}^{(k^\prime)}}= 
		\delta_{j,j^\prime}\delta_{k,k^\prime}.
	\end{split}
	\label{eq:Ve2}
\end{equation}
All together we have that $\frac{\pd}{\pd \epsilon_j^{(k)}} y(p)\d x(p)$ is a differential on the algebraic curve with no pole and with normalizing condition \eqref{eq:Ve2}, this determines $\d\Omega_{\epsilon_j^{(k)}}$ completely to be a basis of normalized first type Abelian differentials. They can be written in terms of the Bergman Kernel $B(p,q)$
\begin{equation}
	\d\Omega_{\epsilon_j^{(k)}}=\frac{\pd}{\pd \epsilon_j^{(k)}} y(p)\d x(p)
	= 2\pi i\d u_j^{(k)} = \oint_{\B_j^{(k)}} B(.,p) .
	\label{eq:Ve3}
\end{equation}

\subsubsection{Variation with respect to the total charge \tops{$T$}{T}}

Imagine now that we change the total charge $T$, the variation of $y(p)\d x(p)$ will be
\begin{equation}
	\frac{\pd}{\pd T} y(p)\d x(p) = \left\{
	\begin{array}{ll}
		\OO{x(p)^{-2}}\d x(p) & \text{ $p\in\X_0$}\\
		\left(\frac{-1}{x(p)}+\OO{x(p)^{-2}}\right)\d x(p) & \text{ $p\in\X_1$}\\
		\left(\frac{1}{x(p)}+\OO{x(p)^{-2}}\right)\d x(p) & \text{ $p\in\X_2$}
	\end{array}\right.
	\label{eq:VT1}
\end{equation}
Indeed by hypothesis the potentials do not depend on $T$, and the behavior at infinity of $W_k(x)$ gives us the above behavior.
This differential has poles only at $\infty_{1,2}$ of order $1$ with residues $+1$, $-1$ respectively.
Also the $\A$ cycles of this differential are zero because the filling fractions do not depend on $T$ either. 
The conclusion is that
\begin{equation}
	\d\Omega_T(p)=\frac{\pd}{\pd T} y(p)\d x(p) = \d S_{\infty_1,\infty_2}(p) = \int_{\infty_2}^{\infty_1} 
	B(. , p),
	\label{eq:VT2}
\end{equation}
which is a normalized third type Abelian differential.

\subsubsection{Variation with respect to the potentials}
Consider now the potentials. The variations with respect to the parameters of the potentials look like
\begin{equation}
	\begin{split}
		\d \Omega_{t_j^{(1)}} (p)\equiv
		\frac{\pd}{\pd t_j^{(1)}} y(p)\d x(p)&=\frac{\pd}{\pd t_j^{(1)}} 
		(U^\prime(p)+\OO{x(p)^{-1}})\d x(p)\\
		&=\left\{\begin{array}{ll}
			\frac{2}{3}j x(p)^{j-1}\d x(p)+\OO{x(p)^{-2}} &\text{for $p\in\X_1$}\\
			-\frac{1}{3}j x(p)^{j-1}\d x(p)+\OO{x(p)^{-2}} &\text{for $p\not\in\X_1$}
		\end{array}\right.\,,\\
		\d \Omega_{t_j^{(2)}} (p)\equiv
		\frac{\pd}{\pd t_j^{(2)}} y(p)\d x(p)&=\frac{\pd}{\pd t_j^{(2)}} 
		(U^\prime(p)+\OO{x(p)^{-1}})\d x(p)\\
		&=(-1)^{j-1}\left\{\begin{array}{ll}
			\frac{2}{3}j x(p)^{j-1}\d x(p)+\OO{x(p)^{-2}} &\text{for $p\in\X_2$}\\
			-\frac{1}{3}j x(p)^{j-1}\d x(p)+\OO{x(p)^{-2}} &\text{for $p\not\in\X_2$}
		\end{array}\right.\,.
	\end{split}
	\label{eq:DIF1}
\end{equation}
This behaviors at $\infty_k$ plus the normalizing conditions $\oint_{\A_i^{(k)}}\d\Omega_j^{(k^\prime)}=0$  determine completely $\d \Omega_j^{(k)}$ as a combination of second kind differentials
\begin{equation}
	\begin{split}
  \d \Omega_j^{(1)}(p) &= \frac{2}{3}\Res_{\infty_1} x(\cdot)^j B(\cdot,p)
  -\frac{1}{3}\Res_{\infty_2,\infty_0} x(\cdot)^j B(\cdot,p),\\
  \d \Omega_j^{(2)}(p) &= 
  (-1)^{j-1}\left(\frac{2}{3}\Res_{\infty_2} x(\cdot)^j B(\cdot,p)
  -\frac{1}{3}\Res_{\infty_1,\infty_0} x(\cdot)^j B(\cdot,p)\right).
	\end{split}
	\label{eq:VP1}
\end{equation}

\paragraph{Explicit vs implicit derivatives of $y(p)\d x(p)$.}
We claim that we can represent $y(p)\d x(p)$ as
\begin{equation}
	\begin{split}
	y(p) \d x(p)=& 2\pi i\sum_{k=1}^2\sum_{j=1}^{d_k-1}\epsilon_j^{(k)} \d u_i^{(k)}(p)
		+T \d S_{\infty_1,\infty_2}(p)+\sum_{k=1}^2\sum_{j}t_j^{(k)}\d\Omega_j^{(k)}(p)\\
		=& \sum_{t_a\in \mathcal{M}} t_a \d\Omega_{t_a}(p),
	\end{split}
	\label{eq:ydx}
\end{equation}
where $t_a$ is a collective notation for the whole set of parameters of our model $\mathcal{M}=\{t_j^{(k)},T,\epsilon_j^{(k)}\}$. To see this, we  provisionally denote by $\eta(p)$ the right-hand-side of (\ref{eq:ydx}); it is seen, from the definition of the various terms that $\eta(p) -y(p) \d x(p)$ has no poles at $\infty_0,\infty_1,\infty_2$ and hence it is a holomorphic differential. On the other hand, from the definition of the filling fractions $\epsilon_j^{(k)}$ it also appears that it has vanishing $\mathcal A$--cycles. Standard theorems then guarantee then that $\eta(p) -y(p)\d x(p)\equiv 0$. Therefore (\ref{eq:ydx}) yields an alternative representation of the fundamental differential $y(p)\d x(p)$. 

On the face of it  appears that $\partial_{t_a} y(p) \d x(p)$ is the same as the coefficient of the corresponding ``time'' $t_a$ in the r.h.s. of  (\ref{eq:ydx}), namely, the ``explicit'' derivative of the r.h.s. of (\ref{eq:ydx}). Note that this is not a trivial statement, since the spectral curve and all the differentials appearing as coefficients of the times in (\ref{eq:ydx}) {\em do depend} (implicitly) on the times as well.

Note that all variations $\d\Omega_{t_a}$ are expressible as an integral operator on the Bergman kernel $B(p,q)$
\begin{equation}
	\begin{split}
		\d \Omega_{t_a} &= \frac{\pd}{\pd t_a} y(p)\d x(p) = \int_{\calC_{t_a}} B(p,\cdot) \Lambda_{t_a}(\cdot),\\
	\end{split}
	\label{eq:dOBK} 
\end{equation} 
where $\calC_{t_a}$ is the integration path associated with the variation with respect to $t_a$ and $\Lambda_{t_a} (p)$ is a meromorphic function on the curve.
For example, looking at the formula above, we have $\calC_{t_j^{1}}=\frac{2}{3}\calC_{\infty_1}-\frac{1}{3}\left(\calC_{\infty_0}+\calC_{\infty_2}\right)$, and 
$\Lambda_{t_j^{(1)}}(p)= x(p)^{j}$, where $\calC_{\infty_k}$ is a small cycle around $\infty_k$.
We thus introduce the integral operators  associated to the various derivatives
\begin{equation}
\begin{array}{c|c|l}
\ds \frac {\partial}{\partial t_a}\square  & & \qquad \mathcal J_a(\square)\\[14pt]
\hline&\\
\ds \frac {\partial}{\partial t_{k}^{(1)}} \square & { \mapsto} &
\ds \left(\frac 23 \Res_{\infty_1}- \frac 1 3 \Res_{\infty_0, \infty_2}\right )  x^j \square\\[14pt]
\ds \frac {\partial}{\partial t_{k}^{(2)}} \square & { \mapsto} &
\ds \left(- \frac 23 \Res_{\infty_2}+ \frac 1 3 \Res_{\infty_0, \infty_1}\right)  (-x)^j \square\\[14pt]
\ds \frac {\partial }{\partial \epsilon_{j}^{(k)}}\square & {\mapsto}  & \ds \oint_{\mathcal B_j}\square\\[14pt]
\ds \frac {\partial}{\partial T} \square  & { \mapsto} & \ds \slint_{\infty_2}^{\infty_1} \square
\end{array}
\label{tableint}
\end{equation}
where the regularized integrals $\ds \slint$ are defined (if necessary) by subtraction of the singular part in the variable $x(p)$ (see \cite{Bert-06.1}) 
With this notation we have --compactly--
\begin{equation}
	y(p) \d x(p)  = \sum_a t_a \mathcal J_a(B(\cdot, p)).
	\label{ydxalt}
\end{equation}

\subsection{Topological Expansion of the Loop Equations}\label{Sec:TE}

The equations we need to work with are equations \eqref{eq:QLE}, \eqref{eq:CLE.b} and the equation 
\begin{equation}
	\sum_{k=0}^2\BK{Y_k(x)}=0.
	\label{eq:Le}
\end{equation}
In terms of connected components we have
\begin{equation}
	\begin{split}
		\sum_{k=0}^2&\BK{Y_k(x)}=0,\\
		\sum_{k=0}^2&\left(\BK{Y_k(x)}^2+\frac{T^2}{N^2}\BK{W_k(x)^2}_c\right)=2\BK{\hat{R}(x)},\\
		\sum_{k=0}^2&\left(\BK{Y_k(x)}^3+3\frac{T^2}{N^2}\BK{Y_k(x)}\BK{W_k(x)^2}_c
		-\frac{T^4}{N^4}\BK{W_k(x)^3}_c\right)=3\BK{D(x)}.
	\end{split}
	\label{eq:Le2}
\end{equation}
To proceed we need to define some more objects: write the topological expansion of our main observables\footnote{Note that whenever the superindex in $W_i^{(h)}(x)$ distinguishes is as a term in topological expansion and thus is different from $W_i(x)$ which is not an averaged quantity.}
\begin{equation}
	\begin{split}
		\BK{Y_i(x)}&=y_i(x)-\sum_{h=1}^\infty \left(\frac{T}{N}\right)^{2h} W_i^{(h)}(x),\\
		\BK{W_i(x_1)W_j(x_2)}_c&=\sum_{h=0}^\infty \left(\frac{T}{N}\right)^{2h} W_{i,j}^{(h)}(x_1,x_2),\\
		\BK{\prod_{k=1}^n W_{i_k}(x_k)}_c&=\sum_{h=0}^\infty \left(\frac{T}{N}\right)^{2h} 
		W_{i_1,\dots,i_n}^{(h)}(x_1,\dots,x_n),\\
		\BK{R(x)}&=\sum_{h=0}^\infty \left(\frac{T}{N}\right)^{2h} R^{(h)}(x),\\
		\BK{D(x)}&=\sum_{h=0}^\infty \left(\frac{T}{N}\right)^{2h} D^{(h)}(x).
	\end{split}
\end{equation}
From the $h$ components of these topological expansions we construct the following multi-differentials on the algebraic curve:
\begin{equation}
\begin{split}
y(p)\d{x(p)}&=y_i(x(p))\d{x(p)} \qquad \forall p\in\X_i, \\
\overline{w}_1^{(0)}(p)&=0,\\
\overline{w}_1^{(h)}(p)&=W_i^{(h)}(x(p))\d{x(p)} \qquad \forall p\in\X_i,\quad h>0 ,\\
\overline{w}_2^{(h)}(p_1,p_2)&=W_{i_1,i_2}^{(h)}(x(p_1),x(p_2))\d{x}(p_1)\d{x}(p_2)\qquad 
\forall p_1 \in\X_{i_1}\text{ and }p_2 \in\X_{i_2},\\
\end{split}
\end{equation}
are the one and two-point resolvent differentials, while in general we define
\begin{equation}
\begin{split}
\overline{w}_n^{(h)}(p_1,\dots,p_n)&=W_{i_1,\dots,i_n}^{(h)}(x(p_{1}),\dots,x(p_{n}))
\prod_{k=1}^{n}\d{x(p_{k})}\qquad
\forall p_k\in\X_{i_k},\,\,(k=1,\dots,n)
\end{split}
\end{equation}
as the $k$-point resolvent differential.
As we can see, all $\overline{w}_k^{(h)}$ are related to the $k$-point resolvent functions except for $\overline{w}_1^{(0)}=0$.
Define the Euler characteristics $\chi$ of $\overline{w}_k^{(h)}$ as\footnote{Note that this is exactly the Euler characteristic of a surface with $h$ holes and $k$ boundaries. This allows us to make connection with the fatgraph interpretation of the topological expansion in which $\overline{w}_k^{(h)}$ represents the generating function of counting numbers for fatgraphs of genus $h$ with $k$ boundaries.}
\begin{equation}
	\chi\left(w_k^{(h)}\right)=2-2h-k.
\label{eq:Euler}
\end{equation}

The sheet structure of this model makes it convenient to use the global loop insertion operator \eqref{eq:LIOP} to write the observables as differentials on the algebraic curve
\begin{equation}
	\overline{w}_n^{(h)}(p_1,\dots,p_n)=-\left(\prod_{k=1}^n
	\d x(p_k)\frac{\partial}{\partial V(p_k)}\right)\F^{(h)},
	\label{eq:dVF}
\end{equation}
where $\F^{(h)}$ is the $h$ order term in the topological expansion of $\F$
\begin{equation}
	\F=\sum_{h=1}^\infty \left(\frac{T}{N}\right)^{2h} \F^{(h)}.
\end{equation}
Finally define the differentials
\begin{equation}
	w_n^{(h)}(p_1,\dotsc,p_n)=\overline{\w}_n^{(h)}(p_1,\dotsc,p_n)
	+\delta_{n,2}\delta_{h,0}\frac{\d x(p_1)\d x(p_2)}{(x(p_1)-x(p_2))^2},
	\label{eq:wkbdef}
\end{equation}
which are just the same as $\overline{w}_k^{(h)}$ above except for $w_2^{(0)}$ which is shifted by the double pole.

\subsubsection{Recursion equations}

If we collect the order $h$ terms of the equations \eqref{eq:Le2} we obtain the following recursion equations for $h> 0$
{\footnotesize
\begin{equation}
	\begin{split}
		\sum_{i=1}^3 y(p^{(i)}) &= 0 \quad,\quad \qquad
		\sum_{i=1}^3 w_1^{(h)}(p^{(i)}) = 0\quad,\\
		\sum_{i=1}^3 y(p^{(i)})\d x(p)w_1^{(h)}(p^{(i)}) &= -R^{(h)}(x(p)) \left(\d x(p)\right)^2
		+\frac{1}{2}\sum_{i=1}^3 \overline{\W}_2^{(h)}(p^{(i)},p^{(i)}),\\
		\sum_{i=1}^3 (y(p^{(i)})\d x(p))^2 w_1^{(h)}(x(p^{(i)})) &= -D^{(h)}(x(p))(\d x(p))^3
		+\sum_{i=1}^3y(p^{(i)})\overline{\W}_2^{(h)}(p^{(i)},p^{(i)})
		-\frac{1}{3}\sum_{i=1}^3\overline{\W}_3^{(h)}(p^{(i)},p^{(i)},p^{(i)}),
	\end{split}
\label{eq:Wsdef}
\end{equation}
}
where
\begin{equation}
	\begin{split}
		\overline{\W}_2^{(h)}(p,q)&=\overline{w}_2^{(h-1)}(p,q)
		+\sum_{m=1}^{h-1}\overline{w}_1^{(m)}(p)\overline{w}_1^{(h-m)}(q),\\
		\text{and } \overline{\W}_3^{(h)}(p,q,r)&=
		\overline{w}_3^{(h-2)}(p,q,r)+	\sum_{m=1}^{h-2}\sum_{n=1}^{h-m-1} 
		\overline{w}_1^{(m)}(p)\overline{w}_1^{(n)}(q)\overline{w}_1^{(h-m-n)}(r)\\
		&+\sum_{m=1}^{h-1} \left(\overline{w}_1^{(m)}(p)\overline{w}_2^{(h-m-1)}(q,r)
		+\overline{w}_1^{(m)}(q)\overline{w}_2^{(h-m-1)}(r,p)
		+\overline{w}_1^{(m)}(r)\overline{w}_2^{(h-m-1)}(p,q)\right).
	\end{split}
	\label{eq:W23h}
\end{equation}
Notice that these are recursion equations indexed by the Euler characteristics $\chi$, in other words, the differentials $w_k^{(h)}$ are expressed in terms of other differentials with higher Euler characteristic $\chi$.\newline
We define as well the corresponding differential $\W_k^{(h)}$ that will be used latter
\begin{equation}
	\begin{split}
		{\W}_2^{(h)}(p,q)&=w_2^{(h-1)}(p,q)
		+\sum_{m=1}^{h-1}w_1^{(m)}(p)w_1^{(h-m)}(q),\\
		{\W}_3^{(h)}(p,q,r)&=
		w_3^{(h-2)}(p,q,r)+
		\sum_{m=1}^{h-2}\sum_{n=1}^{h-m-1}w_1^{(m)}(p)w_1^{(n)}(q)w_1^{(h-m-n)}(r)\\
		&+\sum_{m=1}^{h-1} \left(w_1^{(m)}(p)w_2^{(h-m-1)}(q,r)
		+w_1^{(m)}(q)w_2^{(h-m-1)}(r,p)
		+w_1^{(m)}(r)w_2^{(h-m-1)}(p,q)\right).
	\end{split}
	\label{eq:W23h.b}
\end{equation}
Finally we need to complement these equations with the {\it fixed filling fraction condition} at each order $h$
\begin{equation}
	\oint_{\A_i^{(k)}} w_1^{(h)}(\cdot) = 0.
	\label{eq:hFFC}
\end{equation}
As a corollary we obtain similar conditions for the $n$-point resolvent functions
\begin{equation}
	\oint_{\A_i^{(k)}} w_n^{(h)}(\cdot,p_2,\dots,p_n) = 0.
	\label{eq:hFFC.a}
\end{equation}

\subsubsection{The two-point resolvent correlation function}

The "initial conditions" of the recursion relations above are the only data needed to solve the recursion relations. This data are the fundamental differential $y(p)\d x(p)$ of the spectral curve, the two-point resolvent differential $w_2^{(0)}(p,q)$ and the filling fraction conditions for all $h$. The spectral curve is specified by the moduli, and we can obtain $w_2^{(0)}(p,q)$ by the action of the loop insertion operator on the fundamental differential $y(p)\d x(p)$,
\begin{equation}
	\d x(q)\frac{\pd}{\pd V_1(x(q))} y(p)\d x(p)=
	-\sum_{j=0}^{\infty}\frac{\d \Omega_j^{(1)}(p) \d x(q)}{\x(q)^{j-1}}.
	\label{eq:DIF2}
\end{equation}
Notice that apart from a constant factor in front depending to which $x$-sheet $p$ belongs, the sum in the above formula produce a second order pole in the diagonal $x(p)=x(q)$. This fact, together with the vanishing $\A$ periods condition, indicates that
\begin{equation}
	\begin{split}
	\d x(q)\frac{\pd}{\pd V_1(x(q))} y(p)\d x(p)&=
	-\frac{2}{3}B(q^{(1)},p)+\frac{1}{3}\left(B(q^{(0)},p)+B(q^{(2)},p)\right)\\
	&=-B(q^{(1)},p)+\frac{1}{3}\frac{\d x(q) \d x(p)}{(x(q)-x(p))^2},
	\end{split}
	\label{eq:DIF3}
\end{equation}
where $B(q,p)$ is the so called Bergman kernel. Remember that $q^{(k)}$ satisfy $x(q^{(k)})=x(q)=x$, and in this case we also precise $q^{(k)}\in\X_k$ for convenience. In the last line of \eqref{eq:DIF3} we have used the relation
\begin{equation}
	\sum_{i=0}^2 B(q^{(i)},p)=\frac{\d x(q) \d x(p)}{(x(q)-x(p))^2}.
	\label{eq:BK2}
\end{equation} 
We obtain similar results by applying $\d x(q)\frac{\pd}{\pd V_2(-x(q))}$ or 
$-\d x(q)\left(\frac{\pd}{\pd V_1(x(q))}+\frac{\pd}{\pd V_2(-x(q))}\right)$.
All three operators combined in $\d x(q)\frac{\pd}{\pd V(q)}$ (see \eqref{eq:LIOP}) give
\begin{equation}
	\begin{split}
	\d x(q)\frac{\pd}{\pd V(q)} y(p)\d x(p)
	&=-B(q,p)+\frac{1}{3}\frac{\d x(q) \d x(p)}{(x(q)-x(p))^2}.
	\end{split}
	\label{eq:DIF4}
\end{equation}
From this expression we find that
\begin{equation}
	\begin{split}
		\overline{\w}_2^{(0)}(q,p)\equiv\d x(q)\frac{\pd}{\pd V(q)} w_1^{(0)}(p)&=
		\d x(q)\frac{\pd}{\pd V(q)} (U^\prime(p)-y(p))=\\
		&=B(q,p)-\frac{\d x(q) \d x(p)}{(x(q)-x(p))^2}.\\
	\end{split}
	\label{eq:DIF5}
\end{equation}
Note that \eqref{eq:DIF5} implies $\w_2^{(0)}(q,p)=B(q,p)$.
For latter purposes we write equation \eqref{eq:BK2} in the following form
\begin{equation}
	\begin{split}
		\overline{w}_2^{(0)}(q,p^{(k)})=-\sum_{j\not=k}\w_2^{(0)}(q,p^{(j)}),
	\end{split}
	\label{eq:sh.b}
\end{equation}
which can be trivially extended to
\begin{equation}
	\begin{split}
		\overline{w}_{n+1}^{(h)}(p^{(k)},p_1,\cdots,p_n)=-\sum_{j\not=k}w_{n+1}^{(h)}(p^{(j)},p_1,\cdots,p_n).
	\end{split}
	\label{eq:sh.c}
\end{equation}

\section{Solution of the recursion relations}\label{Sec:SolRR}

The equations \eqref{eq:Wsdef} obtained in the previous section
are in fact non-homogeneous recursion equations. The non-homogeneous terms $R^{(h)}$ and $D^{(h)}$ are unknown meromorphic functions with at most a pole of order 1 and 2 respectively at $x=0$. Although these functions are unknown they are not an impediment to solve the equations.
In order to take advantage of \eqref{eq:Wsdef} we will use the third kind Abelian differential $\d S_{p,o}(q)$ to rewrite $w_1^{(h)}$
\begin{equation}
	\begin{split}
		w_1^{(h)}(q)&=-\Res_{p\to q} \d S_{p,o}(q) w_1^{(h)}(p)\\
		&=\sum_{\alpha\in\Delta}\Res_{p\to\alpha} \d S_{p,o}(q) w_1^{(h)}(p)
	\end{split}
	\label{eq:RBI.2}
\end{equation}
by using the bilinear Riemann identity, the condition \eqref{eq:hFFC} and the properties of $\d S_{p,o}(q)$. 
The sum on the second line runs over all poles of the integrand except $p=q$, since the only poles apart from $p=q$ are the branch-points on the algebraic curve, the set of points $\Delta$ is the set of branch-points of the algebraic curve. We then use \eqref{eq:Wsdef} to express {\it locally} $w_1^{(h)}(p)$ in terms of {\it known} --by recursion hypothesis-- differentials plus the irrelevant terms $R^{(h)}$ and $D^{(h)}$.

In our situation we will distinguish between two possible contributions, the ones coming from simple branch-points (branch-points joining only two sheets of the algebraic curve) and double branch-points joining all three sheets. 

\subsection{Contribution from simple branch-points}

In this section we compute the residue
\begin{equation}
	\Res_{p\to\alpha} \d S_{p,o}(q) w_1^{(h)}(p)
	\label{eq:Rsbp}
\end{equation}
on a simple branch-point $\alpha$. 
Around a simple branch-point the differentials on the algebraic curve can be split into two components $w_1^{(h)}(p)=[w_1^{(h)}(p)]_{1} + [w_1^{(h)}(p)]_{-1}$ satisfying the following property
\begin{equation}
	[w_1^{(h)}(\vartheta^i(p))]_\lambda=\lambda^i\, [w_1^{(h)}(p)]_\lambda.
\end{equation}
Obviously on a simple branch-point $\vartheta^2(p)=p$ so that we can drop the super-index.
In the same way we can write $y(p)\d x(p)=[y(p)\d x(p)]_{1} + [y(p)\d x(p)]_{-1}$.
Suppose the point $p$ is in the neighborhood of the branch-point $\alpha$, it is clear that $\vartheta(p)$ is also in the neighborhood of $\alpha$. The third point on top of $x(p)$ (that we call $\check{p}$) is not in the neighborhood of $\alpha$. The differentials $w_k^{(h)}$ have poles only at branch-points. The terms depending only on $\check{p}$ are holomorphic in $x(p)$ when $p\to\alpha$. With all this we have
\begin{equation}
	\begin{split}
		y(p)\d x(p)=[y(p)\d x(p)]_{1} + [y(p)\d x(p)]_{-1} \quad,\quad 
		& \qquad w_1^{(h)}(p)=[w_1^{(h)}(p)]_{1} + [w_1^{(h)}(p)]_{-1}\quad,\\
		y(\vartheta(p))\d x(p)=[y(p)\d x(p)]_{1} - [y(p)\d x(p)]_{-1} \quad,\quad & \qquad 
		w_1^{(h)}(\vartheta(p))=
		[w_1^{(h)}(p)]_{1} - [w_1^{(h)}(p)]_{-1}\quad,\\
		y(\check{p})\d x(p)=-2[y(p)]_{1}\d x(p)\quad,\quad & 
		\qquad w_1^{(h)}(\check{p})=-2[w_1^{(h)}(p)]_{1}\quad,\\
		\sum_{i=1}^3 y(p^{(i)})\d x(p)w_1^{(h)}(p^{(i)}) &= \frac{3}{2}y(\check{p})\d x(p)w_1^{(h)}(\check{p}) 
		+ 2 [y(p)\d x(p)]_{-1}[w_1^{(h)}(p)]_{-1}\quad,
	\end{split}
\end{equation}
so that in the neighborhood of the branch point $\alpha$
\begin{equation}
	\begin{split}
		\left[w_1^{(h)}(p)\right]_{-1}&= 
		\frac{1}{(y(p)-y(\vartheta(p)))\d x(p)}\left(\sum_{i=1}^3y(p^{(i)})\d x(p)w_1^{(h)}(p^{(i)})-
		\frac{3}{2}y(\check{p})\d x(p)w_1^{(h)}(\check{p})\right).
	\end{split}
\label{eq:W1-1}
\end{equation}
Now we can introduce \eqref{eq:W1-1} into \eqref{eq:Rsbp}\footnote{Notice that $[w_1^{(h)}(p)]_1$ as well as $w_1^{(h)}(\check{p})$ are holomorphic around the branch-point and do not contribute to the residue.}
\begin{equation}
	\begin{split}
		\Res_{p\to\alpha} \d S_{p,o}(q) w_1^{(h)}(p)&
		=\Res_{p\to\alpha}\frac{\d S_{p,o}(q)}{(y(p)-y(\vartheta(p)))\d x(p)}
		\sum_{i=1}^3y(p^{(i)})\d x(p)w_1^{(h)}(p^{(i)})\\
		&=\Res_{p\to\alpha}\frac{\d S_{p,o}(q)}{(y(p)-y(\vartheta(p)))\d x(p)} 
		\frac{1}{2}\sum_{i=1}^3\overline{\W}_2^{(h)}(p^{(i)},p^{(i)})\\
		&=-\Res_{p\to\alpha}\frac{\d S_{p,o}(q)}{(y(p)-y(\vartheta(p)))\d x(p)} \W_2^{(h)}(p,\vartheta(p)).
	\end{split}
\end{equation}
The topological recursive expansion of the loop equation \eqref{eq:Wsdef} has been used in the second equality.
Some terms, included $R^{(h)}(x)$, have disappeared due to the residue operation and the assumption that there is no pole structure in $y(\check{p})\d x(p)$ or $w_1^{(h)}(\check{p})$ when $p\to\alpha$.
Notice also that in the last expression, $\overline{\W}_2^{(h)}$ has been changed into $\W_2^{(h)}$ by using \eqref{eq:sh.c}.

\subsection{Contribution from a branch-point with $n_b=2$}

In this section we compute the residue
\begin{equation}
	\Res_{p\to\alpha} \d S_{p,o}(q) w_1^{(h)}(p)
	\label{eq:Rdbp}
\end{equation}
on branch-points with $n_b=2$. 
The contribution around these branch-points can be found similarly as with the simple branch-points. In this case however we have three components for every differential. 
\begin{equation}
	\begin{split}
		y(p)\d x(p)=[y(p)\d x(p)]_1 + [y(p)\d x(p)]_\th + [y(p)\d x(p)]_{\th^2} \quad, & 
		\quad w_1^{(h)}(p)=[w_1^{(h)}(p)]_1 + [w_1^{(h)}(p)]_\th + [w_1^{(h)}(p)]_{\th^2},\\
		0 = \sum_{i=1}^3 y(p^{(i)})\d x(p) = 3 [y(p)\d x(p)]_1 \quad,& 
		\quad 0 = \sum_{i=1}^3 w_1^{(h)}(p^{(i)}) = 3 [w_1^{(h)}(p)]_1,
	\end{split}
\end{equation}
where $\th=\e{\frac{2 \pi i}{3}}$ is a root of unity.
In order to find $w_1^{(h)}(p)=[w_1^{(h)}(p)]_\th + [w_1^{(h)}(p)]_{\th^2}$ we present the equations
\begin{equation}
	\begin{split}
		\sum_{i=1}^3 y(p^{(i)})\d x(p)w_1^{(h)}(p^{(i)}) =& 
		3([y(p)\d x(p)]_\th [w_1^{(h)}(p)]_{\th^2} + [y(p)\d x(p)]_{\th^2} [w_1^{(h)}(p)]_{\th})\\
		\sum_{i=1}^3 (y(p^{(i)})\d x(p))^2w_1^{(h)}(p^{(i)}) =& 3(([y(p)\d x(p)]_{\th})^2 [w_1^{(h)}(p)]_{\th} 
		+ ([y(p)\d x(p)]_{\th^2})^2 [w_1^{(h)}(p)]_{\th^2})
	\end{split}
\end{equation}
and their solution
\begin{equation}
	\begin{split}
		[w_1^{(h)}(p)]_\th&=\frac{1/3}{([y(p)\d x(p)]_{\th^2})^3-([y(p)\d x(p)]_{\th})^3} 
		\left(([y(p)\d x(p)]_{\th^2})^2 \sum_{i=1}^3y(p^{(i)})\d x(p)w_1^{(h)}(p^{(i)})\right.
		\\&\left.\hspace{200pt}
		-[y(p)\d x(p)]_\th \sum_{i=1}^3 (y(p^{(i)})\d x(p))^2w_1^{(h)}(p^{(i)})\right),\\
		[w_1^{(h)}(p)]_{\th^2}&=\frac{1/3}{([y(p)\d x(p)]_{\th^2})^3-([y(p)\d x(p)]_{\th})^3} 
		\left([y(p)\d x(p)]_{\th^2} \sum_{i=1}^3 (y(p^{(i)})\d x(p))^2w_1^{(h)}(p^{(i)})\right.
		\\&\left.\hspace{200pt}-
		([y(p)\d x(p)]_\th)^2 \sum_{i=1}^3 y(p^{(i)})\d x(p)w_1^{(h)}(p^{(i)})\right),\\
	\end{split}
\end{equation}
which gives the following representation for $w_1^{(h)}(p)$ around a $n_b=2$ branch-point
\begin{equation}
	\begin{split}
		w_1^{(h)}(p)&=[w_1^{(h)}(p)]_\th + [w_1^{(h)}(p)]_{\th^2}\\
		&=\frac{1}{3}\left(
		\frac{([y(p)\d x(p)]_{\th^2})^2-([y(p)\d x(p)]_{\th})^2}{([y(p)\d x(p)]_{\th^2})^3-
		([y(p)\d x(p)]_{\th})^3}
		\sum_{i=1}^3 y(p^{(i)})\d x(p) w_1^{(h)}(p^{(i)}) 
		\right.\\&\hspace{100pt}\left.
		+\frac{[y(p)\d x(p)]_{\th^2}-[y(p)\d x(p)]_{\th}}{([y(p)\d x(p)]_{\th^2})^3-([y(p)\d x(p)]_{\th})^3}
		\sum_{i=1}^3 (y(p^{(i)})\d x(p))^2 w_1^{(h)}(p^{(i)})\right)\\
		&=\frac{1}{3(y(p)-y(\vartheta^1(p)))(y(p)-y(\vartheta^2(p)))\d x(p)^2}
		\left(y(p)\d x(p)\sum_{i=1}^3 y(p^{(i)})\d x(p) w_1^{(h)}(p^{(i)}) 
		\right.\\&\hspace{230pt}\left.+ \sum_{i=1}^3 (y(p^{(i)})\d x(p))^2 w_1^{(h)}(p^{(i)})\right).
	\end{split}
\end{equation}
If we introduce equations \eqref{eq:Wsdef} and define $\omega(p,\vartheta^j(p))=(y(p)-y(\vartheta^j(p)))\d x(p)$, we get the following
{\footnotesize
\begin{equation}
	\begin{split}
		w_1^{(h)}(p)&=\frac{1}{\omega(p,\vartheta^1(p))\omega(p,\vartheta^2(p))}
		\Bigg(-\left(y(p) R^{(h)}(x(p))+D^{(h)}(x(p))\right)(\d x(p))^3\\
		&\qquad+\sum_{i=1}^3\left(\frac{1}{2}y(p)+y(p^{(i)})\right)\d x(p)\overline{\W}_2^{(h)}(p^{(i)},p^{(i)})
		+\sum_{i=1}^3\overline{\W}_3^{(h)}(p^{(i)},p^{(i)},p^{(i)})\Bigg)\\
		&=-\frac{\left(y(p)R^{(h)}(p)+D^{(h)}(p)\right)(\d x(p))^3}
		{\omega(p,\vartheta^1(p))\omega(p,\vartheta^2(p))}
		-\left(\frac{\W_2^{(h)}(p,\vartheta^1(p))}{\omega(p,\vartheta^1(p))}
		+\frac{\W_2^{(h)}(p,\vartheta^2(p))}{\omega(p,\vartheta^2(p))}
		+\frac{\W_3^{(h)}(p,\vartheta^1(p),\vartheta^2(p))}
		{\omega(p,\vartheta^1(p))\omega(p,\vartheta^2(p))}\right),
	\end{split}
\end{equation}
}
where again we used \eqref{eq:sh.c} in the last equality.
When we introduce it into \eqref{eq:Rdbp}, the residue contribution reduces to
\begin{equation}
	\begin{split}
		\Res_{p\to \alpha}\d S_{p,o}(q)w_1^{(h)}(p)&= 
		-\Res_{p\to \alpha}\frac{\d S_{p,o}(q)\W_2^{(h)}(p,\vartheta^1(p))}{\omega(p,\vartheta^1(p))}
		-\Res_{p\to \alpha}\frac{\d S_{p,o}(q)\W_2^{(h)}(p,\vartheta^2(p))}{\omega(p,\vartheta^2(p))}\\
		&-\Res_{p\to \alpha}\frac{\d S_{p,o}(q) \W_3^{(h)}(p,\vartheta^1(p),\vartheta^2(p))}
		{\omega(p,\vartheta^1(p))\omega(p,\vartheta^2(p))},
	\end{split}
\end{equation}
where again the term with $R^{(h)}$ and $D^{(h)}$ disappear due to the lack of pole structure of the integrand.

\subsection{Residue formula for $w_1^{(h)}$}

We can split the set of branch-points as $\Delta=\Delta_1+\Delta_2$, where $\Delta_1$ is the set of simple branch-points and $\Delta_2$ is the set of $n_b=2$ branch-points --in our case there is just one such branch-point--.
Putting together all the contributions we have
\begin{equation}
	\begin{split}
		w_1^{(h)}(q)&=\sum_{\alpha\in\Delta}\Res_{p\to\alpha} \d S_{p,o}(q) w_1^{(h)}(p)\\
		&=-\sum_{i=1}^2\sum_{\alpha\in\Delta_i}\sum_{j=1}^i 
		\Res_{p\to \alpha}\frac{\d S_{p,o}(q)\W_2^{(h)}(p,\vartheta^j(p))}{\omega(p,\vartheta^j(p))}
		-\sum_{\alpha\in\Delta_2}\Res_{p\to \alpha}\frac{\d S_{p,o}(q) 
		\W_3^{(h)}(p,\vartheta^1(p),\vartheta^2(p))}
		{\omega(p,\vartheta^1(p))\omega(p,\vartheta^2(p))}.
	\end{split}
\label{eq:recfor}
\end{equation}
Note that the non-homogenous terms $R^{(h)}$ and $D^{(h)}$ have disappeared under the residue operator. Also we recall the fact that those equations define all the $w_1^{(h)}$ in terms of other such differentials with higher Euler characteristic. 

\subsection{Diagrammatic representation of the residue formula}

There is a very compact diagrammatic representation of the equation above. The diagrammatic representation is a very clear and simple way to write the equation \eqref{eq:recfor}, to understand the important features without getting lost in the details. The reader not interested in the diagrammatic representation can skip this section.
\newline
Define the diagrammatic vertices
\begin{equation}
	\begin{split}
		-\sum_{i=1}^2\sum_{\alpha\in\Delta_i}\sum_{j=1}^i 
		\Res_{p\to \alpha}\frac{\d S_{p,o}(q)}{\omega(p,\vartheta^j(p))}&\qquad\Longrightarrow \qquad
		\begin{array}{l}
			\resizebox{0.2\textwidth}{!}{\input{Vertex2.pstex_t}}
		\end{array},\\
		-\sum_{\alpha\in\Delta_2} 
		\Res_{p\to \alpha}\frac{\d S_{p,o}(q)}{\omega(p,\vartheta^1(p))\omega(p,\vartheta^2(p))}&
		\qquad\Longrightarrow \qquad
		\begin{array}{l}
			\resizebox{0.2\textwidth}{!}{\input{Vertex3.pstex_t}}
		\end{array}	,	
	\end{split}
\end{equation}
where the black dots mark the leg carrying $\vartheta^j(p)$ and it is accompanied by an index when there is ambiguity. We define as well
\begin{equation}
	\begin{split}
		w_n^{(h)}(p_1,\dots,p_n)&\qquad\Longrightarrow \qquad
		\begin{array}{l}
			\resizebox{0.2\textwidth}{!}{\input{wkh.pstex_t}}
		\end{array}
	\end{split}.
\end{equation}
After all these definitions the representation of equation \eqref{eq:recfor} is
\begin{equation}
	\begin{split}
		\begin{array}{l}
			\resizebox{0.2\textwidth}{!}{\input{w1h.pstex_t}}
		\end{array}
		&=
		\begin{array}{l}
			\resizebox{0.2\textwidth}{!}{\input{ww2h.pstex_t}}
		\end{array}
		+\sum_{m=1}^{h-1}
		\begin{array}{l}
			\resizebox{0.2\textwidth}{!}{\input{w1hw1h.pstex_t}}
		\end{array}
		\\&+
		\begin{array}{l}
			\resizebox{0.2\textwidth}{!}{\input{ww3h.pstex_t}}
		\end{array}
		+\sum_{m=1}^{h-2}\sum_{n=1}^{h-m-1}
		\begin{array}{l}
			\resizebox{0.3\textwidth}{!}{\input{w1hw1hw1h.pstex_t}}
		\end{array}
		\\+\sum_{m=1}^{h-1}&\left(
		\begin{array}{l}
			\resizebox{0.2\textwidth}{!}{\input{w1hw2hp.pstex_t}}
		\end{array}+
		\begin{array}{l}
			\resizebox{0.2\textwidth}{!}{\input{w1hw2hq.pstex_t}}
		\end{array}+
		\begin{array}{l}
			\resizebox{0.2\textwidth}{!}{\input{w1hw2hr.pstex_t}}
		\end{array}\right).
	\end{split}
\end{equation}

\subsection{Residue formula for $w_k^{(h)}$}\label{eq:RFwkh}

Define the following multi-differentials
\begin{equation}
	\begin{split}
		\widehat{w}_{k+1}^{(h)}(q,{\bf P_K})&=-\sum_{i=1}^2\sum_{\alpha\in\Delta_i}\sum_{j=1}^i 
		\Res_{t\to \alpha}\frac{\d S_{t,o}(q)\W_{2,k}^{(h)}(t,\vartheta^j(t),{\bf P_K})} 
		{\omega(t,\vartheta^j(t))}\\
		&-\sum_{\alpha\in\Delta_2}\Res_{t\to \alpha}\frac{\d S_{t,o}(q) 
		\W_{3,k}^{(h)}(t,\vartheta^1(t),\vartheta^2(t),{\bf P_K})}
		{\omega(p,\vartheta^1(p))\omega(p,\vartheta^2(p))},
	\end{split}
	\label{eq:recforhat}
\end{equation}
where the notation is ${\bf P_K}= \{p_1,\dots,p_k\}$ and $\W_{2,k}^{(h)}$ and $\W_{3,k}^{(h)}$ are expressed as
{\footnotesize
\begin{equation}
	\begin{split}
		\W_{2,k}^{(h)}(t,t^\prime,{\bf P}_{\bf K})=&w_{k+2}^{(h-1)}(t,t^\prime,{\bf P}_{\bf K})
		+\sum_{{\bf J}\subset{\bf K}}\sum_{m=0}^{h} w_{j+1}^{(m)}(t,{\bf P}_{\bf J})
		w_{k-j+1}^{(h-m)}(t^\prime,{\bf P}_{\bf K/J})\\
		\W_{3,k}^{(h)}(t,t^\prime,t^{\prime\prime},{\bf P}_{\bf K})=&
		w_{3+k}^{(h-2)}(t,t^\prime,t^{\prime\prime},{\bf P}_{\bf K})+
		\sum_{{\bf J}\subset{\bf K}}\sum_{m=0}^{h-1} \left(w_{j+1}^{(m)}(t,{\bf P}_{\bf J})
		w_{k-j+2}^{(h-m-1)}(t^\prime,t^{\prime\prime},{\bf P}_{\bf K/J})\right.\\
		&\left.+w_{j+1}^{(m)}(t^\prime,{\bf P}_{\bf J})
		w_{k-j+2}^{(h-m-1)}(t^{\prime\prime},t,{\bf P}_{\bf K/J})+w_{j+1}^{(m)}(t^{\prime\prime},{\bf P}_{\bf J}) 
		w_{k-j+2}^{(h-m-1)}(t,t^{\prime},{\bf P}_{\bf K/J})\right)\\
		&+\sum_{\substack{{\bf J_1},{\bf J_2}\subset {\bf K}\\{\bf J_1}\cap{\bf J_2}=\emptyset}} 
		\sum_{m=0}^{h-2}\sum_{n=0}^{h-m-1}w_{j_1+1}^{(m)}(t,{\bf P_{J_1}})w_{j_2+1}^{(n)}(t^\prime,{\bf P_{J_2}})
		w_{k-j_1-j_2+1}^{(h-m-n)}(t^{\prime\prime},{\bf P_{K/{J_1\cup J_2}}}),
	\end{split}
	\label{eq:recforW}
\end{equation}
}
with ${\bf K}=\{1,\dots,k\}$ is the set of index values and ${\bf J}$, ${\bf J_1}$ and ${\bf J_2}$ are subsets of $j$, $j_1$ or $j_2$ index values taken from ${\bf K}$ respectively.
Remember that the differential $w_1^{(0)}$ is defined to be equal to zero so that, for example, the double sum in the definitions of $\W_2^{(h)}(t,t^\prime,{\bf P_K})$ does not contain the term ${\bf J=\emptyset},\, m=0$ nor ${\bf J=K},\, m=h$.
It is clear from the definitions above that $\widehat{w}_1^{(h)}(p)=w_1^{(h)}(p)$. In fact, we expect
$\widehat{w}_k^{(h)}(p_1,\dots,p_n)=w_k^{(h)}(p_1,\dots,p_n)$ for every $k$. To prove it is sufficient to check that $\d x(p)\frac{\partial}{\partial V(p)}\widehat{w}_k^{(h)}({\bf P_K})=\widehat{w}_{k+1}^{(h)}({\bf P_K},p)$. 

\subsubsection{The action of the loop insertion operator on the cubic vertex}\label{Sec:LOCV}
Consider the action of the loop insertion operator on the cubic vertex $\begin{array}{l}\includegraphics[scale=0.3]{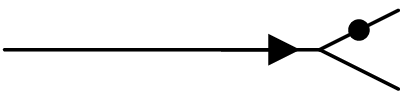} \end{array}$ with a generic $f(t,\vartheta^{(t)})$ attached to it.
{\scriptsize
\begin{equation}
	\begin{split}
		\d x(p)\frac{\partial }{\partial V(p)} &\Bigg(-\sum_{i=1}^2\sum_{\alpha\in\Delta_i}\sum_{j=1}^i 
		\Res_{t\to \alpha}\frac{\d S_{t,o}(q)}{\omega(t,\vartheta^j(t))}f(t,\vartheta^j(t))\Bigg)
		=\\
		&=-\sum_{i=1}^2 \sum_{\alpha\in\Delta_i}\sum_{j=1}^i \Bigg(\Res_{t\to \alpha}
		\frac{f(t,\vartheta^j(t))}{\omega(t,\vartheta^j(t))}
		\d x(p)\frac{\partial \d S_{t,o}(q)}{\partial V(p)}
		-\Res_{t\to \alpha}\frac{\d S_{t,o}(q)f(t,\vartheta^j(t))}{(\omega(t,\vartheta^2(t)))^2}
		\d x(p)\frac{\partial \omega(t,\vartheta^j(t))}{\partial V(p)}\\
		&\hspace{100pt}+\Res_{t\to \alpha}\frac{\d S_{t,o}(q)}{\omega(t,\vartheta^j(t))}\d x(p)
		\frac{\partial f(t,\vartheta^j(t))}{\partial V(p)}\Bigg)\\
		&=\sum_{i=1}^2 \sum_{\alpha\in\Delta_i}\sum_{j=1}^i 
		\Bigg(\Res_{t\to \alpha}\frac{f(t,\vartheta^j(t))}{\omega(t,\vartheta^j(t))}
		\sum_{k=1}^2\sum_{\beta\in\Delta_k}\sum_{l=1}^k\Res_{u\to\beta} 
		\frac{\d S_{u,o}(q)}{\omega(u,\vartheta^l(u))}
		\left(B(u,p)\d S_{t,o}(\vartheta^l(u))+B(\vartheta^l(u),p)\d S_{t,o}(u)\right)\\
		&\hspace{100pt}-\Res_{t\to \alpha}\frac{\d S_{t,o}(q)f(t,\vartheta^j(t))}{(\omega(t,\vartheta^j(t)))^2}
		(B(t,p)-B(\vartheta^j(t),p))
		+\Res_{t\to \alpha}\frac{\d S_{t,o}(q)}{\omega(t,\vartheta^j(t))}\d x(p)
		\frac{\partial f(t,\vartheta^j(t))}{\partial V(p)}\Bigg).
	\end{split}
	\label{eq:dVV}
\end{equation}
}
We will drop from now on the last term involving the action of the loop insertion operator onto $f(a,b)$. We take care of it by chain rule.
after droping this term we can bring last equality to a more meaningful form if we exchange the order of the residues on the first term of the third line. The only difficulty to commute the residues appear whenever $\alpha=\beta\in\Delta_j$, in that case we need to apply the relation
\begin{equation}
	\Res_{t\to\alpha}\Res_{u\to\alpha} = \Res_{u\to\alpha}\Res_{t\to\alpha}
	-\Res_{t\to\alpha}\sum_{k=0}^j\Res_{u\to\vartheta^k(t)}.
	\label{eq:CommRes}
\end{equation}
Two situations appear: when $\alpha\in\Delta_1$ the contribution from the last term of \eqref{eq:CommRes} cancels the last term of \eqref{eq:dVV}, when $\alpha\in\Delta_2$ the cancellation happens as well but some contribution remains giving rise to a term of the type $\begin{array}{l}\includegraphics[scale=0.3]{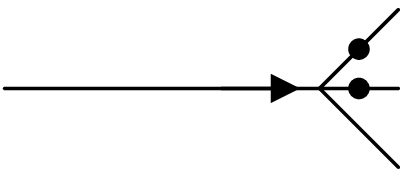} \end{array}$. The result of exchanging the order of the residues is\footnote{In appendix \ref{app:Hrules} there is a detailed account of this calculation.}
{\footnotesize
\begin{equation}
	\begin{split}
		\d x(p)\frac{\partial }{\partial V(p)} &\Bigg(-\sum_{i=1}^2\sum_{\alpha\in\Delta_i}\sum_{j=1}^i 
		\Res_{t\to \alpha}\frac{\d S_{t,o}(q)}{\omega(t,\vartheta^j(t))}f(t,\vartheta^j(t))\Bigg)=\\
		&=-\sum_{k=1}^2\sum_{\beta\in\Delta_k}\sum_{l=1}^k\Res_{u\to\beta}
		\frac{\d S_{u,o}(q)}{\omega(u,\vartheta^l(u))}
		\left(B(u,p)\left(-\sum_{i=1}^2 \sum_{\alpha\in\Delta_i}\sum_{j=1}^i \Res_{t\to\alpha}
		\frac{\d S_{t,o}(\vartheta^l(u))}{\omega(t,\vartheta^j(t))}\right)\right.\\
		&\hspace{100pt}\left.
		+B(\vartheta^l(u),p)\left(-\sum_{i=1}^2 \sum_{\alpha\in\Delta_i}\sum_{j=1}^i \Res_{t\to\alpha}
		\frac{\d S_{t,o}(u)}{\omega(t,\vartheta^j(t))}\right)\right)f(t,\vartheta^j(t))\\
		&-\sum_{\alpha\in\Delta_2} \Res_{t\to\alpha}
		\frac{\d S_{t,o}(q)}{\omega(t,\vartheta^1(t))\omega(t,\vartheta^2(t))}
		\left(B(t,p)f(\vartheta^1(t),\vartheta^2(t))+B(\vartheta^1(t),p)f(\vartheta^2(t),t)
		+B(\vartheta^2(t),p)f(t,\vartheta^1(t))\right).
	\end{split}
	\label{eq:dVVertex2}
\end{equation}
}
Graphically, equation \eqref{eq:dVVertex2} is represented as
\begin{equation}
	\begin{split}
		\d x(p)\frac{\partial }{\partial V(p)}
		\begin{array}{l}
			\resizebox{0.2\textwidth}{!}{\input{Vertex2a.pstex_t}}
		\end{array}&=
		\begin{array}{l}
			\resizebox{0.22\textwidth}{!}{\input{Vertex2b.pstex_t}}
		\end{array}+
		\begin{array}{l}
			\resizebox{0.22\textwidth}{!}{\input{Vertex2c.pstex_t}}
		\end{array}\\
		&+
		\begin{array}{l}
			\resizebox{0.17\textwidth}{!}{\input{Vertex2d.pstex_t}}
		\end{array}+
		\begin{array}{l}
			\resizebox{0.17\textwidth}{!}{\input{Vertex2e.pstex_t}}
		\end{array}+
		\begin{array}{l}
			\resizebox{0.17\textwidth}{!}{\input{Vertex2f.pstex_t}}
		\end{array},
	\end{split}
	\label{eq:Graph2}
\end{equation}
where the differential $w_2^{(0)}(q,p)=B(q,p)$ is represented by the non-arrowed propagator 
$\begin{array}{l}
		\resizebox{0.15\textwidth}{!}{\input{Bpq.pstex_t}}
	\end{array}$.

\subsubsection{The action of the loop insertion operator on the quartic vertex}\label{Sec:ALOQ}
Consider now the action of the loop insertion operator on the quartic vertex $\begin{array}{l} \includegraphics[scale=0.4]{Vertex3bare.eps}\end{array}$
{\footnotesize
\begin{equation}
	\begin{split}
		\d x(p)\frac{\partial }{\partial V(p)} &\Bigg(-\sum_{\alpha\in\Delta_2} 
		\Res_{t\to \alpha}\frac{\d S_{t,o}(q)f(t,\vartheta^1(t),\vartheta^2(t))}
		{\omega(t,\vartheta^1(t))\omega(t,\vartheta^2(t))}
		\Bigg)=\\
		=&-\sum_{\alpha\in\Delta_2} \Bigg(\Res_{t\to \alpha}
		\frac{f(t,\vartheta^1(t),\vartheta^2(t))}{\omega(t,\vartheta^1(t))\omega(t,\vartheta^2(t))}
		\d x(p)\frac{\partial \d S_{t,o}(q)}{\partial V(p)}+
		\Res_{t\to \alpha}\frac{\d S_{t,o}(q)}{\omega(t,\vartheta^1(t))\omega(t,\vartheta^2(t))}
		\d x(p)\frac{\partial f(t,\vartheta^1(t),\vartheta^2(t))}{\partial V(p)}\\
		&\hspace{50pt}-\Res_{t\to \alpha}\frac{\d S_{t,o}(q)f(t,\vartheta^1(t),\vartheta^2(t))} 
		{(\omega(t,\vartheta^1(t))\omega(t,\vartheta^2(t)))^2}
		\left(\omega(t,\vartheta^1(t))\d x(p)\frac{\partial \omega(t,\vartheta^2(t))}{\partial V(p)}
		+\omega(t,\vartheta^2(t))\d x(p)\frac{\partial \omega(t,\vartheta^1(t))}{\partial V(p)}\right)\Bigg)\\
		=&\sum_{\alpha\in\Delta_2} \Bigg(\Res_{t\to \alpha}
		\frac{f(t,\vartheta^1(t),\vartheta^2(t))}{\omega(t,\vartheta^1(t))\omega(t,\vartheta^2(t))}
		\sum_{k=1}^2\sum_{\beta\in\Delta_k}\sum_{l=1}^k\Res_{u\to\beta} 
		\frac{\d S_{u,o}(q)}{\omega(u,\vartheta^l(u))}
		\left(B(u,p)\d S_{t,o}(\vartheta^l(u))+B(\vartheta^l(u),p)\d S_{t,o}(u)\right)\\
		&-\Res_{t\to \alpha}\frac{\d S_{t,o}(q)f(t,\vartheta^1(t),\vartheta^2(t))} 
		{(\omega(t,\vartheta^1(t))\omega(t,\vartheta^2(t)))^2}
		\left(\omega(t,\vartheta^1(t))(B(t,p)-B(\vartheta^2(t),p))
		+\omega(t,\vartheta^2(t))(B(t,p)-B(\vartheta^1(t),p))\right)\\
		&\hspace{50pt}+\Res_{t\to \alpha}\frac{\d S_{t,o}(q)}{\omega(t,\vartheta^1(t))\omega(t,\vartheta^2(t))}
		\d x(p)\frac{\partial f(t,\vartheta^1(t),\vartheta^2(t))}{\partial V(p)}\Bigg).
	\end{split}
	\label{eq:dVVertex3}
\end{equation}}
As with the previous section we will drop the term containing the action of the loop insertion operator on $f(a,b,c)$ and take care of it latter.
After droping the term proceed by applying again the relation \eqref{eq:CommRes} and commute the residues in the first line of the last equality, then the residues of the type $\Res_{t\to\alpha} \sum_{i=1}^j\Res_{u\to\vartheta^i(t)}$ cancel exactly the last line of \eqref{eq:dVVertex3} and we obtain
{\footnotesize
\begin{equation}
	\begin{split}
		\d x(p)\frac{\partial }{\partial V(p)} &\Bigg(-\sum_{\alpha\in\Delta_2}
		\Res_{t\to \alpha}\frac{\d S_{t,o}(q)f(t,\vartheta^1(t),\vartheta^2(t))}
		{\omega(t,\vartheta^1(t))\omega(t,\vartheta^2(t))}
		\Bigg)=\\
		=&-\sum_{k=1}^2\sum_{\beta\in\Delta_k}\sum_{l=1}^k\Res_{u\to\beta}
		\frac{\d S_{u,o}(q)}{\omega(u,\vartheta^l(u))}
		\left(B(u,p)\left(- \sum_{\alpha\in\Delta_2} \Res_{t\to\alpha}
		\frac{\d S_{t,o}(\vartheta^l(u))}{\omega(t,\vartheta^1(t))\omega(t,\vartheta^2(t))}\right)\right.\\
		&\hspace{100pt}\left.
		+B(\vartheta^l(u),p)\left(-\sum_{\alpha\in\Delta_2} \Res_{t\to\alpha}
		\frac{\d S_{t,o}(u)}{\omega(t,\vartheta^1(t))\omega(t,\vartheta^2(t))}\right)\right) 
		f(t,\vartheta^1(t),\vartheta^2(t)).\\
	\end{split}
	\label{eq:dVV3}
\end{equation}}
The graphic representation of this equation is
\begin{equation}
	\begin{split}
		\d x(p)\frac{\partial }{\partial V(p)}
		\begin{array}{l}
			\resizebox{0.2\textwidth}{!}{\input{Vertex3a.pstex_t}}
		\end{array}
		&=
		\begin{array}{l}
			\resizebox{0.22\textwidth}{!}{\input{Vertex3b.pstex_t}}
		\end{array}+
		\begin{array}{l}
			\resizebox{0.22\textwidth}{!}{\input{Vertex3c.pstex_t}}
		\end{array}.
	\end{split}
	\label{eq:Graph3}
\end{equation}

\subsubsection{The loop insertion operator on equation \eqref{eq:recfor}}
If we apply $\d x(p)\frac{\partial }{\partial V(p)}$ to equation \eqref{eq:recforW}, and use equations \eqref{eq:dVVertex2} and \eqref{eq:dVV3} we have
{\scriptsize
\begin{equation}
	\begin{split}
		\d x(p)\frac{\partial }{\partial V(p)}\hat{w}_k^{(h)}({\bf P_K})=&
		-\sum_{k=1}^2\sum_{\beta\in\Delta_k}\sum_{l=1}^k \Res_{u\to\beta}\frac{\d S_{u,0}(p)} 
		{\omega(u,\vartheta^{l}(u))}
		\left\{\left(-\sum_{i=1}^2\sum_{\alpha\in\Delta_i}\sum_{j=1}^i \Res_{t\to\alpha} 
		\frac{\d S_{t,o}(u)} 
		{\omega(t,\vartheta^j(t))}\W_{2,k}^{(h)}(t,\vartheta^j(t),{\bf P_K})\right)B(\vartheta^l(u),(p))\right.\\
		&\hspace{10pt}+\left.\left(-\sum_{i=1}^2\sum_{\alpha\in\Delta_i}\sum_{j=1}^i\Res_{t\to\alpha}
		\frac{\d S_{t,o}(\vartheta^l(u))}{\omega(t,\vartheta^j(t))} 
		\W_{2,k}^{(h)}(t,\vartheta^j(t),{\bf P_K})\right)B(u,p)
		+\d x(p) \frac{\partial }{\partial V(p)}\W_{2,k}^{(h)}(u,\vartheta^l(u),{\bf P_K})\right\}
		\\
		&-\sum_{k=1}^2\sum_{\beta\in\Delta_k}\sum_{l=1}^k \Res_{u\to\beta}
		\frac{\d S_{u,0}(p)}{\omega(u,\vartheta^1(u))\omega(u,\vartheta^2(u))}
		\left(B(u,p)\W_{2,k}^{(h)}(\vartheta^1(u),\vartheta^2(u),{\bf P_K})\right.\\
		&\hspace{80pt}\left.+B(\vartheta^1(u),p)\W_{2,k}^{(h)}(\vartheta^2(u),u,{\bf P_K})+
		B(\vartheta^2(u),p)\W_{2,k}^{(h)}(u,\vartheta^1(u),{\bf P_K})\right)
		\\
		&-\sum_{k=1}^2\sum_{\beta\in\Delta_k}\sum_{l=1}^k \Res_{u\to\beta}\frac{\d S_{u,0}(p)} 
		{\omega(u,\vartheta^{l}(u))}\left\{\left(-\sum_{\alpha\in\Delta_2}\Res_{t\to\alpha}
		\frac{\d S_{t,o}(u)}{\omega(t,\vartheta^1(t))\omega(t,\vartheta^2(t))}
		\W_{3,k}(t,\vartheta^1(t),\vartheta^2(t),{\bf P_K})\right)B(\vartheta^l(u),p)\right.\\
		&\hspace{100pt}+\left(-\sum_{\alpha\in\Delta_2}\Res_{t\to\alpha}
		\frac{\d S_{t,o}(\vartheta^l(u))}{\omega(t,\vartheta^1(t))\omega(t,\vartheta^2(t))}
		\W_{3,k}(t,\vartheta^1(t),\vartheta^2(t),{\bf P_K})\right)B(u,p)\\
		&\hspace{100pt}\left.+\d x(p)\frac{\partial }{\partial V(p)}
		W_{3+k}^{(h)}(t,\vartheta^1(p),\vartheta^2(p),{\bf P_K})\right\}.\\
	\end{split}
\end{equation}
}
This equation simplifies considerably by regrouping some terms
{\footnotesize
\begin{equation}
	\begin{split}
		\d x(p)\frac{\partial }{\partial V(p)}\hat{w}_k^{(h)}({\bf P_K})=&
		-\sum_{k=1}^2\sum_{\beta\in\Delta_k}\sum_{l=1}^k \Res_{u\to\beta}
		\frac{\d S_{u,o}(p)}{\omega(u,\vartheta^l(u))}
		\left\{w_{1+k}^{(h)}(u,{\bf P_K})w_2^{(0)}(\vartheta^l(u),p)+
		w_{1+k}^{(h)}(\vartheta^l(u),{\bf P_K})w_2^{(0)}(u,p)\right.\\
		&\left.\hspace{150pt}+\d x(p)\frac{\partial }{\partial V(p)}
		\W_{2,k}^{(h)}(u,\vartheta^l(u),{\bf P_K})\right\}\\
		&-\sum_{\beta\in\Delta_2}\Res_{u\to\beta}
		\frac{\d S_{u,o}(p)}{\omega(u,\vartheta^1(u))\omega(u,\vartheta^2(u))}
		\left\{w_2^{(0)}(u)\W_{2,k}^{(h)}(\vartheta^1(u),\vartheta^2(u),{\bf P_K})
		+w_2^{(0)}(\vartheta^1(u))\W_{2,k}^{(h)}(\vartheta^2(u),u,{\bf P_K})\right.\\
		&\hspace{100pt}\left.+w_2^{(0)}(\vartheta^2(u))\W_{2,k}^{(h)}(u,\vartheta^1(u),{\bf P_K})
		+\d x(p)\frac{\partial }{\partial V(p)}\W_{3+k}^{(h)}(u,\vartheta^1(u),\vartheta^2(u))\right\},
	\end{split}
	\label{eq:GRE.a}
\end{equation}
}
which is of the form \eqref{eq:recforhat}. One can check that 
{\footnotesize
\begin{equation}
	\begin{split}
		\W_{2,k+1}^{(h)}(u,\vartheta^l(u),{\bf P_K},p)=&w_{1+k}^{(h)}(u,{\bf P_K})w_2^{(0)}(\vartheta^l(u),p)+
		w_{1+k}^{(h)}(\vartheta^l(u),{\bf P_K})w_2^{(0)}(u,p)
		+\d x(p)\frac{\partial }{\partial V(p)}
		\W_{2,k}^{(h)}(u,\vartheta^l(u),{\bf P_K})\\
		\W_{3,k+1}^{(h)}(u,\vartheta^1(u),\vartheta^2(u),{\bf P_K},p)=&
		w_2^{(0)}(u)\W_{2,k}^{(h)}(\vartheta^1(u),\vartheta^2(u),{\bf P_K})
		+w_2^{(0)}(\vartheta^1(u))\W_{2,k}^{(h)}(\vartheta^2(u),u,{\bf P_K})\\
		&\hspace{50pt}+w_2^{(0)}(\vartheta^2(u))\W_{2,k}^{(h)}(u,\vartheta^1(u),{\bf P_K})
		+\d x(p)\frac{\partial }{\partial V(p)}\W_{3+k}^{(h)}(u,\vartheta^1(u),\vartheta^2(u)),
	\end{split}
\end{equation}
}
which finally proves that $\hat{w}_{k+1}^{(h)}({\bf P_K},p)=
\d x(p)\frac{\partial }{\partial V(p)}\hat{w}_{k}^{(h)}({\bf P_K})$.
Consequently we prove that $w_{k}^{(h)}({\bf P_K})=\hat{w}_{k}^{(h)}({\bf P_K})$.
Thus, the residue formula for $w_k^{(h)}({\bf P_K})$ is
\begin{equation}
	\begin{split}
		w_{k+1}^{(h)}(q,{\bf P_K})=&-\sum_{i=1}^2\sum_{\alpha\in\Delta_i}\sum_{j=1}^i 
		\Res_{t\to \alpha}\frac{\d S_{t,o}(q)\W_{2,k}^{(h)}(t,\vartheta^j(t),{\bf P_K})} 
		{\omega(t,\vartheta^j(t))}\\
		&-\sum_{\alpha\in\Delta_2}\Res_{t\to \alpha}\frac{\d S_{t,o}(q) 
		\W_{3,k}^{(h)}(t,\vartheta^1(t),\vartheta^2(t),{\bf P_K})}
		{\omega(p,\vartheta^1(p))\omega(p,\vartheta^2(p))}.
	\end{split}
	\label{eq:recfork}
\end{equation}

\subsubsection{Diagrammatic representation of $w_k^{(h)}$}\label{Sec:DRwkh}

The equation \eqref{eq:recfork} has the following graphical representation 
\begin{equation}
	\begin{split}
		\begin{array}{l}
			\resizebox{0.2\textwidth}{!}{\input{wk+1h.pstex_t}}
		\end{array}&=
		\begin{array}{l}
			\resizebox{0.2\textwidth}{!}{\input{ww2hk.pstex_t}}
		\end{array}+
		\sum_{{\bf J}\subset {\bf K}}\sum_{m=0}^{h-1}
		\begin{array}{l}
			\resizebox{0.2\textwidth}{!}{\input{w1hw1hk.pstex_t}}
		\end{array}\\
		&+
		\begin{array}{l}
			\resizebox{0.2\textwidth}{!}{\input{ww3hk.pstex_t}}
		\end{array}
		+\sum_{{\bf J_1},{\bf J_2}\subset {\bf K}}\sum_{m=0}^{h-2}\sum_{n=0}^{h-m-1}
		\begin{array}{l}
			\resizebox{0.3\textwidth}{!}{\input{w1hw1hw1hk.pstex_t}}
		\end{array}
		\\+\sum_{{\bf J}\subset {\bf K}}\sum_{m=0}^{h-1}&\left(
		\begin{array}{l}
			\resizebox{0.2\textwidth}{!}{\input{w1hw2hpk.pstex_t}}
		\end{array}+
		\begin{array}{l}
			\resizebox{0.2\textwidth}{!}{\input{w1hw2hqk.pstex_t}}
		\end{array}+
		\begin{array}{l}
			\resizebox{0.2\textwidth}{!}{\input{w1hw2hrk.pstex_t}}
		\end{array}\right).
	\end{split}
\end{equation}
Using the diagrammatic equation we can see how the $h$-order $k$-point resolvent differentials $w_k^{(h)}$ combine to construct other such differentials. We can interpret that the residue in the vertices connect two or three free legs (either of the same $w_k^{(h)}$ or of different ones). In the process some handles may be created.

We can also see that they combine in a hierarchical fashion such that the equation \eqref{eq:recfork} form a recursion relation in the Euler characteristic $\chi$. $w_2^{(0)}$ and $w_1^{(0)}$ are the differentials with highest Euler characteristic, they are --as announced earlier-- the initial conditions for the recursion.

\subsection{Variation of $w_k^{(h)}$ with respect to the moduli}

Now the we have found the recursion formula \eqref{eq:recfork} for $w_k^{(h)}$ we can study its variation with respect to the moduli of the problem $t_a$. 
We expect the variation of $w_k^{(h)}$ with respect to $t_a$ to be
\begin{equation}
\begin{split}
	\partial_{t_a}W_k^{(h)}({\bf P_K})=-{\cal J}_a\left(W_{k+1}^{(h)}({\bf P_K},\cdot)\right).
\end{split}
\label{eq:dtWkh}
\end{equation}
We prove this equation by recursion. From the Rauch variational formula we have
{\footnotesize
\begin{equation}
\begin{split}
	\partial_{t_a}B(q,p)=&
	\sum_{i=1}^2\sum_{\alpha\in\Delta_i}\sum_{j=1}^i
	\Res_{t\to\alpha}\frac{\d S_{t,o}(q)}{\omega(t,\vartheta^j(t))}
	\left(\d \Omega_{t_a}(t)B(\vartheta^j(t),p)+\d \Omega_{t_a}(\vartheta^j(t))B(t,p)\right)\\
	=&\sum_{i=1}^2\sum_{\alpha\in\Delta_i}\sum_{j=1}^i
	\Res_{t\to\alpha}\frac{\d S_{t,o}(q)}{\omega(t,\vartheta^j(t))}
	\left({\cal J}_a(B(t,\cdot))B(\vartheta^j(t),p)+{\cal J}_{a}(B(\vartheta^j(t),\cdot))B(t,p)\right)\\
	=&{\cal J}_a\left(\sum_{i=1}^2\sum_{\alpha\in\Delta_i}\sum_{j=1}^i
	\Res_{t\to\alpha}\frac{\d S_{t,o}(q)}{\omega(t,\vartheta^j(t))}
	\left(B(t,\cdot)B(\vartheta^j(t),p)+B(\vartheta^j(t),\cdot)B(t,p)\right)
	\right)\\
	=&-{\cal J}_a\left(W_3^{(0)}(q,p,\cdot)\right),
\end{split}
\label{eq:RVF.1}
\end{equation}
}
which proves \eqref{eq:dtWkh} for $k=2$ and $h=0$. This is the first equation in the recursion.
To prove equation \eqref{eq:dtWkh} for all $k$ and $h$ we need some more results first.
We already have
\begin{equation}
\begin{split}
	\partial_{t_a}B(q,p)=-{\cal J}_a\left(\d x(\cdot)\frac{\partial}{\partial V(\cdot)}B(q,p)\right).
\end{split}
\label{eq:RVF.2}
\end{equation}
Integrating this equation we have
\begin{equation}
\begin{split}
	\partial_{t_a}\d S_{u,o}(q)=-{\cal J}_a\left(\d x(\cdot)\frac{\partial}{\partial V(\cdot)}\d S_{u,o}(q)\right).
\end{split}
\label{eq:RVF.3}
\end{equation}
We also need
\begin{equation}
\begin{split}
	\partial_{t_a}\frac{1}{\omega(t,\vartheta^j(t))}
	=&\frac{-1}{\omega(t,\vartheta^j(t))^2}\partial_{t_a}\omega(t,\vartheta^j(t))\\
	=&\frac{-1}{\omega(t,\vartheta^j(t))^2}\left(B(t,\cdot)-B(\vartheta^j(t),\cdot)\right)\\
	=&{\cal J}_a\left(-\frac{B(t,\cdot)-B(\vartheta^j(t),\cdot)}{\omega(t,\vartheta^j(t))^2}\right)\\
	=&-{\cal J}_a\left(\d x(\cdot)\frac{\partial}{\partial V(\cdot)}\frac{1}{\omega(t,\vartheta^j(t))}\right).
\end{split}
\label{eq:RVF.4}
\end{equation}
We have now enough information to apply $\partial_{t_a}$ to equation \eqref{eq:recfork}
{\scriptsize
\begin{equation}
\begin{split}
		\partial_{t_a}&w_{k+1}^{(h)}(q,{\bf 
		P_K})=\partial_{t_a}\left(-\sum_{i=1}^2\sum_{\alpha\in\Delta_i}\sum_{j=1}^i 
		\Res_{t\to \alpha}\frac{\d S_{t,o}(q)\W_{2,k}^{(h)}(t,\vartheta^j(t),{\bf P_K})} 
		{\omega(t,\vartheta^j(t))}-\sum_{\alpha\in\Delta_2}\Res_{t\to \alpha}\frac{\d S_{t,o}(q) 
		\W_{3,k}^{(h)}(t,\vartheta^1(t),\vartheta^2(t),{\bf P_K})}
		{\omega(p,\vartheta^1(p))\omega(p,\vartheta^2(p))}\right)\\
		=&\sum_{i=1}^2\sum_{\alpha\in\Delta_i}\sum_{j=1}^i 
		\Res_{t\to \alpha}\left(
		\partial_{t_a}\left(\frac{\d S_{t,o}(q)}
		{\omega(t,\vartheta^j(t))}\right)\W_{2,k}^{(h)}(t,\vartheta^j(t),{\bf P_K})
		+\frac{\d S_{t,o}(q)}{\omega(t,\vartheta^j(t))}\partial_{t_a}\W_{2,k}^{(h)}(t,\vartheta^j(t),{\bf P_K})
		\right)\\
		&-\sum_{\alpha\in\Delta_2}\Res_{t\to \alpha}\left(\partial_{t_a}
		\left(\frac{\d S_{t,o}(q)}{\omega(p,\vartheta^1(p))\omega(p,\vartheta^2(p))}\right)
		\W_{3,k}^{(h)}(t,\vartheta^1(t),\vartheta^2(t),{\bf P_K})+
		\frac{\d S_{t,o}(q)}{\omega(p,\vartheta^1(p))\omega(p,\vartheta^2(p))}
		\partial_{t_a}\W_{3,k}^{(h)}(t,\vartheta^1(t),\vartheta^2(t),{\bf P_K})\right).
\end{split}
\label{eq:dtWkh.1}
\end{equation}
}
By recursion hypothesis we assume that
\begin{equation}
\begin{split}
	\partial_{t_a}W_{k^\prime}^{(h^\prime)}({\bf P_K})=-{\cal J}_a\left(W_{k^\prime+1}^{(h^\prime)}({\bf 
	P_K},\cdot)\right),
\end{split}
\label{eq:dtWkh.2}
\end{equation}
where $\chi\left(W_{k^\prime}^{(h^\prime)}\right)>1-2h-k$. With that assumption we prove that
\begin{equation}
\begin{split}
	\partial_{t_a}\W_{2,k}^{(h)}(a,b,{\bf P_K})=&-{\cal J}_a\left(\d x(\cdot)\frac{\partial}{\partial V(\cdot)}
	\W_{2,k}^{(h)}(a,b,{\bf P_K})\right),\\
	\partial_{t_a}\W_{3,k}^{(h)}(a,b,c,{\bf P_K})=&-{\cal J}_a\left(\d x(\cdot)\frac{\partial}
	{\partial V(\cdot)}
	\W_{3,k}^{(h)}(a,b,c,{\bf P_K})\right).\\
\end{split}
\label{eq:dtWkh.3}
\end{equation}
Using this result and \eqref{eq:RVF.2}, \eqref{eq:RVF.3} and \eqref{eq:RVF.4} into equation \eqref{eq:dtWkh} we have
{\footnotesize
\begin{equation}
\begin{split}
	\partial_{t_a}w_{k+1}^{(h)}(q,{\bf P_K})=&-{\cal J}_a\left(\d x(\cdot)\frac{\partial}{\partial V(\cdot)}
	\left(-\sum_{i=1}^2\sum_{\alpha\in\Delta_i}\sum_{j=1}^i 
		\Res_{t\to \alpha}\frac{\d S_{t,o}(q)\W_{2,k}^{(h)}(t,\vartheta^j(t),{\bf P_K})} 
		{\omega(t,\vartheta^j(t))}\right)
		\right.\\&\hspace{30pt}\left.
		+\d x(\cdot)\frac{\partial}{\partial V(\cdot)}\left(-\sum_{\alpha\in\Delta_2}\Res_{t\to \alpha}
		\frac{\d S_{t,o}(q) 
		\W_{3,k}^{(h)}(t,\vartheta^1(t),\vartheta^2(t),{\bf P_K})}
		{\omega(p,\vartheta^1(p))\omega(p,\vartheta^2(p))}\right)\right)\\
		=&-{\cal J}_a\left(\d x(\cdot)\frac{\partial}{\partial V(\cdot)} W_{k+1}^{(h)}(q,{\bf P_K})\right)\\
		=&-{\cal J}_a\left(W_{k+2}^{(h)}(q,{\bf P_K},\cdot)\right),
\end{split}
\label{eq:dtWkh.4}
\end{equation}
}
which proves equation \eqref{eq:dtWkh}.\newline
\Remark{Notice that in all these calculations the ${\cal J}_a$ integral operators commute with the residues at the branch-points. This is normal since any of the contours in the definitions of ${\cal J}_a$ never approach the branch-points.}

\section{Topological expansion of the free energy}\label{TEFE}

The topological expansion of the free energy $\F=\sum_{h=0}^{\infty}\left(\frac{T}{N}\right)^{2h}\F^{(h)}$ is related with the $k$-point resolvent differentials $w_k^{(h)}$ by equation \eqref{eq:dVF}. Consider for the moment the simplest case $k=1$
\begin{equation}
	w_1^{(h)}(p)=-\d x(p)\frac{\partial}{\partial V(p)}\F^{(h)}.
	\label{eq:F-w1}
\end{equation}
By inverting the loop insertion operator we can find the coefficients $\F^{(h)}$ of the free energy topological expansion.
\newline
In this section we want to prove that there is an operator that we call $H_{x(p)}$ such that
\begin{equation}
	\begin{split}
	H_{x(p)}\left[w_{k+1}^{(h)}({\bf P_K},p)\right]&=(2-2h-k)w_k^{(h)}({\bf P_K})\\
	&=\chi\left(w_{k+1}^{(h)}\right)\, w_k^{(h)}({\bf P_K})
	\end{split}
	\label{eq:Hwkh}
\end{equation}
for all $w_k^{(h)}$ with $\chi\leq 0$. 
We want to prove as well that 
\begin{equation}
	\F^{(h)}=\frac{1}{2h-2}H_{x(p)}\left[w_1^{(h)}(p)\right]\qquad\text{for } h>1,
	\label{eq:Hw1h}
\end{equation}
and so that the $H_{x(p)}$ operator is indeed related to the inverse of the loop insertion operator on all resolvent differentials with $\chi<0$.

\subsection{The $H_{x(p)}$ operator}
Consider the operators defined in \eqref{tableint} and construct from them the $H_{x(p)}$-operator
\begin{equation}
	\begin{split}
		H_{x(p)}(\psi(p))&=-\sum_{a\in{\cal M}} t_a {\cal J}_a(\psi(p)),
	\end{split}
	\label{eq:HopDef}
\end{equation}
where $\psi(p)$ is a meromorphic differential on the algebraic curve with pole structure only at the branch-points without residue. The $H_{x(p)}$ operator is an essential piece of the whole construction of the topological expansion of the free energy, and as such has appeared very often in the literature \cite{ChEyOr-06.1,EyOr-07.1,EyPr-08.1}. It has been proved in the literature that the $H_{x(p)}$ operator can be written as well as
\begin{equation}
		H_{x(p)}(\psi(p))=-\sum_{\alpha}\Res_{p\to\alpha}\Psi(p)\psi(p),
\label{eq:HopDef.b}
\end{equation}
where $\alpha$ are all the poles of $\psi(p)$, and $\Psi(p)$ is a primitive of $y(p)\d x(p)$
\begin{equation}
	\Psi(p)=\int_{o}^p y(\tilde{p}) \d x(\tilde{p}).
\end{equation}
In equation \eqref{ydxalt} we saw that $H_{x(p)}(B(p,q))=-y(q)\d x(q)$. For our purposes this result is better expressed as 
\begin{equation}
	H_{x(p)}\left[w_2^{(0)}(p,q)\right]=-y(q)\d x(q).
	\label{eq:Hw20}
\end{equation}
This is not of the form \eqref{eq:Hwkh} because $w_2^{(0)}$ has a positive $\chi$.
Notice that $H_{x(p)}$ is an integral operator that acts only on the $p$-variable.

We use the equation \eqref{eq:recfork} to prove \eqref{eq:Hwkh} by recurrence. First we prove that it is true for the first term of the recursion, in our case this term will be $w_3^{(0)}$. Next we prove \eqref{eq:Hwkh}  for generic $(k,h)$ assuming it to be true for resolvent differentials with higher $\chi$. 

\subsection{The $H_{x(p)}$ operator acting on $w_3^{(0)}$}

The first equation of the recursion \eqref{eq:recfork} is for $w_3^{(0)}$ we have to apply $H_{x(p)}$ to that equation to check equation \eqref{eq:Hwkh} on $w_3^{(0)}$ 
\begin{equation}
	\begin{split}
		H_{x(p)}\left[w_3^{(0)}(p_1,p_2,q)\right]=&
		H_{x(p)}\left[-\sum_{i=1}^2\sum_{\alpha\in\Delta_i}\sum_{j=1}^i
		\Res_{t\to\alpha}\frac{\d S_{t,o}(p_1)}{\omega(t,\vartheta^j(t))}
		\left(B(t,p_2)B(\vartheta^j(t),p)+B(\vartheta^j(t),p_2)B(t,p)\right)\right]\\
		=&-\sum_{i=1}^2\sum_{\alpha\in\Delta_i}\sum_{j=1}^i
		\Res_{t\to\alpha}\frac{\d S_{t,o}(p_1)}{\omega(t,\vartheta^j(t))}
		\left(B(t,p_2)y(\vartheta^j(t))\d x(t)+B(\vartheta^j(t),p_2)y(t)\d x(t)\right)\\
		=&-\sum_{i=1}^2\sum_{\alpha\in\Delta_i}\sum_{j=1}^i
		\Res_{t\to\alpha}\frac{\frac{1}{2}\d S_{t,\vartheta^j(t)}(p_1)}{\omega(t,\vartheta^j(t))}
		B(\vartheta^j(t),p_2)y(t)\d x(t)\\
		=&0.
	\end{split}
\end{equation}
Each of the residues on the third line vanish due to the lack of pole structure of the integrand. Notice that this result satisfies \eqref{eq:Hwkh} because $\chi\left(w_3^{(0)}\right)=0$.

\subsection{The $H_{x(p)}$ operator acting on $w_k^{(h)}$ for $k\geq2$ and $\chi\leq 0$}\label{Sec:Hwkh.2}

Consider $w_{k+2}^{(h)}$ with $\chi=1-2h-k$:
\begin{equation}
	\begin{split}
		H_{x(p)}\left[w_{k+2}^{(h)}(q,{\bf P_K},p)\right]=&
		H_{x(p)}\left[-\sum_{i=1}^2\sum_{\alpha\in\Delta_i}\sum_{j=1}^i 
		\Res_{t\to \alpha}\frac{\d S_{t,o}(q)\W_{2,k}^{(h)}(t,\vartheta^j(t),{\bf P_K},p)} 
		{\omega(t,\vartheta^j(t))}\right.\\
		&\left.-\sum_{\alpha\in\Delta_2}\Res_{t\to \alpha}\frac{\d S_{t,o}(q) 
		\W_{3,k}^{(h)}(t,\vartheta^1(t),\vartheta^2(t),{\bf P_K},p)}
		{\omega(p,\vartheta^1(p))\omega(p,\vartheta^2(p))}\right].
	\end{split}
	\label{eq:Hwkh.a}
\end{equation}
Next we introduce equation \eqref{eq:recforW} to make the differentials $w_2^{(0)}$ explicit\footnote{Note that the only $k$-point resolvent differential in $\W_{r,k}^{(h)}$ that do not satisfy \eqref{eq:Hwkh} is $w_2^{(0)}$. Making them explicit we can treat them separately.} and express \eqref{eq:Hwkh.a} as the sum of three terms
{\footnotesize
\begin{equation}
	\begin{split}
		H_{x(p)}\left[w_{k+2}^{(h)}(q,{\bf P_K},p)\right]=&
		H_{x(p)}\Bigg[-\sum_{i=1}^2\sum_{\alpha\in\Delta_i}\sum_{j=1}^i 
		\Res_{t\to \alpha}\frac{\d S_{t,o}(q)} 
		{\omega(t,\vartheta^j(t))}\left(w_2^{(0)}(t,p)w_{1+k}^{(h)}(\vartheta^j(t),{\bf P_K})+
		w_2^{(0)}(\vartheta^j(t),p)w_{1+k}^{(h)}(t,{\bf P_K})\right)\\
		&\hspace{50pt}-\sum_{\alpha\in\Delta_2}\Res_{t\to \alpha}\frac{\d S_{t,o}(q)}
		{\omega(p,\vartheta^1(p))\omega(p,\vartheta^2(p))}
		\left(w_2^{(0)}(t,p)\W_{2,k}^{(h)}(\vartheta^1(t),\vartheta^2(t),{\bf P_K})\right.\\
		&\hspace{80pt}\left.+w_2^{(0)}(\vartheta^1(t),p)\W_{2,k}^{(h)}(\vartheta^2(t),t,{\bf P_K})+
		w_2^{(0)}(\vartheta^2(t),p)\W_{2,k}^{(h)}(t,\vartheta^1(t),{\bf P_K})\right)\Bigg]\\
		&+H_{x(p)}\left[-\sum_{i=1}^2\sum_{\alpha\in\Delta_i}\sum_{j=1}^i 
		\Res_{t\to \alpha}\frac{\d S_{t,o}(q)}{\omega(t,\vartheta^j(t))}
		\d x(p) \frac{\partial}{\partial V(p)}\W_{2,k}^{(h)}(t,\vartheta^j(t),{\bf P_K})\right]\\
		&+H_{x(p)}\left[-\sum_{\alpha\in\Delta_2}\Res_{t\to \alpha}\frac{\d S_{t,o}(q)}
		{\omega(p,\vartheta^1(p))\omega(p,\vartheta^2(p))} 
		\d x(p)\frac{\partial}{\partial V(p)}W_{3,k}^{(h)}(t,\vartheta^1(t),\vartheta^2(t),{\bf P_K})\right].
	\end{split}
	\label{eq:Hwkh.b}
\end{equation}
}
Let us treat each one of those three terms separately.
The first one needs to be further expanded using equation \eqref{eq:recfork} for $w_{1+k}^{(h)}(a,{\bf P_K})$
{\scriptsize
\begin{equation}
	\begin{split}
		H_{x(p)}&\left[-\sum_{i=1}^2\sum_{\alpha\in\Delta_i}\sum_{j=1}^i 
		\Res_{t\to \alpha}\frac{\d S_{t,o}(q)} 
		{\omega(t,\vartheta^j(t))}\left(w_2^{(0)}(t,p)w_{1+k}^{(h)}(\vartheta^j(t),{\bf P_K})+
		w_2^{(0)}(\vartheta^j(t),p)w_{1+k}^{(h)}(t,{\bf P_K})\right)\right.\\
		&-\sum_{\alpha\in\Delta_2}\Res_{t\to \alpha}\frac{\d S_{t,o}(q)}
		{\omega(p,\vartheta^1(p))\omega(p,\vartheta^2(p))}
		\left(w_2^{(0)}(t,p)\W_{2,k}^{(h)}(\vartheta^1(t),\vartheta^2(t),{\bf P_K})
		\right.\\&\hspace{80pt}\left.\left.
		+w_2^{(0)}(\vartheta^1(t),p)\W_{2,k}^{(h)}(\vartheta^2(t),t,{\bf P_K})+
		w_2^{(0)}(\vartheta^2(t),p)\W_{2,k}^{(h)}(t,\vartheta^1(t),{\bf P_K})\right)\right]\\
		=&H_{x(p)}\left[-\sum_{i=1}^2\sum_{\alpha\in\Delta_i}\sum_{j=1}^i
		\Res_{t\to \alpha}\frac{\d S_{t,o}(q)} {\omega(t,\vartheta^j(t))}
		\left(w_2^{(0)}(t,p)\left(-\sum_{k=1}^2\sum_{\beta\in\Delta_k}\sum_{l=1}^k\Res_{u\to\beta}
		\frac{\d S_{u,0}(\vartheta^j(t))}{\omega(u,\vartheta^l(u))}\W_{2,k}^{(h)}(u,\vartheta^l(u),{\bf P_K})\right)
		\right.\right.
		\\&\hspace{80pt}\left.
		+w_2^{(0)}(\vartheta^j(t),p)\left(-\sum_{k=1}^2\sum_{\beta\in\Delta_k}\sum_{l=1}^k\Res_{u\to\beta}
		\frac{\d S_{u,0}(t)}{\omega(u,\vartheta^l(u))} 
		\W_{2,k}^{(h)}(u,\vartheta^l(u),{\bf P_K})\right)\right)\\
		&-\sum_{\alpha\in\Delta_2}\Res_{t\to \alpha}\frac{\d S_{t,o}(q)}
		{\omega(p,\vartheta^1(p))\omega(p,\vartheta^2(p))}
		\left(w_2^{(0)}(t,p)\W_{2,k}^{(h)}(\vartheta^1(t),\vartheta^2(t),{\bf P_K})
		\right.\\&\hspace{80pt}\left.
		+w_2^{(0)}(\vartheta^1(t),p)\W_{2,k}^{(h)}(\vartheta^2(t),t,{\bf P_K})+
		w_2^{(0)}(\vartheta^2(t),p)\W_{2,k}^{(h)}(t,\vartheta^1(t),{\bf P_K})\right)\Bigg]\\
		&+H_{x(p)}\left[-\sum_{i=1}^2\sum_{\alpha\in\Delta_i}\sum_{j=1}^i
		\Res_{t\to \alpha}\frac{\d S_{t,o}(q)} {\omega(t,\vartheta^j(t))}
		\left(w_2^{(0)}(t,p)\left(-\sum_{\beta\in\Delta_2}\Res_{u\to\beta}
		\frac{\d S_{u,0}(\vartheta^j(t))}{\omega(u,\vartheta^1(u))\omega(u,\vartheta^2(u))}
		\W_{3,k}^{(h)}(u,\vartheta^1(u),\vartheta^2(u),{\bf P_K})\right)
		\right.\right.\\&\hspace{80pt}\left.\left.
		+w_2^{(0)}(\vartheta^j(t),p)\left(-\sum_{\beta\in\Delta_k}\Res_{u\to\beta}
		\frac{\d S_{u,0}(t)}{\omega(u,\vartheta^1(u))\omega(u,\vartheta^2(u))} 
		\W_{3,k}^{(h)}(u,\vartheta^1(u),\vartheta^2(u),{\bf P_K})\right)\right)\right].
	\end{split}
	\label{eq:A34}
\end{equation}
}
By applying equation \eqref{eq:Hw20}, commuting the residues and after some lengthy algebra (see appendix \ref{app:Hrules} for a more detailed report of this computation) we get the following contribution
\begin{equation}
	\begin{split}
		&\sum_{k=1}^2\sum_{\beta\in\Delta_k}\sum_{l=1}^i \Res_{u\to\beta}
		\frac{\d S_{u,o}(q)}{\omega(u,\vartheta^l(u))} \W_{2,k}^{(h)}(u,\vartheta^l(u),{\bf P_K})\\
		&+2\sum_{\beta\in\Delta_2} \Res_{u\to\beta} 
		\frac{\d S_{u,o}(q)}{\omega(u,\vartheta^1(u),\vartheta^2(u))}
		\W_{3,k}(u,\vartheta^1(u),\vartheta^2(u),{\bf P_K}).
	\end{split}
	\label{eq:Hwkh.c}
\end{equation}

The second and third terms of \eqref{eq:Hwkh.b} are much simpler. Notice that by the recursion hypothesis we assume \eqref{eq:Hwkh} to be true for $\chi(w_{k^\prime}^{(h^\prime)})>(1-2h-k)$. It can be seen that all the resolvent differentials in the expression $\d x(p)\frac{\partial}{\partial V(p)}\W_{r,k}^{(h)}$ have
an index $\chi\geq r-2h-k$ for $r=2,3$ and so we can use \eqref{eq:Hwkh} for them. Applying repeatedly \eqref{eq:Hwkh} we have
\begin{equation}
	\begin{split}
		H_{x(p)}\left[\d x(p)\frac{\partial}{\partial V(p)}\W_{2,K}^{(h)}(a,b,{\bf P_K})\right]=&
		-(2h+k-2)\W_{2,K}^{(h)}(a,b,{\bf P_K}),\\
		H_{x(p)}\left[\d x(p)\frac{\partial}{\partial V(p)}\W_{3,K}^{(h)}(a,b,c,{\bf P_K})\right]=&
		-(2h+k-3)\W_{3,K}^{(h)}(a,b,c,{\bf P_K}),\\
	\end{split}
	\label{eq:Hwkh.d}
\end{equation}
Collecting all the contributions from \eqref{eq:Hwkh.c} and \eqref{eq:Hwkh.d} back into \eqref{eq:Hwkh.b} we have 
\begin{equation}
	\begin{split}
		H_{x(p)}\left[w_{2+k}^{(h)}(q,{\bf P_K},p)\right]=& 
		-(2h+(k+1)-2)\Bigg(
		-\sum_{k=1}^2\sum_{\beta\in\Delta_k}\sum_{l=1}^i \Res_{u\to\beta}
		\frac{\d S_{u,o}(q)}{\omega(u,\vartheta^l(u))} \W_{2,k}^{(h)}(u,\vartheta^l(u),{\bf P_K})\\
		&-\sum_{\beta\in\Delta_2} \Res_{u\to\beta} 
		\frac{\d S_{u,o}(q)}{\omega(u,\vartheta^1(u),\vartheta^2(u))}
		\W_{3,k}(u,\vartheta^1(u),\vartheta^2(u),{\bf P_K})\Bigg)\\
		=&-(2h+(k+1)-2)w_{k+1}^{(h)}(q,{\bf P_K}),
	\end{split}
\end{equation}
which by recursion proves equation \eqref{eq:Hwkh}.

\subsection{The $H_{x(p)}$ operator acting on $w_1^{(h)}$ for $\chi<0$: Residue formula for $\F^{(h)}$}

Our goal in this section is to prove equation \eqref{eq:Hw1h}. 

Consider the following hierarchy of constants (depending only on the moduli of the matrix model)
\begin{equation}
	\tF^{(h)}=\frac{1}{2h-2}H_{x(p)}\left[w_1^{(h)}(p)\right]
	=\frac{-1}{2h-2}\sum_\alpha \Res_{p\to\alpha} \Psi(p)w_1^{(h)}(p)
\end{equation}
for $h>1$, where as usual $\alpha$ are the branch-points of the algebraic curve. We have used here the form \eqref{eq:HopDef.b} of the $H_{x(p)}$ operator.
The claim is that these constants $\tF^{(h)}=\F^{(h)}$ are equal to the coefficients of the topological expansion of the free energy. As a first check we act with the loop insertion operator on $\tF^{(h)}$
\begin{equation}
	\begin{split}
		(2h-2)\d{x}(p)\frac{\pd}{\pd V(p)} \tF^{(h)} &= -\d{x}(p)\frac{\pd}{\pd V(p)} 
		\sum_\alpha\Res_{q\to\alpha} \Psi(q)w_1^{(h)}(q)\\
		&=-\sum_\alpha\Res_{q\to\alpha} \left(\d{x}(p)\frac{\pd}{\pd V(p)} \Psi(q)\right)w_1^{(h)}(q)
		-\sum_\alpha\Res_{q\to\alpha} \Psi(q)\left(\d{x}(p)\frac{\pd}{\pd V(p)} w_1^{(h)}(q)\right).
	\end{split}
\label{eq:dVtF} 
\end{equation} 
The action of the loop insertion operator on the first term is
\begin{equation}
	\begin{split}
		\d{x}(p)\frac{\pd}{\pd V(p)} \Psi(q)&=\int_o^q \d{x}(p)\frac{\pd}{\pd V(p)} y(r)\d x(r) \\
		&= - \int_{o}^{q} \left(B(p,r)-\frac{1}{3}\frac{\d x(p)\d x(r)}{(x(p)-x(r))^2}\right)\\
		&= - \d S_{q,o}(p) + \frac{\d{x}(p)}{3}\left(\frac{1}{x(p)-x(q)}-\frac{1}{x(p)-x(o)}\right)
	\end{split}
\end{equation} 
and so the result of the residue is
\begin{equation}
	\begin{split}
		&\sum_\alpha\Res_{q\to\alpha} \left(\frac{\pd}{\pd V(p)} \Psi(q)\right)w_1^{(h)}(q)\\
		&=- w_1^{(h)}(p) - \sum_{i=0}^2 w_1^{(h)}(p^{(i)})\\
		&=- w_1^{(h)}(p)
	\end{split}
\end{equation} 
after moving contours and integrating.
The second term of \eqref{eq:dVtF} is
\begin{equation}
	\begin{split}
		&\sum_\alpha\Res_{q\to\alpha} \Psi(q)\d{x}(p)\frac{\pd}{\pd V(p)} w_1^{(h)}(q)\\
		&=-H_{x(q)}\left[w_2^{(h)}(q,p)\right]\\
		&=(2h-1) w_1^{(h)}(p).
	\end{split}
\end{equation} 
And collecting both contributions
\begin{equation}
	\d{x}(p)\frac{\pd}{\pd V(x)} \tF^{(h)} = - w_1^{(h)}.
	\label{eq:dVF2} 
\end{equation} 
This is telling us that $\tF^{(h)}-\F^{(h)}$ may only depend on $T$ and the filling fractions $\epsilon$.

In fact we can work out any variation with respect to the parameters $t_a\in\calM$, 
\begin{equation}
	\begin{split}
		(2h-2)\frac{\pd}{\pd t_a}\tF^{(h)}&= -
		\sum_\alpha \Res_{p\to\alpha} \left(\frac{\pd}{\pd t_a} \Psi(p)\right) w_1^{(h)}(p)
		-\sum_\alpha \Res_{p\to\alpha} \Psi(p)\left(\frac{\pd}{\pd t_a} w_1^{(h)}(p)\right) \\
		&=-\sum_\alpha \Res_{p\to\alpha} \left(\int_{r=o}^{r=p} \mathcal J_a(B(r,\cdot)) w_1^{(h)}(p)\right)
		+\sum_\alpha \Res_{p\to\alpha} \Psi(p) \mathcal J_a \left(\w_2^{(h)}(p,\cdot)\right) \\
		&=-\mathcal J_a\left( \sum_\alpha \Res_{p\to\alpha} \d S_{p,o}(\cdot)w_1^{(h)}(p)\ri) 
		-\mathcal J_a\left( H_{x(p)} \w_2^{(h)}(p,\cdot)\ri) \\
		&=-\mathcal J_a\left(w_1^{(h)}(\cdot)\ri)
		+ (2h-1)\mathcal J_a(w_1^{(h)}(\cdot))\\
		&=(2h-2)\mathcal J_a\left(w_1^{(h)}{\cdot}\ri),\\
		\Rightarrow\frac{\pd}{\pd t_a}\tF^{(h)}&=\mathcal J_a( w_1^{(h)}(\cdot)),
	\end{split}
\label{eq:dtFh}
\end{equation} 
where we have used the relation $\frac{\partial}{\partial_{t_a}}w_k^{(h)}=-{\cal J}_a\left(w_{k+1}^{(h)}\right)$ proved above. These are expected equations for $\F^{(h)}$ so we get to the conclusion that $\tF^{(h)}-\F^{(h)}$ does not depend on any parameter of the model and so
\begin{equation}
	\F^{(h)}=\frac{1}{2h-2}H_{x(p)}\left[w_1^{(h)}(p)\right].
\end{equation}

\subsection{Homogeneity of $\F^{(h)}$}

Now we can check the homogeneity equation for $\F^{(h)}$
\begin{equation}
	\begin{split}
		\sum_{a\in{\cal M}}t_a\frac{\partial}{\partial t_a}\F^{(h)}=&
		\sum_{a} t_a {\cal J}_a\left(w_1^{(h)}(\cdot)\right)\\
		=&-H_{x(p)}\left[w_1^{(h)}(p)\right],
	\end{split}
\end{equation}
which implies $\sum_{a}t_a\partial_{t_a}\F^{(h)}=(2h-2)\F^{(h)}$

\subsection{Commutation relation of $H_{x(p)}$ and $\d x(q)\frac{\partial}{\partial V(q)}$}

In the space of $\chi<0$ resolvent multi-differentials these two operators satisfy
\begin{equation}
	\left[H_{x(p)},\d x(q)\frac{\partial}{\partial V(q)}\right]=-\id,
	\label{eq:HdVCommRel.a}
\end{equation}
where $\id$ is the identity operator. Indeed, define $w_0^{(h)}=-\F^{(h)}$ and consider the commutator of the $H_{x(p)}$ operator and the loop insertion operator acting on a generic resolvent multi-differential $w_{k+1}^{(h)}$ with $k\geq 0$ and $\chi<0$,
\begin{equation}
	\begin{split}
	\left[H_{x(p)},\d x(q)\frac{\partial}{\partial V(q)}\right] w_{k+1}^{(h)}(p,{\bf P_K})=&
	 H_{x(p)}\left[\d x(q)\frac{\partial}{\partial V(q)}w_{k+1}^{(h)}(p,{\bf P_K})\right]
	 -\d x(q)\frac{\partial}{\partial V(q)}H_{x(p)}\left[w_{k+1}^{(h)}(p,{\bf P_K})\right]\\
	 =& -(2h+(k+1)-2)w_{k+1}^{(h)}(q,{\bf P_K})+(2h+k-2)w_{k+1}^{(h)}(q,{\bf P_K})\\
	 =& -w_{k+1}^{(h)}(q,{\bf P_K}),
	\end{split}
	\label{eq:HdVCommRel.b}
\end{equation}
which proves the statement.

\subsection{Computation of leading and next to leading orders of the free energy.}

The leading order and the next to leading order of the topological expansion of the free energy cannot be found through formulas like \eqref{eq:Hw1h}, they have to be computed independently instead. Fortunately in our case not much is changed from the case already studied in \cite{BePr-09.1}. Indeed, The leading term $\F^{(0)}$ (as with all $w_k^{(0)}({\bf P_K})$) is not changed by the convergence of the two branch-points, and as for $\F^{(1)}$, the recursion formula for $w_1^{(1)}(p)$ is identical to the usual Eynard-Orantin formula, and thus $\F^{(1)}$ remains unchanged except for the presence of a hard edge.
For the sake of completeness we include them here without its derivation, which can be found in the literature (see \cite{BePr-09.1}).

\subsubsection{Formula for $\F^{(0)}$}
Following \cite{BePr-09.1} we have
\begin{equation}
	\F^{(0)}=\sum_{a}t_a {\cal J}_a(y(p)\d x(p)) = \sum_{a,b}t_a t_b {\cal J}_a\otimes{\cal J}_b(B(p,q)),
\end{equation}
where the operators ${\cal J}_a$ are defined by equation \eqref{tableint}. The reader will find more information in \cite{BePr-09.1} and \cite{Bert-06.1}

\subsubsection{Formula for $\F^{(1)}$}
The first order contribution to the topological expansion of the free energy is almost the same as the one found in \cite{BePr-09.1}
\begin{equation}
	\F^{(1)}= -\frac{1}{24}\ln({\tau_{B_x}}^{12} (y^{-2}(\beta))^8\prod_{\alpha\in\Delta_1}y^\prime(\alpha)).
\end{equation}
where $y^{-2}(\beta)$ is the coefficient of $z(p)^{-2}$ in the jet expansion of $y(p)$ around $\beta\in\Delta_2$,\footnote{It has been already pointed out elsewhere that $y(x)\propto x^{-2/3}$ as $x\to 0$ whenever $x=0$ becomes a hard edge for both $M_1$ and $M_2$ (see for example \cite{BeGeSz-08.2,BePr-09.1}), thus the jet expansion must begin at order $z(p)^{-2}$} $y(p)=y^{-2}(\beta)z(p)^{-2}+y^{-1}(\beta)z(p)^{-1}+y^0(\beta)+\OO{z(p)}$ where $z(p)$ is a local coordinate around $\beta$.
The only difference with the expression in \cite{BePr-09.1} is that the product over the branch-points does not include the hard edge at $\beta$ whose contribution is given by the $(y^{-2}(\beta))^8$ term. The derivation of this term follows exactly the same lines as that of the other terms.
The contribution of the double branch point $\beta$ to $w_1^{(1)}$ is
\begin{equation}
\begin{split}
	-\Res_{p\to\beta} \d S_{p,o}(q)\left(\frac{B(p,\vartheta^1(p))}{\omega(p,\vartheta^1(p))}
	+\frac{B(p,\vartheta^2(p))}{\omega(p,\vartheta^2(p))}\right)&=\frac{1}{9}\frac{B(\beta,q)}{y^{-2}(\beta)},
\end{split}
\end{equation}
where $B(\beta,q)=\left.\frac{B(p,q)}{\d z(p)}\right|_{p\to\beta}$ and $z(p)$ is a local coordinate around $\beta$.
Looking at the jet expansion of the equation \eqref{eq:DIF4} we obtain at leading order the relation
\begin{equation}
\begin{split}
	3\d x(q)\frac{\partial}{\partial V(q)} y^{-2}(\beta) = - B(\beta,q))
\end{split}
\end{equation}
This proves the contribution $(y^{-2}(\beta))^8$ in $\F^{(1)}$.

\section{Conclusion}\label{Sec:Conc}

In this article the topological expansion for the free energy and all $k$-point resolvent correlation functions of the Cauchy matrix model with one non-simple (but non-critical) branch-point is successfully derived. This model lies right outside the reach of the standard topological expansion where only simple branch-points are allowed.
The new and larger structure that is uncovered here contains the Eynard-Orantin topological expansion formulas and extends them to a more general situation.

Though the results of this article involve only branch-points with $n_b=2$, the possibility of generalization to the situation where the multiple branch-points are of any order seems straightforward. Those equations are not written yet but an immediate guess can be formulated from the results here and will be precisely derived in a future work.

\appendix

\section{Formal Matrix Models}\label{App:FMM}

In this section we present a simplified definition of Formal Matrix Models. For more complete definitions see \cite{Ey-06.1,Or-07.1}.

Consider the matricial $S(M_1,M_2)$ action. The action of the Cauchy matrix model is
\begin{equation}
\begin{split}
	S(M_1,M_2)&=\frac{N}{T}\tr{V_1(M_1)}+\frac{N}{T}\tr{V_2(M_2)}+\tr{\ln (M_1\otimes\id+\id\otimes M_2)}.
\end{split}
\end{equation}
The formal matrix model is defined as a controlled perturbative expansion of the action around one of its minima. For this purpose we define $M_1^*$ and $M_2^*$ as the value of the matrices on one of these minima. Thanks to the $U(N)$ invariance of the action (and the measure) we can supose $M_i^*$ to be diagonal and to avoid complications that would only obscure the definition will present only the one-cut situation (or genus zero case) in which $M_i^*$ is proportional to the identity. We refer to the above cited works for a more complete definition. 
Around this particular minima we have
\begin{equation}
\begin{split}
	S(M_1,M_2)=S(M_1^*,M_2^*)+\frac{N t_2^{(1)}}{T}(\delta M_1)^2+\frac{N t_2^{(2)}}{T}(\delta M_2)^2
	+\delta S(M_1^*,M_2^*,\delta M_1,\delta M_2),
\end{split}
\end{equation}
where under the assumptions above we have $\delta M_i= M_i-m_i^*\id$ and $m_i^*$ is the value of the unique eigenvalue of $M_i^*$ and $t_2^{(i)}$ is the coefficient corresponding to the quadratic term in the action.
For simplicity we drop the dependence on $M_i^*$ of $\delta S$.
Consider the finite gaussian integral, it is proven that such an integral is just a finite polynomial in $\frac{T}{N}$
\begin{equation}
\begin{split}
	\int \d M_1 \d M_2 \left(\delta S(M_1-m_1^*\id,M_2-m_2^*\id) \right)^n 
	\e{-\frac{N}{T}\left(t_2^{(1)}(M_1-m_1^*\id)^2+t_2^{(2)}(M_2-m_2^*\id)^2\right)}&
	=\sum_l \left(\frac{T}{N}\right)^{l} A_{n,l}.
\end{split}
\end{equation}
Note the combinatorial interpretation of this integral in terms of Feynman graphs gives meaning to the quantity $\frac{T}{N}$ as a means of grouping together terms with different topology, thus the name topological expansion.
We call formal matrix model the sum
\begin{equation}
\begin{split}
	\e{-S(M_1^*,M_2^*)}&
	\sum_{n=0}^{\infty}\frac{(-1)^n}{n!}\int \d M_1 \d M_2 \left(\delta S(M_1-m_1^*\id,M_2-m_2^*\id) \right)^n 
	\e{-\frac{N}{T}\left(t_2^{(1)}(M_1-m_1^*\id)^2+t_2^{(2)}(M_2-m_2^*\id)^2\right)}\\
	&=\e{-S(M_1^*,M_2^*)}\sum_{l}\left(\frac{T}{N}\right)^l\sum_{n}\frac{(-1)^n}{n!}A_{n,l},
\end{split}
\label{eq:Form}
\end{equation}
where the $n$-sum has a finite number of terms for every value of $l$ but the whole $l$-sum is divergent.
Note that, would we be allowed to exchange the integration and $n$-summation, we would obtain the regular convergent matrix integral
\begin{equation}
\begin{split}
	\int \d M_1 \d M_2 \e{-S(M_1,M_2)}.
\end{split}
\label{eq:Conv}
\end{equation}
As a means of simplifying notation we use expression \eqref{eq:Conv} instead of \eqref{eq:Form} while explicitely stating that the nature of our research involves only formal matrix models.

\section{Loop Equations and definitions}\label{App:LE&D}

\subsection{General Form}
The loop equations are a glorified version of the following fact:
suppose we have a multivariate integral on a domain $\Omega$,   $\int_\Omega f(\vec x) \d  x$ and a vector field $\dot {\vec x} = \vec h(\vec x)$. 
Then the action of the vector field on the integrand induces the integral of a ``total derivative'', namely
\be
\frac {\d {}}{\d t}   \int_{\Omega} f (\vec x)\d{} ^n x = \int_{\Omega}\left( \vec h \cdot \nabla f +f  {\rm div} (\vec h)\ri)\d{}^n x = \int_{\Omega} {\rm div}\left(f \vec h \ri) = -\int_{\pa \Omega} f(\vec x) \vec h(\vec x) \d{}^{n-1} x,
\ee 
where the last integral is the ``flux'' of the vector on the boundary of $\Omega$. If either $f$ or $\vec h$ vanish on (or are tangential to) the boundary, then the variation is zero and thus yields identities amongst different integrals. In our case $\vec  x$ are the two matrices and the domain of integration $\Omega$ is the cone of positive matrices. 
Consider the vector field induced by the infinitesimal change of variable
\be
M_j \to M_j+\ep \delta M_j\ .\label{vectf}
\ee
The partition function $\ZC$ is an integral on the space of positive definite matrices; the infinitesimal variation of the integrand under (\ref{vectf}) is the divergence of a vector
\begin{equation}
\begin{split}
M_1 &\to M_1+\ep \delta M_1,\\
\ZC &\to \ZC+\ep \delta \ZC + \OO{\ep^2}.
\end{split}
	\label{eq:LE1}
\end{equation}
In order to have a vanishing variation the vector field must vanish on the set of matrices with at least one zero-eigenvalue. Under this condition we will have  $\delta \ZC=0$.
From the explicit form of $\ZC$ \eqref{eq:CMM}, $\delta \ZC=0$ is expressed as
\begin{equation}
	\BK{J(\delta M_1)}-\frac{N}{T}\BK{\tr{\delta M_1 V_1^\prime(M_1)}}
	-\BK{\Tr{1}{\delta M_1 \Tr{2}{\frac{1}{M_1\otimes\id +\id\otimes M_2}}}}=0.
	\label{eq:LE2}
\end{equation}
$J(\delta M_1)$ is the factor coming from the Jacobian of the change of variables. This is the trace of the matrix $\frac{\pd{} (\delta M_1)_{i,j}}{\pd{} (M_1)_{k,l}}$, i.e.
\begin{equation}
	J(\delta M_1)=\sum_{i,j=1}^N\frac{\pd{} (\delta M_1)_{i,j}}{\pd{} (M_1)_{i,j}}.
	\label{eq:Jac}
\end{equation}

\subsection{Quadratic and Cubic loop equations}

The equations \eqref{eq:QLE.1} and \eqref{eq:CLE.a} were already derived in \cite{BePr-09.1}. We are going to provide only the necessary definitions to read the equations and the infinitesimal change of variables originating the loop equation. We refer the reader to the cited article for details.

For the quadratic loop equation \eqref{eq:QLE.1} we need the changes of variables\begin{equation}
 \begin{split}
  \delta M_1&=\frac{T}{N}\left(\frac{1}{x-M_1}-\frac{1}{x}\right)\\
  \text{and }\delta M_2&=\frac{T}{N}\left(\frac{1}{-x-M_2}-\frac{1}{-x}\right)
 \end{split}
\end{equation} 
and the corresponding Jacobians
\begin{equation}
	\begin{split}
		J(\delta M_1)=\left(\frac{T}{N}\tr{\frac{1}{x-M_1}}\right)^2\\
		\text{and }J(\delta M_2)=\left(\frac{T}{N}\tr{\frac{1}{-x-M_2}}\right)^2.
	\end{split}
	\label{eq:MLE.2}
\end{equation}
and the definitions of the following objects (apart from $W_i$, $Y_i$ and $U_i$) 
\begin{equation*}
 \begin{split}
  P_1(x)&=\frac{T}{N}\tr{\frac{V^\prime_1(x)-V^\prime_1(M_1)}{x-M_1}},\\
  P_2(x)&=-\frac{T}{N}\tr{\frac{V^\prime_2(-x)-V_2^\prime(M_2)}{x+M_2}},\\
  I_1&=\frac{T}{N}\tr{V_1^\prime(M_1)}+\frac{T^2}{N^2}\tr{\frac{1}{M_1\otimes\id+\id\otimes M_2}},\\
  I_2&=-\frac{T}{N}\tr{V_2^\prime(M_2)}-\frac{T^2}{N^2}\tr{\frac{1}{M_1\otimes\id+\id\otimes M_2}}.
 \end{split}
\end{equation*} 
If we put all this together into \eqref{eq:LE2}, after some simplifications we have
\begin{equation}
	\BK{(Y_1(x))^2}+\BK{(Y_2(x))^2}+\BK{Y_1(x)Y_2(x)} = \BK{R(x)},
\end{equation}
where
\begin{equation}
	\begin{split}
		R(x)&=\frac{1}{3}(V_1^\prime(x)^2+V_2^\prime(-x)^2-V_1^\prime(x)V_2^\prime(-x))-P_1(x)-{P}_2(x)\\
    &=(U^\prime_1(x))^2+(U^\prime_2(x))^2+U^\prime_1(x)U^\prime_2(x)-P_1(x)-{P}_2(x).
	\end{split}
\end{equation}
To get the cubic loop equation \eqref{eq:CLE.a} we need to consider
\begin{equation}
 \begin{split}
  \delta M_1&=\frac{T^2}{N^2}\left(\frac{1}{x-M_1}-\frac{1}{x}\right)
  \left(\frac{T}{N}\tr{\frac{1}{x-M_1}}-\frac{T}{N}\tr{\frac{1}{-x-M_2}}-V_1^\prime(x)
  +V_2^\prime(-x)\right),\\
  \delta M_2&=\frac{T^2}{N^2}\left(\frac{1}{-x-M_2}-\frac{1}{-x}\right)
  \left(\frac{T}{N}\tr{\frac{1}{-x-M_2}}-\frac{T}{N}\tr{\frac{1}{x-M_1}}-V_2^\prime(-x)+V_1^\prime(x)\right),
 \end{split}
\end{equation} 
and the corresponding Jacobians
{\footnotesize
\begin{equation}
	\begin{split}
		J(\delta M_1)&=\left(\frac{T}{N}\tr{\frac{1}{\xi-M_1}}\right)^3
		-\left(\frac{T}{N}\tr{\frac{1}{\xi-M_1}}\right)^2\frac{T}{N}\tr{\frac{1}{\eta-M_2}}
		-(V_1^\prime(\xi)-V_2^\prime(\eta))\left(\frac{T}{N}\tr{\frac{1}{\xi-M_1}}\right)^2\\
		&\qquad +{\frac{T^3}{N^3}}\tr{\frac{1}{\left(\xi-M_1\right)^3}} 
		-{\frac{T^3}{N^3}}\frac{1}{\xi}\tr{\frac{1}{\left(\xi-M_1\right)^2}},\\
		J(\delta M_2)&=\left(\frac{T}{N}\tr{\frac{1}{\eta-M_2}}\right)^3
		-\left(\frac{T}{N}\tr{\frac{1}{\eta-M_2}}\right)^2\frac{T}{N}\tr{\frac{1}{\xi-M_1}}
		-(V_2^\prime(\eta)-V_1^\prime(\xi))\left(\frac{T}{N}\tr{\frac{1}{\eta-M_2}}\right)^2\\
		&\qquad +{\frac{T^3}{N^3}}\tr{\frac{1}{\left(\eta-M_2\right)^3}} 
		-\frac{T^3}{N^3}\frac{1}{\eta}\tr{\frac{1}{\left(\eta-M_2\right)^2}}.
	\end{split}
\end{equation}}
We also need to define the objects
\begin{equation*}
 \begin{split}
  Q_1(x)&=
  \frac{T^2}{N^2}\Tr{1}{\frac{V_1^\prime(x)-V_1^\prime(M_1)}{x-M_1}
  \Tr{2}{\frac{1}{M_1\otimes\id+\id\otimes M_2}}},\\
  Q_2(x)&=
  -\frac{T^2}{N^2}\Tr{2}{\frac{V_2^\prime(-x)-V_2^\prime(M_2)}{x+M_2}
  \Tr{1}{\frac{1}{M_1\otimes\id+\id\otimes M_2}}},\\
  S_1(x)&=\frac{T^2}{N^2}\Tr{1}{\frac{V_1^\prime(M_1)}{x-M_1}
  \Tr{2}{\frac{1}{M_1\otimes\id+\id\otimes M_2}}},\\
  S_2(x)&=-\frac{T^2}{N^2}\Tr{2}{\frac{V_2^\prime(M_2)}{x+M_2}
  \Tr{1}{\frac{1}{M_1\otimes\id+\id\otimes M_2}}}.
 \end{split}
\end{equation*} 
Putting all together we find three cubic equations, one for each $Y_i$
\begin{equation}
\begin{split}
	\BK{Y_k(\x)^3}-\BK{R(\x)Y_k(\x)}
	-{\frac{T^2}{N^2}}\left(\frac{1}{2}\frac{\d{}^2}{\d{\x}^2}+\frac{1}{\x}\frac{\d{}}{\d{\x}}\right)
	\BK{W_k(\x)}&=\BK{D(\x)}\,,\quad \text{for $k=0,1,2$},
\end{split}
\end{equation}
where 
\begin{equation}
\begin{split}
	\BK{D(\x)}&=-\BK{Y_1(\x)^2{Y}_2(\x)+Y_1(\x){Y}_2(\x)^2}\\
	&=-U^\prime_0(\x)U^\prime_1(\x)U^\prime_2(\x)-U^\prime_1(\x)\BK{{P}_2(\x)}-U^\prime_2(\x)\BK{P_1(\x)}
	-\frac{1}{\x}\left(U^\prime_1(\x)\BK{{I}_2}+U^\prime_2\BK{I_1}\right)\\
	&\qquad +\BK{Q_1(\x)+{Q}_2(\x)}+\BK{W_1(\x)^2{W}_2(\x)}+\BK{W_1(\x){W}_2(\x)^2}+\BK{S_1(\x)}+\BK{{S}_2(\x)}.
\end{split}
	\label{Aeq:DDef}
\end{equation}

\subsection{Pole Structure of $D(x)$}

In order to find explicitly the pole structure at $x=0$ we need a new loop equation not derived in \cite{BePr-09.1}. Consider the following change of variables
{\footnotesize
\begin{equation}
 \begin{split}
  \delta M_1 =& \frac{T^3}{N^3}\left(\frac{1}{x-M_1}-\frac{1}{x}-\frac{M_1}{x^2}\right)
  \Tr{2}{\frac{1}{M_1\otimes\id+\id\otimes M_2}},\\
  \delta M_2 =& \frac{T^3}{N^3}\left(\frac{1}{-x-M_2}-\frac{1}{-x}-\frac{M_2}{x^2}\right)
  \Tr{1}{\frac{1}{M_1\otimes\id+\id\otimes M_2}},
 \end{split}
\end{equation} 
}
with the corresponding Jacobians
{\tiny
\begin{equation}
 \begin{split}
  J(\delta M_1)=& \frac{T^3}{N^3}\Bigg(\tr{\frac{1}{x-M_1}}\Tr{1}{\frac{1}{x-M_1}
  \Tr{2}{\frac{1}{M_1\otimes\id+\id\otimes M_2}}}
  -\Tr{2}{\Tr{1}{\frac{1}{x-M_1\otimes\id}\frac{1}{M_1\otimes\id+\id\otimes M_2}}
  \Tr{1}{\frac{1}{M_1\otimes\id+\id\otimes M_2}}}\\
  &+\frac{1}{x}\Tr{2}{\Tr{1}{\frac{1}{M_1\otimes\id+\id\otimes M_2}}
  \Tr{1}{\frac{1}{M_1\otimes\id+\id\otimes M_2}}}
  -\frac{1}{x^2}\Tr{2}{M_2\Tr{1}{\frac{1}{M_1\otimes\id+\id\otimes M_2}}
  \Tr{1}{\frac{1}{M_1\otimes\id+\id\otimes M_2}}}\Bigg),\\
  J(\delta M_2)=& \frac{T^3}{N^3}\Bigg(\tr{\frac{1}{-x-M_2}}\Tr{2}{\frac{1}{-x-M_2}
  \Tr{1}{\frac{1}{M_1\otimes\id+\id\otimes M_2}}}
  -\Tr{1}{\Tr{2}{\frac{1}{-x-\id\otimes M_2}\frac{1}{M_1\otimes\id+\id\otimes M_2}}
  \Tr{2}{\frac{1}{M_1\otimes\id+\id\otimes M_2}}}\\
  &+\frac{1}{-x}\Tr{1}{\Tr{2}{\frac{1}{M_1\otimes\id+\id\otimes M_2}}
  \Tr{2}{\frac{1}{M_1\otimes\id+\id\otimes M_2}}}
  -\frac{1}{x^2}\Tr{1}{M_1\Tr{2}{\frac{1}{M_1\otimes\id+\id\otimes M_2}}
  \Tr{2}{\frac{1}{M_1\otimes\id+\id\otimes M_2}}}\Bigg).
 \end{split}
\end{equation} 
}
Introducing this change of variables into \eqref{eq:LE2} we get (after some algebra and lots of cancellations)
{\scriptsize
\begin{equation}
 \begin{split}
  \BK{W_1(x)^2W_2(x)}&+\BK{W_1(x)W_2(x)^2} +\BK{S_1(x)}+\BK{S_2(x)}=\\
  &=\frac{1}{x}\left(\BK{\frac{T^2}{N^2}\Tr{1}{V_1^\prime(M^1)\Tr{2}{\frac{1}{M_1\otimes\id+\id\otimes M_2}}}}
  -\BK{\frac{T^2}{N^2}\Tr{2}{V_2^\prime(M^2)\Tr{1}{\frac{1}{M_1\otimes\id+\id\otimes M_2}}}}\right)\\
  &+\frac{1}{x^2}\left(\BK{\frac{T^2}{N^2}\Tr{1}{M_1V_1^\prime(M^1)
  \Tr{2}{\frac{1}{M_1\otimes\id+\id\otimes M_2}}}}
  +\BK{\frac{T^2}{N^2}\Tr{2}{M_2V_2^\prime(M^2)\Tr{2}{\frac{1}{M_1\otimes\id+\id\otimes M_2}}}}\right)\\
  &=\frac{1}{x}\BK{K_1}+\frac{1}{x^2}\BK{K_2}.
 \end{split}
\end{equation}  
}
This loop equation specifies the pole structure of $\BK{D(x)}$ which amounts to a degree two pole at most.

\section{Definitions on the algebraic curve}\label{App:AC}

In this appendix we will give a brief review of several definitions of objects on the algebraic curve.  The notions are relatively standard and we refer for example to \cite{FarkasKra, Faybook}.
\subsection{The algebraic curve}\label{sSec:AC}

Given an algebraic curve $E(x,y)=0$ (irreducible polynomial equation of degree $d_1$ in $x$ and $d_2$ in $y$), its points $(x,y)$ parametrize the points $p$ of a Riemann surface $\Sigma$. Choosing $x\in\C\cup\{0\}$ ($y\in\C\cup\{0\}$) as the base coordinate there are $d_2$ ($d_1$) solutions $y_i(x)$ ($x_i(y)$) of the algebraic curve. Those functions define $d_2$ $x$-sheets ($d_1$ $y$-sheets) which form a chart on the Riemann surface. 
Since, for every point $p\in\Sigma$ there is a pair $(x,y)$ that solve $E(x,y)=0$ we can define two meromorphic functions $x(p),y(p)$ on $\Sigma$.
For a generic $p=p^{(0)}\in\Sigma$ there are $d_2-1$ other points $p^{(i)},\,i=1,\dotsc,d_2-1$ such that $x(p^{(i)})=x(p)$. Each one of these points lies in a different $x$-sheet. In general, on these points we have $y(p^{(i)})\not=y(p^{(j)})$ for all pairs $i,j$.
The same definitions can be made based on $y$-sheets but we do not use it in this work.

\subsection{The branch points, branch cuts and locally conjugated points}\label{sSec:BP}

We will consider curves where $\d x(p)$ and $\d y(p)$ never vanish at the same point $p$ (i.e. without cusps).
The points where $\d x(p)=0$ are called branch points and are denoted by $\alpha_i$. 
We only  consider algebraic curves where the zeros of $\d x(p)$ are simple except at $x=0$ where we have a double zero.
On simple branch points, the sheets are connected pairwise by {\it cuts} that go from one branch point to another, every branch point having only one attached cut. The branch point at $x=0$ joins the three sheets together and belongs to two different cuts. We call $\Delta_1$ the ensemble of simple branch-points and $\Delta_2$ the ensemble of double branch points
In a neighborhood $U_\alpha$ of any branch-point $\alpha\in\Delta_i$, given a point $p\in U_\alpha$ there are $i$ points $p^{(1)},\dots,p^{(i)}\in \Sigma$ such that $x(p^{(j)}) = x(p)$ for $j=1,\dots,i$. 
The map $\vartheta_\alpha^{i}: U_\alpha \mapsto U_\alpha$ is the map defined moving the point $x(p)\in\C$ around the point $x(\alpha)=a$ a total $i$ times ending at the same value $x$ and looking at the corresponding trajectory on $U_\alpha$.
We will order the points $p^{(i)}$ defined above as $p^{(i)}=\vartheta_\alpha^i(p)$.
Note that the operation $\vartheta_\alpha^j$ is defined only local and has no intrinsic meaning at the global level. Obviously, for $\alpha\in\Delta_i$ we have the following properties:
\begin{equation}
\begin{split}
\vartheta_\alpha^0(p)&=\vartheta_\alpha^i=p  \\
\vartheta^j(\vartheta^k(p))&=\vartheta^{j+k}(p)=p^{(j+k \text{ mod } i)}
\end{split}
\end{equation}

\subsection{Cycles and genus}\label{sSec:CG}

In every Riemann surface we can choose a basis of cycles for the homology group. These cycles can be chosen to form a canonical basis, i.e. two types of cycles $\A_i,\B_i$ independent which satisfy the intersection conditions independent which satisfy the intersection conditions
\begin{equation}
	\begin{split}
		\A_i\cdot\A_j&=0=\B_i\cdot\B_j,\\
		\A_i\cdot\B_j&=-\B_j\cdot\A_i=\delta_{i,j}.
	\end{split}
	\label{eq:CCC}
\end{equation}  
A Riemann surface of genus $g$ has exactly $g$ pairs of conjugated cycles $\A_i,\B_i$.
Usually one defines the homotopy group by choosing related cycles $\overline{\A}$ and $\overline{\B}$. These cycles are equivalent to the previous ones with the constraint that all pass through a common point $p_0$. Cutting $\Sigma$ along $\overline{\A}$ and $\overline{\B}$ defines $\overline{\Sigma}$ which is isomorphic to an open region of the complex plane $\C$. This is called the canonical dissection of $\Sigma$

\subsection{Abelian differentials.}\label{sSec:AB}

There are three type of meromorphic differentials on an algebraic curve.
The {\bf differentials of the first kind} (holomorphic)  form a vector space of dimension $g$ and we denote them by $\d u_i$. We can always choose a basis $\d u_i(p)$ such that it satisfy
\begin{equation}
	\oint_{\A_j}\d u_i(p)=\delta_{i,j}.
	\label{eq:AD1}
\end{equation}
With this normalization we define the Riemann period matrix as the matrix of $\B$ periods
\begin{equation}
	\tau_{i,j}=\oint_{\B_i}\d u_j(p).
	\label{eq:RPM}
\end{equation}
This matrix is symmetric and its imaginary part is positive definite $\Im(\tau)>0$.
These differentials $\d u_i(p)$ are called normalized Abelian differentials of the first kind.

The Abelian {\bf differentials of the second kind} are defined as meromorphic differentials with poles but with zero residues. A basis for these differentials may be called $\d\Omega_k^{(q)}(p)$ which is a meromorphic differential with a pole at $p=q$ without residue normalized as follows
\begin{equation}
	\begin{split}
		\d\Omega_k^{(q)}(p)&\sim (-kz_q(p)^{-k-1}+\OO{1})\d z_q(p),\\
		\oint_{\A_i} \d\Omega_k^{(q)}(p) & =0,
	\end{split}
\end{equation}
in some local parameter $z_q(p)$ such that $z_q(q)=0$.

Finally the {\bf differentials of the  third kind} are meromorphic differentials that have only poles of first order. Again we will choose a basis for these differentials consisting of differentials $\d S_{q,r}(p)$ with only two poles of order one, say at $p=q,r$, with residue $1,-1$ respectively, and with vanishing $\A$ cycles, i.e. normalized following
\begin{equation}
	\begin{split}
		\d S_{q,r}(p)&\sim \left(\frac{1}{z_q(p)}+\OO{1}\right)\d z_q(p),\\
		\d S_{q,r}(p)&\sim \left(\frac{-1}{z_r(p)}+\OO{1}\right)\d z_r(p),\\
		\oint_{\A_i} \d S_{q,r}(p) & =0,
	\end{split}
\end{equation}
where the local coordinates are defined as above.
The differentials in this basis is called normalized Abelian differentials of the third kind.
\subsection{The fundamental bi-differential and relations with the Abelian differentials}

We also introduce the fundamental bi-differential (Bergman kernel) $B(p,q)$ which is a meromorphic bi-differential with poles only at $p=q$ of order two with zero residue, and normalized in the following way
\begin{equation}
	\begin{split}
		B(p,q)&\sim \frac{\d z(p)\d z(q)}{(z(p)-z(q))^2}+\OO{1},\\
		\oint_{\A_i}B(p,q)&=0 \qquad \text{for $i=1,\dotsc,g$}.
	\end{split}
\end{equation}
This bi-differential is fundamentally connected with the Abelian differentials presented above.
\begin{equation}
	\begin{split}
		B(p,q)&=B(q,p),\\
		\d f(p)&=\Res_{q\to p}B(p,q) f(q),\\
		\sum_{i}B(p^{(i)},q)&=\frac{\d x(q)\d x(p)}{(x(p)-x(q))^2},\\
		\oint_{\B_j}B(p,\cdot)&=2\pi i \d u_j(p),\\
		\d S_{q,r}(p)&=\int_r^q B(p,\cdot),\\
		\d \Omega_k^{(r)}(p) &= \Res_{q\to p} B(p,q) z_r(q)^{-k}=-\Res_{q\to r} B(p,q) z_r(q)^{-k},\\
		\oint_{\B_j}\d S_{q,r}(\cdot)&=2\pi i \int_r^q \d u_j(\cdot),\\
		\int_{t=p}^{t=q} \d S_{s,r}(t) &= \int_{t=r}^{t=s} \d S_{q,p}(t).
	\end{split}
	\label{eq:DifProp} 
\end{equation}

In all these equations, the integrations paths that are not closed cycles are constrained to not cross the $\A$ and $\B$ cycles, in other words belongs to the cut algebraic curve along the cycles.

\subsection{Riemann Bi-linear indetity}

Consider $\eta$ and $\omega$ two meromorphic differentials. Consider also $\Omega$, one of the anti-derivatives of $\omega$ based on a point $p_0$
\begin{equation}
	\Omega=\int_{p_0}^p \omega.
\end{equation}
Under these definitions the Riemann bilinear identity takes the form
\begin{equation}
	\sum_{\alpha}\Res_{p\to\alpha}\Omega(p)\eta(p) = 
	\frac{1}{2\pi i}\sum_{i,k}\left(\oint_{\A_i^{(k)}}\omega(p)\oint_{\B_i^{(k)}}\eta(p)
	-\oint_{\B_i^{(k)}}\omega(p)\oint_{\A_i^{(k)}}\eta(p)\right).
	\label{eq:RBI}
\end{equation}
This expression is the key for \eqref{eq:RBI.2}.

\subsection{Rauch Variational Formula}

Changes on the algebraic curve induce changes on the objects we have defined so far. The Rauch variational formula expresses how these changes affect the Bergman Kernel. 
Consider a generic variation $\delta_{\d\Omega}$,
\begin{equation}
\begin{split}
	\delta_{\d\Omega}B(p,q)&=\sum_{i=1}^2\sum_{\alpha\in\Delta_i}\sum_{j=1}^i
	\Res_{t\to\alpha}\frac{\d\Omega(t)B(t,p)B(t,q)}{\d y(t)\d x(t)}\\
	&=\sum_{i=1}^2\sum_{\alpha\in\Delta_i}\sum_{j=1}^i
	\Res_{t\to\alpha}\frac{\d S_{t,\vartheta^{j}{t}}\d\Omega(t)B(t,p)}{\omega(t,\vartheta^j(t))}.
\end{split}
\end{equation}
The situations in which we are interested are the variations with respect to the moduli of our model, in particular the with respect to the potential. In that case we have the variation can be written as
\begin{equation}
\begin{split}
	\d x(r)\frac{\partial}{\partial V(r)}B(p,q)
	&=-\sum_{i=1}^2\sum_{\alpha\in\Delta_i}\sum_{j=1}^i
	\Res_{t\to\alpha}\frac{\frac{1}{2}\d S_{t,\vartheta^j(t)}(q)}{\omega(t,\vartheta^j(t))}
	\left[B(t,r)B(\vartheta^j(t),p)+B(\vartheta^j(t),r)B(t,p)\right]\\
	&=-\sum_{i=1}^2\sum_{\alpha\in\Delta_i}\sum_{j=1}^i
	\Res_{t\to\alpha}\frac{\d S_{t,o}(q)}{\omega(t,\vartheta^j(t))}
	\left[B(t,r)B(\vartheta^j(t),p)+B(\vartheta^j(t),r)B(t,p)\right].
\end{split}
\end{equation}
It is now easy to find the variation of other objects like the third type Abelian differential $\d S_{a,b}(p)$
\begin{equation}
\begin{split}
	\d x(r)\frac{\partial}{\partial V(r)}\d S_{u,o}(q)
	&=-\sum_{i=1}^2\sum_{\alpha\in\Delta_i}\sum_{j=1}^i
	\Res_{t\to\alpha}\frac{\d S_{t,o}(q)}{\omega(t,\vartheta^j(t))}
	\left[B(t,r)\d S_{u,o}(\vartheta^j(t))+B(\vartheta^j(t),r)\d S_{u,o}(t)\right].
\end{split}
\end{equation}

\section{Details of the computations on sections \ref{eq:RFwkh} and \ref{Sec:Hwkh.2}}\label{app:Hrules}

Some of the calculations of sections \ref{eq:RFwkh} and \ref{Sec:Hwkh.2} are cumbersome. This appendix contains the details of these calculations.

\subsection{Calculation on section \ref{Sec:LOCV}}

The action of the loop insertion operator on the cubic vertex 
(see equation \eqref{eq:dVV}) consists of two main terms
{\scriptsize
\begin{equation}
	\begin{split}
		\d x(p)\frac{\partial}{\partial V(p)} &\Bigg(-\sum_{i=1}^2\sum_{\alpha\in\Delta_i}\sum_{j=1}^i 
		\Res_{t\to \alpha}\frac{\d S_{t,o}(q)}{\omega(t,\vartheta^j(t))}f(t,\vartheta^j(t))\Bigg)
		=\\
		&=\sum_{i=1}^2 \sum_{\alpha\in\Delta_i}\sum_{j=1}^i 
		\Bigg(\Res_{t\to \alpha}\frac{f(t,\vartheta^j(t))}{\omega(t,\vartheta^j(t))}
		\sum_{k=1}^2\sum_{\beta\in\Delta_k}\sum_{l=1}^k\Res_{u\to\beta} 
		\frac{\d S_{u,o}(q)}{\omega(u,\vartheta^l(u))}
		\left(B(u,p)\d S_{t,o}(\vartheta^l(u))+B(\vartheta^l(u),p)\d S_{t,o}(u)\right)\\
		&\hspace{100pt}-\Res_{t\to \alpha}\frac{\d S_{t,o}(q)f(t,\vartheta^j(t))}{(\omega(t,\vartheta^j(t)))^2}
		(B(t,p)-B(\vartheta^j(t),p))\Bigg),
	\end{split}
	\label{eq:A1.0}
\end{equation}
}
where we have overseen the action of the loop insertion operator on the function $f(a,b)$. It is important to notice as well that the function $f(a,b)$ is symmetric under the permutation of its variables by hypothesis.
We need to commute the residues of the second line. When the both residues are located on the same branch-point we use the relation 
\begin{equation}
	\Res_{t\to\alpha}\Res_{u\to\alpha} = \Res_{u\to\alpha}\Res_{t\to\alpha}
	-\Res_{t\to\alpha}\sum_{k=0}^i\Res_{u\to\vartheta^k(t)},
\end{equation}
where $\alpha\in\Delta_i$. Residues around different branch-points commute without the need of extra terms.
When the branch point $\alpha$ belongs to $\Delta_1$ we obtain
{\scriptsize
\begin{equation}
\begin{split}
	\Res_{t\to \alpha} \frac{f(t,\vartheta(t))}{\omega(t,\vartheta(t))}&
	\Res_{u\to\alpha} \frac{\d S_{u,o}(q)}{\omega(u,\vartheta(u))}
	\left(B(u,p)\d S_{t,o}(\vartheta(u))+B(\vartheta(u),p)\d S_{t,o}(u)\right)=\\
	=&\Res_{u\to\alpha}\frac{\d S_{u,o}(q)}{\omega(u,\vartheta(u))}\left[B(u,p)
	\Res_{t\to \alpha}\d S_{t,o}(\vartheta(u))\frac{f(t,\vartheta(t))}{\omega(t,\vartheta(t))}+
	B(\vartheta(u),p)\Res_{t\to \alpha}\d S_{t,o}(u)\frac{f(t,\vartheta(t))}{\omega(t,\vartheta(t))}\right]\\
	&-\Res_{t\to \alpha}\frac{f(t,\vartheta(t))}{\omega(t,\vartheta(t))}
	\Res_{u\to t,\vartheta(t)}\frac{\d S_{u,o}(q)}{\omega(u,\vartheta(u))}
	\left[B(u,p)\d S_{t,o}(\vartheta(u))+B(\vartheta(u),p)\d S_{t,o}(u)\right].
\end{split}
\label{eq:A1.1}
\end{equation}
}
we evaluate the residues in the last line and we find
{\scriptsize
\begin{equation}
\begin{split}
	-\Res_{t\to \alpha}\frac{f(t,\vartheta(t))}{\omega(t,\vartheta(t))}&
	\Res_{u\to t,\vartheta(t)}\frac{\d S_{u,o}(q)}{\omega(u,\vartheta(u))}
	\left[B(u,p)\d S_{t,o}(\vartheta(u))+B(\vartheta(u),p)\d S_{t,o}(u)\right]=\\
	=&-\Res_{t\to \alpha}\frac{f(t,\vartheta(t))}{\omega(t,\vartheta(t))}
	\left[-\frac{\d S_{\vartheta(t),o}(q)}{\omega(t,\vartheta(t))}B(\vartheta(t),p)+
	\frac{\d S_{t,o}(q)}{\omega(t,\vartheta(t))}B(\vartheta(t),p)\right]\\
	=&\Res_{t\to \alpha}\frac{\d S_{t,o}(q)f(t,\vartheta(t))}{\omega(t,\vartheta(t))^2}
	(B(t,p)-B(\vartheta(t),p)).
\end{split}
\nonumber
\label{eq:A1.2}
\end{equation}
}
When the branch-point under consideration belongs to $\Delta_2$ the commutation becomes more involved
{\scriptsize
\begin{equation}
\begin{split}
	\sum_{j=1}^2\Res_{t\to \alpha} &\left(\frac{f(t,\vartheta^j(t))}{\omega(t,\vartheta^j(t))}\right)
	\sum_{l=1}^2\Res_{u\to\alpha} \frac{\d S_{u,o}(q)}{\omega(u,\vartheta^l(u))}
	\left(B(u,p)\d S_{t,o}(\vartheta^l(u))+B(\vartheta^l(u),p)\d S_{t,o}(u)\right)=\\
	=&\sum_{l=1}^2\Res_{u\to\alpha}\frac{\d S_{u,o}(q)}{\omega(u,\vartheta^l(u))}
	\left[B(u,p)\sum_{i=1}^2\Res_{t\to \alpha}
	\frac{\d S_{t,o}(\vartheta^l(u))f(t,\vartheta^j(t))}{\omega(t,\vartheta^j(t))}
	+B(\vartheta^l(u),p)\sum_{i=1}^2\Res_{t\to \alpha}
	\frac{\d S_{t,o}(u)f(t,\vartheta^j(t))}{\omega(t,\vartheta^j(t))}\right]\\
	&\hspace{30pt}-\sum_{j=1}^2\Res_{t\to \alpha}\left(\frac{f(t,\vartheta^j(t))}{\omega(t,\vartheta^j(t))}\right)
	\sum_{l=1}^2\Res_{\begin{subarray}{c}u\to t\\u\to\vartheta^{-l}(t)\end{subarray}} 
	\frac{\d S_{u,o}(q)}{\omega(u,\vartheta^l(u))}
	\left(B(u,p)\d S_{t,o}(\vartheta^l(u))+B(\vartheta^l(u),p)\d S_{t,o}(u)\right).
\end{split}
\label{eq:A1.3}
\end{equation}
}
We evaluate the residues of the last line and simplify all the terms and we get
{\scriptsize
\begin{equation}
\begin{split}
	-\sum_{j=1}^2\Res_{t\to \alpha}&\left(\frac{f(t,\vartheta^j(t))}{\omega(t,\vartheta^j(t))}\right)
	\sum_{l=1}^2\Res_{\begin{subarray}{c}u\to t\\u\to\vartheta^{-l}(t)\end{subarray}} 
	\frac{\d S_{u,o}(q)}{\omega(u,\vartheta^l(u))}
	\left(B(u,p)\d S_{t,o}(\vartheta^l(u))+B(\vartheta^l(u),p)\d S_{t,o}(u)\right)\\
	=&-\sum_{j=1}^2\Res_{t\to \alpha}\left(\frac{f(t,\vartheta^j(t))}{\omega(t,\vartheta^j(t))}\right)
	\sum_{l=1}^2\left(
	-\frac{\d S_{\vartheta^l(t),o}(q)}{\omega(t,\vartheta^l(t))}B(\vartheta^l(t),p)
	+\frac{\d S_{t,o}(q)}{\omega(t,\vartheta^l(t))}B(\vartheta^l(t),p)\right)=\\
	=&-\Res_{t\to \alpha}\left(
	\frac{\d S_{t,\vartheta^2(t)}(q)f(t,\vartheta^1(t))B(\vartheta^2(t),p)}
	{\omega(t,\vartheta^1(t))\omega(t,\vartheta^2(t))}+
	\frac{\d S_{t,\vartheta^1(t)}(q)f(t,\vartheta^2(t))B(\vartheta^1(t),p)}
	{\omega(t,\vartheta^1(t))\omega(t,\vartheta^2(t))}\right)\\
	&\hspace{50pt}-\sum_{j=1}^2\Res_{t\to \alpha}
	\frac{\d S_{t,\vartheta^j(t)}(q)f(t,\vartheta^j(t))B(\vartheta^j(t),p)}{\omega(t,\vartheta^j(t))^2}\\
	=&-\Res_{t\to \alpha}
	\frac{\d S_{t,o}(q)}{\omega(t,\vartheta^1(t))\omega(t,\vartheta^2(t))}
	\left(B(t,p)f(\vartheta^1(t),\vartheta^2(t))+B(\vartheta^1(t),p)f(\vartheta^2(t),t)
	+B(\vartheta^2(t),p)f(t,\vartheta^1(t))\right)\\
	&\hspace{50pt}+\sum_{j=1}^2\Res_{t\to \alpha}
	\frac{\d S_{t,o}(q)f(t,\vartheta^j(t))}{\omega(t,\vartheta^j(t))^2}(B(t,p)-B(\vartheta^j(t),p)).	
\end{split}
\label{eq:A1.4}
\end{equation}
}
Putting all the contributions together, equation \eqref{eq:A1.0} becomes
{\scriptsize
\begin{equation}
\begin{split}
	\d x(p)\frac{\partial}{\partial V(p)} &\Bigg(-\sum_{i=1}^2\sum_{\alpha\in\Delta_i}\sum_{j=1}^i 
	\Res_{t\to \alpha}\frac{\d S_{t,o}(q)}{\omega(t,\vartheta^j(t))}f(t,\vartheta^j(t))\Bigg)=\\
	=&\sum_{k=1}^2\sum_{\beta\in\Delta_k}\sum_{l=1}^k\Res_{u\to\beta}\frac{\d S_{u,o}(q)}{\omega(u,\vartheta^l(u))}
	\Bigg[B(u,p)\sum_{i=1}^2 \sum_{\alpha\in\Delta_i}\sum_{j=1}^i
	\Res_{t\to \alpha}\frac{\d S_{t,o}(\vartheta^l(u))f(t,\vartheta^j(t))}{\omega(t,\vartheta^j(t))}\\
	&\hspace{140pt}+B(\vartheta^l(u),p)\sum_{i=1}^2 \sum_{\alpha\in\Delta_i}\sum_{j=1}^i
	\Res_{t\to \alpha}\frac{\d S_{t,o}(u)f(t,\vartheta^j(t))}{\omega(t,\vartheta^j(t))}\Bigg]\\
	&-\sum_{\alpha\in\Delta_2}\Res_{t\to \alpha}
	\frac{\d S_{t,o}(q)}{\omega(t,\vartheta^1(t))\omega(t,\vartheta^2(t))}
	\bigg(B(t,p)f(\vartheta^1(t),\vartheta^2(t))+B(\vartheta^1(t),p)f(\vartheta^2(t),t)
	+B(\vartheta^2(t),p)f(t,\vartheta^1(t))\bigg),
\end{split}
\label{eq:A1.5}
\end{equation}
}
where some terms from \eqref{eq:A1.2} and \eqref{eq:A1.4} have cancelled the last term of equation \eqref{eq:A1.0}.

\subsection{Calculation of section \ref{Sec:ALOQ}}

The action of the loop insertion operator on the quartic vertex (see equation \eqref{eq:dVVertex3}) is composed of two main terms
{\scriptsize
\begin{equation}
	\begin{split}
		\d x(p)\frac{\partial}{\partial V(p)} &\Bigg(-\sum_{\alpha\in\Delta_2} 
		\Res_{t\to \alpha}\frac{\d S_{t,o}(q)f(t,\vartheta^1(t),\vartheta^2(t))}
		{\omega(t,\vartheta^1(t))\omega(t,\vartheta^2(t))}
		\Bigg)=\\
		=&\sum_{\alpha\in\Delta_2} \Bigg(\Res_{t\to \alpha}
		\frac{f(t,\vartheta^1(t),\vartheta^2(t))}{\omega(t,\vartheta^1(t))\omega(t,\vartheta^2(t))}
		\sum_{k=1}^2\sum_{\beta\in\Delta_k}\sum_{l=1}^k\Res_{u\to\beta} 
		\frac{\d S_{u,o}(q)}{\omega(u,\vartheta^l(u))}
		\left(B(u,p)\d S_{t,o}(\vartheta^l(u))+B(\vartheta^l(u),p)\d S_{t,o}(u)\right)\\
		&-\Res_{t\to \alpha}\frac{\d S_{t,o}(q)f(t,\vartheta^1(t),\vartheta^2(t))} 
		{(\omega(t,\vartheta^1(t))\omega(t,\vartheta^2(t)))^2}
		\left(\omega(t,\vartheta^1(t))(B(t,p)-B(\vartheta^2(t),p))
		+\omega(t,\vartheta^2(t))(B(t,p)-B(\vartheta^1(t),p))\right)\Bigg),
	\end{split}
	\label{eq:A2.0}
\end{equation}}
where we do not consider the action of the loop insertion operator on the function is a symmetric function $f(a,b,c)$.
We must commute the double residue on the second line. The only situation that deserves some attention is when both branch-points coincide
{\scriptsize
\begin{equation}
\begin{split}
	\Res_{t\to \alpha}&\frac{f(t,\vartheta^1(t),\vartheta^2(t))}{\omega(t,\vartheta^1(t))\omega(t,\vartheta^2(t))}
	\sum_{l=1}^2\Res_{u\to\beta}\frac{\d S_{u,o}(q)}{\omega(u,\vartheta^l(u))}
	\left(B(u,p)\d S_{t,o}(\vartheta^l(u))+B(\vartheta^l(u),p)\d S_{t,o}(u)\right)=\\
	=&\sum_{l=1}^2\Res_{u\to\beta}\frac{\d S_{u,o}(q)}{\omega(u,\vartheta^l(u))}
	\left[B(u,p)\Res_{t\to \alpha}
	\frac{\d S_{t,o}(\vartheta^l(u))}{\omega(t,\vartheta^1(t))\omega(t,\vartheta^2(t))}
	+B(\vartheta^l(u),p)\Res_{t\to \alpha}
	\frac{\d S_{t,o}(u)}{\omega(t,\vartheta^1(t))\omega(t,\vartheta^2(t))}\right]f(t,\vartheta^1(t),\vartheta^2(t))\\
	&\hspace{20pt}
	-\Res_{t\to \alpha}\frac{f(t,\vartheta^1(t),\vartheta^2(t))}{\omega(t,\vartheta^1(t))\omega(t,\vartheta^2(t))}
	\sum_{l=1}^2\Res_{\begin{subarray}{c}u\to t\\ u\to\vartheta^{-l}(t)\end{subarray}}
	\frac{\d S_{u,o}(q)}{\omega(u,\vartheta^l(u))}
	\left(B(u,p)\d S_{t,o}(\vartheta^l(u))+B(\vartheta^l(u),p)\d S_{t,o}(u)\right).
\end{split}
\label{eq:A2.1}
\end{equation}
}
If we evaluate the residues in the last line and simplify the expression we get
{\scriptsize
\begin{equation}
\begin{split}
	-\Res_{t\to \alpha}\frac{f(t,\vartheta^1(t),\vartheta^2(t))}{\omega(t,\vartheta^1(t))\omega(t,\vartheta^2(t))}&
	\sum_{l=1}^2\Res_{\begin{subarray}{c}u\to t\\ u\to\vartheta^{-l}(t)\end{subarray}}
	\frac{\d S_{u,o}(q)}{\omega(u,\vartheta^l(u))}
	\left(B(u,p)\d S_{t,o}(\vartheta^l(u))+B(\vartheta^l(u),p)\d S_{t,o}(u)\right)\\
	=&-\Res_{t\to \alpha}\frac{f(t,\vartheta^1(t),\vartheta^2(t))}{\omega(t,\vartheta^1(t))\omega(t,\vartheta^2(t))}
	\sum_{l=1}^2\left(-\frac{\d S_{\vartheta^l(t),o}(q)}{\omega(t,\vartheta^l(t))}B(\vartheta^l(t),p)+
	\frac{\d S_{t,o}(q)}{\omega(t,\vartheta^l(t))}B(\vartheta^l(t),p)\right)\\
	=&-\Res_{t\to \alpha}\frac{f(t,\vartheta^1(t),\vartheta^2(t))}{\omega(t,\vartheta^1(t))\omega(t,\vartheta^2(t))}
	\sum_{l=1}^2\frac{\d S_{t,o}(q)}{\omega(t,\vartheta^l(t))}B(\vartheta^l(t),p)\\
	&\hspace{15pt}-\Res_{t\to \alpha}\frac{\d S_{t,o}(q)f(t,\vartheta^1(t),\vartheta^2(t))B(t,p)}
	{\omega(\vartheta^1(t),\vartheta^2(t))}
	\left(\frac{1}{\omega(t,\vartheta^2(t))^2}-\frac{1}{\omega(t,\vartheta^1(t))^2}\right),
\end{split}
\label{eq:A2.2}
\end{equation}
}
and using the relation
{\scriptsize
\begin{equation}
	\frac{1}{\omega(t,\vartheta^2(t))^2}-\frac{1}{\omega(t,\vartheta^1(t))^2}=
	\frac{\omega(\vartheta^1(t),\vartheta^2(t))(\omega(t,\vartheta^2(t))+\omega(t,\vartheta^2(t)))}
	{\omega(t,\vartheta^1(t))^2\omega(t,\vartheta^2(t))^2}
\end{equation}
}
we get the final expression
{\scriptsize
\begin{equation}
\begin{split}
	-\Res_{t\to \alpha}\frac{f(t,\vartheta^1(t),\vartheta^2(t))}{\omega(t,\vartheta^1(t))\omega(t,\vartheta^2(t))}&
	\sum_{l=1}^2\Res_{\begin{subarray}{c}u\to t\\ u\to\vartheta^{-l}(t)\end{subarray}}
	\frac{\d S_{u,o}(q)}{\omega(u,\vartheta^l(u))}
	\left(B(u,p)\d S_{t,o}(\vartheta^l(u))+B(\vartheta^l(u),p)\d S_{t,o}(u)\right)\\
	=&\Res_{t\to \alpha}\frac{f(t,\vartheta^1(t),\vartheta^2(t))}{\omega(t,\vartheta^1(t))\omega(t,\vartheta^2(t))}
	\sum_{l=1}^2\frac{\d S_{t,o}(q)}{\omega(t,\vartheta^l(t))}(B(t,p)-B(\vartheta^l(t),p)).
\end{split}
\label{eq:A2.3}
\end{equation}
}
Note that this cancels exactly the last line of \eqref{eq:A2.0}. Putting all contributions together the result is
{\scriptsize
\begin{equation}
	\begin{split}
		\d x(p)\frac{\partial}{\partial V(p)} &\Bigg(-\sum_{\alpha\in\Delta_2} 
		\Res_{t\to \alpha}\frac{\d S_{t,o}(q)f(t,\vartheta^1(t),\vartheta^2(t))}
		{\omega(t,\vartheta^1(t))\omega(t,\vartheta^2(t))}
		\Bigg)=\\
		=&\sum_{l=1}^2\Res_{u\to\beta}\frac{\d S_{u,o}(q)}{\omega(u,\vartheta^l(u))}
	\left[B(u,p)\Res_{t\to \alpha}
	\frac{\d S_{t,o}(\vartheta^l(u))}{\omega(t,\vartheta^1(t))\omega(t,\vartheta^2(t))}
	+B(\vartheta^l(u),p)\Res_{t\to \alpha}
	\frac{\d S_{t,o}(u)}{\omega(t,\vartheta^1(t))
	\omega(t,\vartheta^2(t))}\right]f(t,\vartheta^1(t),\vartheta^2(t)).
	\end{split}
	\label{eq:A2.4}
\end{equation}}

\subsection{Calculation of section \ref{Sec:Hwkh.2}}

The right hand side of equation \eqref{eq:A34} is composed of two terms, we are going to treat each of them independently. 
\subsubsection{First term}
The first term is
{\scriptsize
\begin{equation}
\begin{split}
	H_{x(p)}&\Bigg[-\sum_{i=1}^2\sum_{\alpha\in\Delta_i}\sum_{j=1}^i
		\Res_{t\to \alpha}\frac{\d S_{t,o}(q)} {\omega(t,\vartheta^j(t))}
		\left(w_2^{(0)}(t,p)\left(-\sum_{k=1}^2\sum_{\beta\in\Delta_k}\sum_{l=1}^k\Res_{u\to\beta}
		\frac{\d S_{u,0}(\vartheta^j(t))}{\omega(u,\vartheta^l(u))}f(u,\vartheta^l(u))\right)\right.
		\\&\hspace{120pt}
		\left.+w_2^{(0)}(\vartheta^j(t),p)\left(-\sum_{k=1}^2\sum_{\beta\in\Delta_k}\sum_{l=1}^k\Res_{u\to\beta}
		\frac{\d S_{u,0}(t)}{\omega(u,\vartheta^l(u))} 
		f(u,\vartheta^l(u))\right)\right)\\
		&-\sum_{\alpha\in\Delta_2}\Res_{t\to \alpha}\frac{\d S_{t,o}(q)}
		{\omega(p,\vartheta^1(p))\omega(p,\vartheta^2(p))}
		\left(w_2^{(0)}(t,p)f(\vartheta^1(t),\vartheta^2(t))
		+w_2^{(0)}(\vartheta^1(t),p)f(\vartheta^2(t),t)+
		w_2^{(0)}(\vartheta^2(t),p)f(t,\vartheta^1(t))\right)\Bigg]\\
		=&\sum_{i=1}^2\sum_{\alpha\in\Delta_i}\sum_{j=1}^i
		\Res_{t\to \alpha}\frac{\d S_{t,o}(q)} {\omega(t,\vartheta^j(t))}
		\left(y(t)\d x(t)\left(-\sum_{k=1}^2\sum_{\beta\in\Delta_k}\sum_{l=1}^k\Res_{u\to\beta}
		\frac{\d S_{u,0}(\vartheta^j(t))}{\omega(u,\vartheta^l(u))}f(u,\vartheta^l(u))\right)\right.
		\\&\hspace{120pt}
		\left.+y(\vartheta^j(t))\d x(t)\left(-\sum_{k=1}^2\sum_{\beta\in\Delta_k}\sum_{l=1}^k\Res_{u\to\beta}
		\frac{\d S_{u,0}(t)}{\omega(u,\vartheta^l(u))} 
		f(u,\vartheta^l(u))\right)\right)\\
		&+\sum_{\alpha\in\Delta_2}\Res_{t\to \alpha}\frac{\d S_{t,o}(q)}
		{\omega(p,\vartheta^1(p))\omega(p,\vartheta^2(p))}
		\left(y(t)\d x(t)f(\vartheta^1(t),\vartheta^2(t))
		+y(\vartheta^1(t))\d x(t)f(\vartheta^2(t),t)+
		y(\vartheta^2(t))\d x(t)f(t,\vartheta^1(t))\right).
\end{split}
\label{eq:A3.0}
\end{equation}
}
The next step is to commute the pairs of residues. When the two residues are evaluated on the same branch-point we use again formula \eqref{eq:CommRes} put into another form
\begin{equation}
	\Res_{t\to\alpha}\Res_{u\to\alpha} = \Res_{u\to\alpha}\Res_{t\to\alpha}
	+\Res_{u\to\alpha}\sum_{k=0}^i\Res_{t\to\vartheta^k(t)}\quad.
	\label{eq:CommRes2}
\end{equation}
Two situations occur depending on the branching number of the branch-point. When $\alpha\in\Delta_1$ we have
{\scriptsize
\begin{equation}
\begin{split}
		-\Res_{t\to \alpha}\frac{\d S_{t,o}(q)} {\omega(t,\vartheta(t))}&
		\left(y(t)\d x(t)\left(\Res_{u\to\alpha}
		\frac{\d S_{u,0}(\vartheta(t))}{\omega(u,\vartheta(u))}f(u,\vartheta(u))\right)
		+y(\vartheta(t))\d x(t)\left(\Res_{u\to\alpha}
		\frac{\d S_{u,0}(t)}{\omega(u,\vartheta(u))}f(u,\vartheta(u))\right)\right)=\\
		=&-\Res_{u\to\alpha}\frac{f(u,\vartheta(u))}{\omega(u,\vartheta(u))}
		\Res_{t\to\alpha}\frac{\d S_{t,o}(q)\d x(t)}{\omega(t,\vartheta(t))}
		\left(y(t)\d S_{u,o}(\vartheta(t))+y(\vartheta(t))\d S_{u,o}(t)\right)\\
		&-\Res_{u\to\alpha}\frac{f(u,\vartheta(u))}{\omega(u,\vartheta(u))}\Res_{t\to u,\vartheta(u)}
		\frac{\d S_{t,o}(q)\d x(t)}{\omega(t,\vartheta(t))}
		\left(y(t)\d S_{u,o}(\vartheta(t))+y(\vartheta(t))\d S_{u,o}(t)\right).
\end{split}
\label{eq:A3.1}
\end{equation}
}
The first line gives zero due to the $\Res_{t\to\alpha}$. 
Evaluating the residues on the second line gives
{\scriptsize
\begin{equation}
\begin{split}
	-\Res_{u\to\alpha}\frac{f(u,\vartheta(u))}{\omega(u,\vartheta(u))}
	\left(-\frac{\d S_{\vartheta(u),o}(q)\d x(u)}{\omega(u,\vartheta(u))}y(\vartheta(u))
	+\frac{\d S_{u,o}(q)\d x(u)}{\omega(u,\vartheta(u))}y(\vartheta(u))\right)=
	\Res_{u\to\alpha}\frac{\d S_{u,o}(q)}{\omega(u,\vartheta(u))}f(u,\vartheta(u)).
\end{split}
\label{eq:A3.2}
\end{equation}
}
When the branch-point has branching number $n_b=2$ the calculation becomes more involved,
{\scriptsize
\begin{equation}
\begin{split}
	\sum_{j=1}^2\Res_{t\to \alpha}&\frac{\d S_{t,o}(q)} {\omega(t,\vartheta^j(t))}
	\left(y(t)\d x(t)\left(-\sum_{l=1}^2\Res_{u\to\alpha}
	\frac{\d S_{u,0}(\vartheta^j(t))}{\omega(u,\vartheta^l(u))}f(u,\vartheta^l(u))\right)
	+y(\vartheta^j(t))\d x(t)\left(-\sum_{l=1}^2\Res_{u\to\alpha}
	\frac{\d S_{u,0}(t)}{\omega(u,\vartheta^l(u))}f(u,\vartheta^l(u))\right)\right)\\
	&+\Res_{t\to \alpha}\frac{\d S_{t,o}(q)}{\omega(p,\vartheta^1(p))\omega(p,\vartheta^2(p))}
	\left(y(t)\d x(t)f(\vartheta^1(t),\vartheta^2(t))
	+y(\vartheta^1(t))\d x(t)f(\vartheta^2(t),t)+	y(\vartheta^2(t))\d x(t)f(t,\vartheta^1(t))\right).
\end{split}
\label{eq:A3.3}
\end{equation}
}
To commute the residues of the first line we use equation \eqref{eq:CommRes2} again
{\scriptsize
\begin{equation}
\begin{split}
	-\sum_{j=1}^2\Res_{t\to \alpha}\frac{\d S_{t,o}(q)} {\omega(t,\vartheta^j(t))}&
	\left(y(t)\d x(t)\left(\sum_{l=1}^2\Res_{u\to\alpha}
	\frac{\d S_{u,0}(\vartheta^j(t))}{\omega(u,\vartheta^l(u))}f(u,\vartheta^l(u))\right)
	+y(\vartheta^j(t))\d x(t)\left(\sum_{l=1}^2\Res_{u\to\alpha}
	\frac{\d S_{u,0}(t)}{\omega(u,\vartheta^l(u))}f(u,\vartheta^l(u))\right)\right)\\
	=&-\sum_{l=1}^2\Res_{u\to\alpha}\frac{f(u,\vartheta^l(u))}{\omega(u,\vartheta^l(u))}
	\sum_{j=1}^2\Res_{t\to \alpha}\frac{\d S_{t,o}(q)\d x(t)} {\omega(t,\vartheta^j(t))}
	\left(\d S_{u,0}(\vartheta^j(t))y(t)+y(\vartheta^j(t))\d S_{u,0}(t)\right)\\
	&-\sum_{l=1}^2\Res_{u\to\alpha}\frac{f(u,\vartheta^l(u))}{\omega(u,\vartheta^l(u))}
	\sum_{j=1}^2\Res_{t\to u,\vartheta^{-j}(u)}\frac{\d S_{t,o}(q)\d x(t)} {\omega(t,\vartheta^j(t))}
	\left(\d S_{u,0}(\vartheta^j(t))y(t)+y(\vartheta^j(t))\d S_{u,0}(t)\right).
\end{split}
\label{eq:A3.4}
\end{equation}
}
As before, the first line is zero due to the $\Res_{t\to\alpha}$. The second term, after evaluating the residues is
{\scriptsize
\begin{equation}
\begin{split}
	-\sum_{l=1}^2\Res_{u\to\alpha}\frac{f(u,\vartheta^l(u))}{\omega(u,\vartheta^l(u))}&
	\sum_{j=1}^2\Res_{t\to u,\vartheta^{-j}(u)}\frac{\d S_{t,o}(q)\d x(t)} {\omega(t,\vartheta^j(t))}
	\left(\d S_{u,0}(\vartheta^j(t))y(t)+y(\vartheta^j(t))\d S_{u,0}(t)\right)\\
	=&-\sum_{l=1}^2\Res_{u\to\alpha}\frac{f(u,\vartheta^l(u))}{\omega(u,\vartheta^l(u))}
	\sum_{j=1}^2\left(-\frac{\d S_{\vartheta^j(u),o}(q)\d x(u)}{\omega(u,\vartheta^j(u))}y(\vartheta^j(u))
	+\frac{\d S_{u,o}(q)\d x(u)}{\omega(u,\vartheta^j(u))}y(\vartheta^j(u))\right)\\
	=&-\sum_{l=1}^2\Res_{u\to\alpha}\frac{f(u,\vartheta^l(u))}{\omega(u,\vartheta^l(u))^2}
	\left(-\d S_{\vartheta^l(u),o}(q)y(\vartheta^l(u))+\d S_{u,o}(q)y(\vartheta^l(u))\right)\\
	&\hspace{20pt}-\Res_{u\to\alpha}\frac{\d S_{u,o}(q)\d x(u)}{\omega(u,\vartheta^1(u))\omega(u,\vartheta^2(u))}
	\left(y(\vartheta^1(u))f(u,\vartheta^2(u))+y(\vartheta^1(u))f(u,\vartheta^2(u))\right)\\
	&\hspace{20pt}+\Res_{u\to\alpha}\frac{\d x(u)}{\omega(u,\vartheta^1(u))\omega(u,\vartheta^2(u))}
	\left(\d S_{\vartheta^1(u),o}(q)y(\vartheta^1(u))f(u,\vartheta^2(u))+
	\d S_{\vartheta^2(u),o}(q)y(\vartheta^2(u))f(u,\vartheta^1(u))\right)\\
	=&+\sum_{l=1}^2\Res_{u\to\alpha}\frac{\d S_{u,o}(q)}{\omega(u,\vartheta^l(u))}f(u,\vartheta^l(u))
	-\Res_{u\to\alpha}\frac{\d S_{u,o}(q)\d x(u)}{\omega(u,\vartheta^1(u))\omega(u,\vartheta^2(u))}
	\left(y(\vartheta^1(u))f(u,\vartheta^2(u))+y(\vartheta^2(u))f(u,\vartheta^1(u))\right)\\
	&+\Res_{u\to\alpha}\frac{y(u)\d x(u)f(\vartheta^1(u),\vartheta^2(u))\d S_{u,o}(q)}
	{\omega(\vartheta^1(u),\vartheta^2(u))}
	\left(-\frac{1}{\omega(u,\vartheta^1(u))}+\frac{1}{\omega(u,\vartheta^2(u))}\right)\\
	=&\sum_{l=1}^2\Res_{u\to\alpha}\frac{\d S_{u,o}(q)}{\omega(u,\vartheta^l(u))}f(u,\vartheta^l(u))\\
	&-\Res_{u\to\alpha}\frac{\d S_{u,o}(q)\d x(u)}{\omega(u,\vartheta^1(u))\omega(u,\vartheta^2(u))}
	\left(y(\vartheta^1(u))f(u,\vartheta^2(u))+y(\vartheta^2(u))f(u,\vartheta^1(u))
	+y(u)f(\vartheta^1(u),\vartheta^2(u))\right).
\end{split}
\label{eq:A3.5}
\end{equation}
}
Notice that the last term cancel exactly the last term in \eqref{eq:A3.0}. 
Putting together contributions from the $\Delta_1$ and $\Delta_2$ branch-points we have
{\scriptsize
\begin{equation}
\begin{split}
	H_{x(p)}&\Bigg[-\sum_{i=1}^2\sum_{\alpha\in\Delta_i}\sum_{j=1}^i
	\Res_{t\to \alpha}\frac{\d S_{t,o}(q)} {\omega(t,\vartheta^j(t))}
	\left(w_2^{(0)}(t,p)\left(-\sum_{k=1}^2\sum_{\beta\in\Delta_k}\sum_{l=1}^k\Res_{u\to\beta}
	\frac{\d S_{u,0}(\vartheta^j(t))}{\omega(u,\vartheta^l(u))}f(u,\vartheta^l(u))\right)\right.
	\\&\hspace{120pt}
	\left.+w_2^{(0)}(\vartheta^j(t),p)\left(-\sum_{k=1}^2\sum_{\beta\in\Delta_k}\sum_{l=1}^k\Res_{u\to\beta}
	\frac{\d S_{u,0}(t)}{\omega(u,\vartheta^l(u))} 
	f(u,\vartheta^l(u))\right)\right)\\
	&-\sum_{i=1}^2\sum_{\alpha\in\Delta_i}\sum_{j=1}^i\Res_{t\to \alpha}\frac{\d S_{t,o}(q)}
	{\omega(p,\vartheta^1(p))\omega(p,\vartheta^2(p))}
	\left(w_2^{(0)}(t,p)f(\vartheta^1(t),\vartheta^2(t))
	+w_2^{(0)}(\vartheta^1(t),p)f(\vartheta^2(t),t)+
	w_2^{(0)}(\vartheta^2(t),p)f(t,\vartheta^1(t))\right)\Bigg]\\
	&\hspace{50pt}=\sum_{l=1}^2\Res_{u\to\alpha}\frac{\d S_{u,o}(q)}{\omega(u,\vartheta^l(u))}f(u,\vartheta^l(u)).
\end{split}
\label{eq:A3.6}
\end{equation}
}
\subsubsection{Second term}
The second term is
{\scriptsize
\begin{equation}
\begin{split}
	H_{x(p)}\Bigg[-\sum_{i=1}^2\sum_{\alpha\in\Delta_i}\sum_{j=1}^i&
	\Res_{t\to \alpha}\frac{\d S_{t,o}(q)} {\omega(t,\vartheta^j(t))}
	\Bigg(w_2^{(0)}(t,p)\left(-\sum_{\beta\in\Delta_2}\Res_{u\to\beta}
	\frac{\d S_{u,0}(\vartheta^j(t))}{\omega(u,\vartheta^1(u))\omega(u,\vartheta^2(u))}
	f(u,\vartheta^1(u),\vartheta^2(u))\right)
	\\&\hspace{80pt}
	+w_2^{(0)}(\vartheta^j(t),p)\left(-\sum_{\beta\in\Delta_k}\Res_{u\to\beta}
	\frac{\d S_{u,0}(t)}{\omega(u,\vartheta^1(u))\omega(u,\vartheta^2(u))} 
	f(u,\vartheta^1(u),\vartheta^2(u))\right)\Bigg)\Bigg]\\
	=&\sum_{i=1}^2\sum_{\alpha\in\Delta_i}\sum_{j=1}^i
	\Res_{t\to \alpha}\frac{\d S_{t,o}(q)} {\omega(t,\vartheta^j(t))}
	\left(y(t)\d x(t)\left(-\sum_{\beta\in\Delta_2}\Res_{u\to\beta}
	\frac{\d S_{u,0}(\vartheta^j(t))}{\omega(u,\vartheta^1(u))\omega(u,\vartheta^2(u))}
	f(u,\vartheta^1(u),\vartheta^2(u))\right)
	\right.\\&\hspace{80pt}\left.
	+y(\vartheta^j(t))\d x(t)\left(-\sum_{\beta\in\Delta_k}\Res_{u\to\beta}
	\frac{\d S_{u,0}(t)}{\omega(u,\vartheta^1(u))\omega(u,\vartheta^2(u))} 
	f(u,\vartheta^1(u),\vartheta^2(u))\right)\right).
\end{split}
\label{eq:A3.7}
\end{equation}
}
Again, to commute the residues we just need to consider coinciding branch-points
{\scriptsize
\begin{equation}
\begin{split}
	-\sum_{j=1}^2\Res_{t\to \alpha}&\frac{\d S_{t,o}(q)\d x(t)} {\omega(t,\vartheta^j(t))}
	\left(y(t)\Res_{u\to\alpha}\frac{\d S_{u,0}(\vartheta^j(t))f(u,\vartheta^1(u),\vartheta^2(u))}
	{\omega(u,\vartheta^1(u))\omega(u,\vartheta^2(u))}
	+y(\vartheta^j(t))\Res_{u\to\alpha}
	\frac{\d S_{u,0}(t)f(u,\vartheta^1(u),\vartheta^2(u))}{\omega(u,\vartheta^1(u))\omega(u,\vartheta^2(u))} 
	\right)\\
	=&-\Res_{u\to\alpha}\frac{f(u,\vartheta^1(u),\vartheta^2(u))}{\omega(u,\vartheta^1(u))\omega(u,\vartheta^2(u))}
	\sum_{j=1}^2\Res_{t\to \alpha}\frac{\d S_{t,o}(q)\d x(t)} {\omega(t,\vartheta^j(t))}
	\left(y(t)\d S_{u,0}(\vartheta^j(t))+y(\vartheta^j(t))\d S_{u,0}(t)\right)\\
	&-\Res_{u\to\alpha}\frac{f(u,\vartheta^1(u),\vartheta^2(u))}{\omega(u,\vartheta^1(u))\omega(u,\vartheta^2(u))}
	\sum_{j=1}^2\Res_{t\to u,\vartheta^{-j}(u)}\frac{\d S_{t,o}(q)\d x(t)} {\omega(t,\vartheta^j(t))}
	\left(y(t)\d S_{u,0}(\vartheta^j(t))+y(\vartheta^j(t))\d S_{u,0}(t)\right).
\end{split}
\label{eq:A3.8}
\end{equation}
}
The first line is equal to zero once more and the second becomes
{\scriptsize
\begin{equation}
\begin{split}
	-\Res_{u\to\alpha}&\frac{f(u,\vartheta^1(u),\vartheta^2(u))}{\omega(u,\vartheta^1(u))\omega(u,\vartheta^2(u))}
	\sum_{j=1}^2\frac{y(\vartheta^j(u))\d x(u)} {\omega(u,\vartheta^j(u))}
	\left(-\d S_{\vartheta^j(u),o}(q)+\d S_{u,o}(q)\right)=\\
	=&-\Res_{u\to\alpha}
	\frac{\d S_{u,o}(q)f(u,\vartheta^1(u),\vartheta^2(u))\d x(u)}{\omega(u,\vartheta^1(u))\omega(u,\vartheta^2(u))}
	\sum_{j=1}^2\frac{y(\vartheta^j(u))}{\omega(u,\vartheta^j(u))}\\
	&\hspace{20pt}+\Res_{u\to\alpha}
	\frac{\d S_{u,o}(q)f(u,\vartheta^1(u),\vartheta^2(u))\d x(u)}{\omega(\vartheta^1(u),\vartheta^2(u))}
	y(u)\left(\frac{1}{\omega(u,\vartheta^1(u))^2}-\frac{1}{\omega(u,\vartheta^2(u))^2}\right)\\
	=&\Res_{u\to\alpha}
	\frac{\d S_{u,o}(q)f(u,\vartheta^1(u),\vartheta^2(u))}{\omega(u,\vartheta^1(u))\omega(u,\vartheta^2(u))}
	\sum_{j=1}^2\frac{(y(u)-y(\vartheta^j(u)))\d x(u)}{\omega(u,\vartheta^j(u))}\\
	=&2\Res_{u\to\alpha}
	\frac{\d S_{u,o}(q)f(u,\vartheta^1(u),\vartheta^2(u))}{\omega(u,\vartheta^1(u))\omega(u,\vartheta^2(u))}.
\end{split}
\label{eq:A3.9}
\end{equation}
}

\bibliographystyle{unsrt}
\bibliography{Biblio}
\end{document}